\documentclass[11pt, hidelinks]{article}

\usepackage{graphicx, bbm, setspace, longtable, lscape, verbatim, dcolumn, natbib, multirow, hhline, xr, amsmath, authblk, fullpage, amsthm, amssymb, ifthen, mathrsfs, float, xcolor, authblk, tikz, etoolbox, wrapfig, subcaption, bibentry, eurosym, amsthm, pdfpages, rotating, adjustbox, endnotes, mathtools, thmtools, thm-restate, hyperref, mdframed}
\declaretheorem[name=Proposition]{proposition}

\declaretheorem[name=Lemma]{lemma}
\declaretheorem[name=Corollary]{corollary}

\declaretheoremstyle[
headfont=\normalfont\bfseries, 
bodyfont = \normalfont,
qed=$\square$
]{simpleQED}

\declaretheoremstyle[
headfont=\normalfont\bfseries, 
bodyfont = \normalfont,
]{simple}
\newmdtheoremenv{algo}{Algorithm}

\nobibliography*

\newcommand{\hideText}[1]{}

\newcommand{\indicator}[1]{\mathbbm{1}\{#1\}}

\newcommand{\iid}{$i.i.d.$}

\DeclareMathOperator*{\argmin}{arg\,min}

\DeclareMathOperator*{\barBreak}{\,|\,}

\DeclareMathOperator*{\E}{E}

\definecolor{deepBlue}{rgb}{0.2, .2, .7}
\definecolor{LightGrey}{rgb}{0.9, .9, .9}

\newcolumntype{g}{D{.}{.}{3}}

\newcommand{\cites}[1]{\citeauthor{#1}'s \citeyearpar{#1}}

\newcommand{\citeauthors}[1]{\citeauthor{#1}'s}

\ifdefined\largePage
\fi

\newcommand{\aq}{\text{and}\quad}
\newcommand{\wq}{\text{where}\quad}

\newcommand{\q}[1]{
	\newcount\quadCount
	\quadCount=#1
	\loop
	\quad 
	\advance \quadCount -1
	\ifnum \quadCount>0	
	\repeat
}

\newcommand{\vertiiSingle}{\vert\kern-0.25ex\vert}

\newcommand{\vertii}[1]{
	{\vert\kern-0.25ex\vert #1 \vert\kern-0.25ex\vert}
}

\newcommand{\vertiii}[1]{
	{\vert\kern-0.25ex\vert\kern-0.25ex\vert #1 
		\vert\kern-0.25ex\vert\kern-0.25ex\vert}
}

\newcommand{\vertiib}[1]{
	{\left\vert\kern-0.25ex\left\vert #1 
		\right\vert\kern-0.25ex\right\vert}
}
\newcommand{\vertiiib}[1]{
	{\left\vert\kern-0.25ex\left\vert\kern-0.25ex\left\vert #1 
		\right\vert\kern-0.25ex\right\vert\kern-0.25ex\right\vert}
}
\usepackage{algorithm}
\usepackage{algpseudocode}
\newtheorem{thm}{Theorem}

\begin{document}
\author{Robert L. Bray}
\affil{Kellogg School of Management, Northwestern University}

\title{An $\Omega(\log(N)/N)$ Lookahead is Sufficient \\ to Bound Costs in the Overloaded Loss Network}
\maketitle

\begin{abstract}
	\singlespacing \noindent
	I study the simplest model of revenue management with reusable resources: admission control of two customer classes into a loss queue. This model's long-run average collected reward has two natural upper bounds: the deterministic relaxation and the full-information offline problem. With these bounds, we can decompose the costs faced by the online decision maker into (i) the \emph{cost of variability}, given by the difference between the deterministic value and the offline value, and (ii) the \emph{cost of uncertainty}, given by the difference between the offline value and the online value. \cite{Xie2025} established that the sum of these two costs is $\Theta(\log N)$, as the number of servers, $N$, goes to infinity. I show that we can entirely attribute this $\Theta(\log N)$ rate to the cost of uncertainty, as the cost of variability remains $O(1)$ as $N \rightarrow \infty$. In other words, I show that anticipating future fluctuations is sufficient to bound operating costs---smoothing out these fluctuations is unnecessary. In fact, I show that an $\Omega(\log(N)/N)$ lookahead window is sufficient to bound operating costs.
\end{abstract}

%%%%%%%%%%%% Notes

%See notes_on_lower_bound.txt for notes on creating a lower bound.

%Don't forget to work on paper titled "kwong"! This is an important follow-up. Just to remind you: how Kwong's paper does it is totally screwed up, and he's comparing apples and oranges.

% Ask Zizoue to be a reviewer. Or perhaps Christos or Cong Shi at Miami?

%%%%%%%%%%%%

\section{Introduction}

\cite{Xie2025} sharply characterized the regret of the overloaded loss queuing network. The simplest version of their model comprises $N$ servers that service two populations: high-type customers ($H$) who pay a premium, and low-type customers ($L$) who pay a discount. The system prioritizes $H$ jobs, and uses its excess capacity to satisfy $L$ jobs (demand and supply grow in proportion, with the $H$- and $L$-job arrival rates scaling with $N$). Unfortunately, it's difficult to anticipate whether accepting a given $L$ job will necessitate turning away a future $H$ job, so the system controller will invariably incur both Type I errors, rejecting $L$ customers who would not pre-empt future $H$ customers, and Type II errors, accepting $L$ jobs who do pre-empt future $H$ customers. 

\citeauthors{Xie2025} analysis suggests that the cost of these errors is $\Theta(\log N)$: benchmarking the model to an LP fluid approximation, in which customers flow deterministically through the system, they find that ``in the overloaded regime, log $N$ captures precisely the cost of stochasticity." However, it's not clear what ``stochasticity" means in this context, as it could refer to Type-I and -II errors, or to more intrinsic volatility. For examples of the latter mechanism, note that the system will reject at least one $H$ job when $N + 1$ type-$H$ jobs arrive before any of them depart, and that the system will have at least one idle server between the departure of one job and the arrival of the next. These inefficiencies do not plague the fluid approximation, which smears out such fluctuations, but they also do not represent ``mistakes" that the system controller would later regret. \citet[p. 2098]{Xie2025} conclude their article by conceding that the absence of variability in their benchmark might account for their $\Theta(\log N)$ performance gap:
\begin{quote}
	It is important that we benchmarked our policy against an LP upper bound (a deterministic counterpart). In the dynamic stochastic knapsack setting (finite horizon with nonreusable resources), the offline decision maker---one that sees the future realization of demand---was used as a benchmark; see \cite{Arlotto2019} and \cite{Vera2021a}. The offline objective value is a tighter upper bound than the LP. In that setting, the offline decision maker solves an LP with a random right-hand side. In the case of reusable resources, the offline problem---in which arrivals and service times are known to the decision maker---is a complicated dynamic program and, hence, difficult to use as a benchmark. We use the LP, instead. It is natural to ask whether (i) one can do better relative to the offline upper bound ... [This question was] answered affirmatively in the dynamic stochastic knapsack setting. ... Indeed---with nondegeneracy---this offline version is O(1) from the LP, whereas we prove that no online policy can achieve sublogarithmic regret.
\end{quote}

This quotation is difficult to interpret out of context, so let me explain. The authors begin their article by delineating between two types of resource allocation problems: stochastic knapsack problems, in which customers consume resources permanently, and reusable resource problems, in which customers occupy resources only while they're in service. They explain that for stochastic knapsack problems, the standard benchmarking model is not the fluid approximation, but rather the offline decision problem, in which all arrival and service times are apparent upfront. This offline problem is simple for stochastic knapsacks---in which case it is ``an LP with a random right-hand side"---but is intractable for reusable resources---in which case it is ``a complicated dynamic program." To avoid this complexity, they benchmark against the fluid approximation, rather than the offline problem. However, they concede that doing so could yield an artificially large performance gap, since ``The offline objective value is a tighter upper bound than the [fluid] LP." After all, the offline--online gap is only $O(1)$ for the analogous stochastic knapsack problem \citep{Arlotto2019}.

Whether the fluid LP provides a competitive benchmark matters, as the relative tightness of the offline and LP bounds indicates what drives costs: variability or uncertainty. Indeed, the offline problem is simply the online problem stripped of its uncertainty (but not its variability). Hence, if the offline value is near the online value, then the $\Theta(\log N)$ gap \citeauthor{Xie2025} identify must arise from irreducible variability; conversely, if the offline value is near the fluid value, then the gap must arise from sample path unpredictability. 

In an article that inspired this work, \citet[p. 2091]{Spencer2014} detail the benefit of distinguishing between variability-induced and uncertainty-induced costs:
\begin{quote}
	Two important ingredients often make the design and analysis of a queueing system difficult: the demands and resources can be both variable and unpredictable. Variability refers to the fact that the arrivals of demands or the availability of resources can be highly volatile and nonuniformly distributed across the time horizon. Unpredictability means that such nonuniformity ``tomorrow" is unknown to the decision maker ``today," and she is obliged to make allocation decisions only based on the state of the system at the moment, and some statistical estimates of the future.
	
	While the world will remain volatile as we know it, in many cases, the amount of unpredictability about the future may be reduced thanks to forecasting technologies and the increasing accessibility of data. For instance:
	
	(1) advance booking in the hotel and textile industries allows forecasting of demands ahead of time [9];
	
	(2) the availability of monitoring data enables traffic controllers to predict the traffic pattern around potential bottlenecks [18]; 
	
	(3) advance scheduling for elective surgeries could inform care providers several weeks before the intended appointment [12].
	
	In all of these examples, future demands remain \emph{exogenous} and variable, yet the decision maker is revealed with (some of) their realizations.
	
	\emph{Is there significant performance gain to be harnessed by ``looking into the future?"}
\end{quote}

\citeauthor{Spencer2014} outline a new research program: quantifying the future information required to ensure operating costs remain bounded as traffic intensity increases. Almost all of the work in this vein corresponds to some version of the M/M/1 queue \citep[e.g., see][]{Xu2015, Xu2016, delana2021proactive}. I extend the paradigm to a completely new setting: the loss queuing network described above. This latter model has more complex dynamics, as the state of the loss network (the number of jobs in service) influences the service rate, whereas the state of the M/M/1 queue (the number of jobs waiting) does not influence the service rate. Further, the constant drift of the overloaded M/M/1 queue permits a simple acceptance rule: admit a job if the queue will at some point return to the current level when all future jobs are accepted. The loss queuing network admits no such rule, because the drift in its idle server count switches from positive to negative as customers depart the system, which makes the system positive recurrent.

We can gauge the relative complexity of the loss network and the admission-controlled M/M/1 by comparing their simplest Markovian representations. As \citet[p. 2097]{Spencer2014} explain, the key difficulty introduced by future information is the loss of memorylessness:
\begin{quote}
	One challenge that arises as one tries to move beyond the online setting is that policies with lookahead typically do not admit a clean Markov description, and hence common techniques for analyzing Markov decision problems do not easily apply. 
\end{quote}
In effect, the offline problem has an infinite-dimensional state space, as all future arrivals and service times are relevant. Analyzing the system thus first requires identifying a filtration under which the offline dynamics evolve according to a finite-state Markov chain. The Markov chain induced by \cites{Spencer2014} offline policy is simply a random walk that increments with probability $\frac{1-p}{\lambda + 1 - p}$ and decrements with probability $\frac{\lambda}{\lambda + 1 - p}$. In contrast, the Markov chain associated with my offline policy evolves according to the law
\begin{align*}
	\mathcal{L}\big(\xi \barBreak \xi \in \mathcal{S}_{\mathsf{U}^{\ell}}(\tau_{1}^{\ell}<\tau_{\lfloor\sqrt{N}\rfloor}^{\ell})
	\cap
	\mathcal{S}_{\mathsf{P}^{\ell}\cap\mathsf{W}^{\ell}}(\hat{\tau}_{\lfloor\sqrt{N}\rfloor-1}^{\ell}<\hat{\tau}_{0}^{\ell})
	\cap
	\mathcal{S}_{\mathsf{P}^{\ell}\cap\mathsf{W}^{\ell c}}(\tau_{\lfloor\sqrt{N}\rfloor}^{\ell}<\tau_{1}^{\ell}) \big),
\end{align*}
whose characterization requires the introduction of more than a dozen auxiliary objects, such as event $\mathsf{P}^{\ell} \equiv \big\{a^{\ell} < \underleftarrow{\tau}_{\lfloor \sqrt{N}\rfloor}^{\ell} , \ \ a^{\ell}\wedge s^{\ell} < \underleftarrow{\tau}_{\lfloor \sqrt{N}\rfloor-1}(t^{\ell}\wedge(t^{\ell}-a^{\ell}+s^{\ell}))\big\}$. 

Other works on loss networks with future information primarily focus on advance reservations, whereby customers specify service start and end times upon arrival, and the controller decides whether to block out time for them. My work differs fundamentally from this appointment-scheduling literature. For example, the performance gap relative to the fluid approximation is $O(\log N)$ in my setting, but is $O(N^{1/2 + \delta})$, for any $\delta > 0$, in \cites{chen2017revenue} case.

\section{Model}
Consider a sequence of models indexed by the number of servers, $N \in \mathbb{N}$. Every model in the sequence corresponds to the same four primitives: arrival processes $A_{H} \equiv (A_{H}(j), j \in \mathbb{N}_{+})$ and $A_{L} \equiv (A_{L}(j), j \in \mathbb{N}_{+})$, and service processes $S_{H} \equiv (S_{H}(j), j \in \mathbb{N}_{+})$ and $S_{L} \equiv (S_{L}(j), j \in \mathbb{N}_{+})$. The processes subscripted with $H$ correspond to high-type customers, and those subscripted with $L$ correspond to low-type customers. The four processes are independent of one another, each comprising \iid\ Exp(1) random variables. 

The type-$i \in \{H, L\}$ inter-arrival time process of the $N$-server system equals baseline process $A_{i}$ divided by $\lambda_{i}N$, for some $\lambda_{i} > 0$. For example, $A_{i}(j)/(\lambda_{i}N)$ denotes the number of hours between the $(j-1)$th and $j$th type-$i$ job arrival, and $\mathcal{A}_{i}(\lambda_{i}Nt)$ denotes the cumulative number of type-$i$ jobs to arrive by time $t$, where $\mathcal{A}_{i}(t) \equiv \sum_{j=1}^{\infty} \indicator{A_{i}(1) + \cdots + A_{i}(j) \le t}$ is a standard Poisson count process.

The type-$i \in \{H, L\}$ service time process of the $N$-server system equals baseline process $S_{i}$ divided by $\mu > 0$. For example, $S_{i}(j)/\mu$ denotes the number of hours required to complete the $j$th type-$i$ job accepted into service. Note that $S_{i}(j)$ corresponds to the $j$th job to \emph{receive service}, not to the $j$th job to \emph{arrive}: fixing the realized service times upfront prevents the offline agent from cherry picking the quick jobs. 

Service is non-preemptive: once begun, a job will run to completion. However, not all jobs will begin, as there is no queue to store waiting requests, and the system is overloaded, satisfying
\begin{align*} 
	\lambda_H < \mu < \lambda_H + \lambda_L.
\end{align*}
The inequalities above ensure that the system can generally accommodate the inflow of type-$H$ jobs, but not the inflow of all jobs. Accordingly, the controller must weed out some jobs. Upon each job arrival, the controller decides, on the spot, whether to accept it. Let $\pi_{i}(t) \in \{0, 1\}$ indicate whether the controller would accept a job of type $i \in \{H, L\}$ at time $t$, under admissions policy $\pi$. The long-run average reward in the $N$-server system under policy $\pi$ is
\begin{align*}
	\mathcal{R}_{\pi}^{N} & \equiv \liminf_{T\rightarrow \infty} \E\Big(\frac{1}{T} \int_{t=0}^{T} r_{H}\pi_{H}(t) d\mathcal{A}_{H}(\lambda_{H}Nt) + r_{L}\pi_{L}(t) d\mathcal{A}_{L}(\lambda_{L}Nt)\Big).
\end{align*}

The online decision maker chooses the best policy from $\Pi_{\text{on}}^{N}$, the set of history-adapted policies for the $N$-server system, and thus yields long-run average reward $\mathcal{R}_{\text{on}}^{N} \equiv \sup_{\pi\in\Pi_{\text{on}}^{N}} \mathcal{R}_{\pi}^{N}$. In contrast, if the system controller could observe the values in $A_H$, $A_L$, $S_{H}$, and $S_{L}$ that will resolve in the next $W > 0$ hours then they would receive long-run average reward $\mathcal{R}_{W}^{N} \equiv \sup_{\pi\in\Pi_{W}^{N}} \mathcal{R}_{\pi}^{N}$, where $\Pi_{W}^{N}$ is the set of policies whose time-$t$ actions are measurable with respect to the information available up to time $t+W$. And if the system controller could observe processes $A_H$, $A_L$, $S_{H}$, and $S_{L}$ upfront then they would receive long-run average reward $\mathcal{R}_{\text{off}}^{N} \equiv \sup_{\pi\in\Pi_{\text{off}}^{N}} \mathcal{R}_{\pi}^{N}$, where $\Pi_{\text{off}}^{N}$ is the family of $A_H$-, $A_L$-, $S_{H}$-, and $S_{L}$-dependent policies. Finally, if the system controller could remove all volatility from the system, then they would receive fluid-approximation reward 
\begin{align}
	\mathcal{R}_{\text{fluid}}^{N} \equiv  N r_H\lambda_H+ N r_L (\mu - \lambda_H). \label{eq:fluidReward}	
\end{align}
The following lemma formally relates these four rewards.

\begin{lemma}[Regret decomposition]\label{l:cc}
	The difference between the fluid-approximation reward and the best achievable reward has distinct volatility, long-run uncertainty, and short-run uncertainty components:
	\begin{align*}
		\mathcal{R}_{\emph{fluid}}^{N} - \mathcal{R}_{\emph{on}}^{N} &= \mathcal{C}_{\emph{vol}}^{N} + \mathcal{C}_{\emph{long}}^{N} + \mathcal{C}_{\emph{short}}^{N}, \\
		\wq \mathcal{C}_{\emph{vol}}^{N} &\equiv \mathcal{R}_{\emph{fluid}}^{N}-\mathcal{R}_{\emph{off}}^{N} \ge 0, \\
		\mathcal{C}_{\emph{long}}^{N} & \equiv \mathcal{R}_{\emph{off}}^{N}-\mathcal{R}_{W}^{N} \ge 0 ,\\
		\aq \mathcal{C}_{\emph{short}}^{N} & \equiv \mathcal{R}_{W}^{N}-\mathcal{R}_{\emph{on}}^{N} \ge 0 .
	\end{align*}
\end{lemma}

This result is largely tautological---the only non-trivial claim it makes is $\mathcal{C}_{\text{vol}}^{N} \ge 0$. But this decomposition gives useful context for the following result by \cite{Xie2025}: 
\begin{align}
	\mathcal{R}_{\text{fluid}}^{N}-\mathcal{R}_{\text{on}}^{N}=\Theta(\log N), \ \text{as}\ N \rightarrow \infty.\label{eq:OmegaGap2}
\end{align}
In light of Lemma \ref{l:cc}, this result raises a natural question: what accounts for the $\Theta(\log N)$ discrepancy between the actual reward and its fluid approximation? Does the gap arise from uncertainty resolving with less than $w$ period's notice ($\mathcal{C}_{\text{short}}^{N}$), from uncertainty resolving with more than $w$ period's notice ($\mathcal{C}_{\text{long}}^{N}$), or from anticipatable volatility ($\mathcal{C}_{\text{vol}}^{N}$)?

\section{Results}

A priori, it's not clear whether the gap between the online and fluid values stems from the uncertainty associated with the finite but timely set of signals that correspond to the interval $[t, t+W]$, or to the uncertainty associated with the infinite but deferred set that corresponds to the horizon after $t+W$. Or uncertainty might not even be the problem---perhaps accommodating the helter-skelter arrivals and departures requires a $\log N$ buffer of idle servers, even when they are perfectly forecastable. My first results rule out this latter possibility. 

\begin{thm}[Bounded volatility cost]\label{thm:main} 
	The cost of variability remains bounded in the number of servers: $\mathcal{C}_{\emph{vol}}^{N} = O(1)$ as $N \rightarrow \infty$.
\end{thm} 

\begin{corollary}[Logarithmic uncertainty cost]\label{cor:shortRun} 
	The total cost of uncertainty grows logarithmically in the number of servers: $\mathcal{C}_{\emph{long}}^{N} + \mathcal{C}_{\emph{short}}^{N} = \Theta(\log N)$ as $N \rightarrow \infty$.
\end{corollary}

Theorem \ref{thm:main} establishes the equivalence of the fluid and offline benchmarks in the loss network. I initially found this result counterintuitive, as the hourly number of type-$H$ and type-$L$ jobs arriving and departing have $\Theta(\sqrt{N})$ standard deviations, yet the expected number of idle servers required to synchronize these ever-more-volatile processes remains bounded. Put differently, demeaned arrival processes $\mathcal{A}_{H}(t) - \lambda_{H}Nt$ and $\mathcal{A}_{L}(t) - \lambda_{L}Nt$ become infinitely jagged as $N \to \infty$, and yet the offline controller can dovetail these unboundedly spiky functions with a bounded budget of idleness. How could a finite buffer withstand an infinite volatility? Well, it can’t. It turns out that my server-idleness framing was incorrect: the real stabilizing force isn't the servers waiting for work, but the type-$L$ jobs destined for rejection. Every hour, the system rejects $\Theta(N)$ type-$L$ jobs, and that $\Theta(N)$ surplus smooths out the $O(\sqrt{N})$ spikes in the work in progress---the excess flow filling in the irregularities of the arrival and departure processes, like peanut butter spreading into the nooks and crannies of an English muffin.

The preceding results establish that the $\Theta(\log N)$ gap between $\mathcal{R}_{\text{fluid}}^{N}$ and $\mathcal{R}_{\text{on}}^{N}$ stems from uncertainty, not volatility. And the following results more specifically attribute the gap to the uncertainty that resolves in the next $O(\log(N)/N)$ hours.
\begin{thm}[Bounded long-run uncertainty cost]\label{thm:lastMin} 
	The cost of long-run uncertainty is bounded in the number of servers when the lookahead window is $\Omega(\log(N)/N)$: $\mathcal{C}_{\emph{long}}^{N} = O(1)$ as $N \rightarrow \infty$, when $W \ge \frac{8(7\lambda_H+\mu)(\lambda_H+\mu)(\lambda_H+\lambda_L+\mu)^2}{\lambda_{H}(\mu-\lambda_H)^2(\lambda_H+\lambda_L-\mu)^2} \frac{\log(N)}{N}$.
\end{thm} 

\begin{corollary}[Logarithmic short-run uncertainty cost]\label{cor:shortRun2} 
	The cost of short-run uncertainty grows logarithmically in the number of servers when the lookahead window is $\Omega(\log(N)/N)$: $\mathcal{C}_{\emph{short}}^{N} = \Theta(\log N)$ as $N \rightarrow \infty$, when $W \ge \frac{8(7\lambda_H+\mu)(\lambda_H+\mu)(\lambda_H+\lambda_L+\mu)^2}{\lambda_{H}(\mu-\lambda_H)^2(\lambda_H+\lambda_L-\mu)^2} \frac{\log(N)}{N}$.
\end{corollary} 

Combining Theorems \ref{thm:main} and \ref{thm:lastMin} establishes my titular claim: an $\Omega(\log(N)/N)$ lookahead yields a bounded operating cost, as $N \rightarrow \infty$. In other words, the $\Theta(\log N)$ regret intrinsic to the loss network arises from uncertainty that resolves with $O(\log(N)/N)$ hours' notice. And since $\Theta(N)$ customers arrive per hour, anticipating the movements of the next $O(\log N)$ customers would suffice to bound costs.

\section{Bounding Volatility Cost}

\subsection{Algorithm}

I will prove Theorem \ref{thm:main} with the Promise of Future Idleness (PFI) algorithm. This algorithm depends on the future sample path of the hypothetical reject-all-$L$-jobs idle-server process, which specifies what the number of idle servers would be if---starting now---the controller were to reject the current and all future $L$ jobs and accept all $H$ jobs (whenever idle servers are available). More specifically, the algorithm references random variable $\tau_{y}(t)$, which is the length of time after $t$ required for the reject-all-$L$-jobs idle-server process to reach $y$ (so that $t+\tau_{y}(t)$ is the hitting time of level $y$). The PFI policy accepts an $H$ job if $0 < \tau_{0}$---i.e., if there is at least one idle server---and it accepts an $L$ job if $\tau_{\lfloor \sqrt{N}\rfloor}(t) < \tau_{1}(t)$---i.e., if rejecting the proposed $L$ job and all jobs like it would make the number of idle servers reach $\lfloor \sqrt{N}\rfloor$ before $1$.

Since it sets $\pi_{H}(t) = \indicator{0 < \tau_{0}(t)}$ and $\pi_{L}(t) = \indicator{\tau_{\lfloor \sqrt{N}\rfloor}(t) < \tau_{1}(t)}$, the PFI algorithm has long-run average reward
\begin{align}
	\mathcal{R}_{\text{PFI}}^{N} & \equiv \liminf_{T\rightarrow \infty}\E\Big(\frac{1}{T}\int_{t=0}^{T} r_{H}\indicator{0 < \tau_{0}(t)} d\mathcal{A}_{H}(\lambda_{H}Nt) + r_{L} \indicator{\tau_{\lfloor \sqrt{N}\rfloor}(t) < \tau_{1}(t)} d\mathcal{A}_{L}(\lambda_{L}Nt) \Big). \label{eq:defOfRFI}
\end{align}

\begin{algorithm}[h!!]
	\renewcommand{\thealgorithm}{}
	\floatname{algorithm}{Algorithm}\caption{Promise of Future Idleness (PFI)}
	\begin{algorithmic}[1]
		\State Input server count $N$, time horizon $T$, and processes $A_{H}$, $A_{L}$, $S_{H}$, and $S_{L}$.
		\State 
		Define $\tau_{y}$ as the length of time required for the reject-all-$L$-jobs idle-server process to reach $y \in \{0, 1, \lfloor \sqrt{N}\rfloor\}$, given the preceding acceptances and service completions.
		\label{State:wait}
		\State Wait until the next event. If the event occurs after time $T$ then end.
		\State If the next event is a type-$H$ arrival then go to Step \ref{State:high}, if it is a type-$L$ arrival then go to Step \ref{State:low}, and if it is a service completion then go to Step \ref{State:wait}.
		\label{State:checkType}
		\item If $0 < \tau_{0}$ then accept the job, otherwise reject it. Go to Step \ref{State:wait}.\label{State:high}
		\item If $\tau_{\lfloor \sqrt{N}\rfloor} < \tau_{1}$ then accept the job, otherwise reject it. Go to Step \ref{State:wait}.\label{State:low}
	\end{algorithmic}
\end{algorithm}

\subsection{System Dynamics} \label{s:dynamicsPFI}

Characterizing how the system evolves under the PFI algorithm will require distilling infinite processes $A_{H}$, $A_{L}$, $S_{H}$, and $S_{L}$ to a finite Markov chain. Identifying a collection of variables that capture the distribution of the future as a function of the past will be tricky, as the acceptance policies are forward-looking, which makes the past depend on the future (which, in turn, depends on the past).

We begin by working with the embedded discrete-time chain. Define $t^{\ell}$ as the time immediately before the $\ell$th action (either a job arrival or departure). At this time, the system's probability law is governed by $\Omega^{\ell} \equiv (A_{H}^{\ell}, A_{L}^{\ell}, S_{H}^{\ell}, S_{L}^{\ell}, R^{\ell})$, where $R^{\ell}$ is a vector that stores the residual processing times of the jobs in service at time $t^{\ell}$, and $A_{H}^{\ell}$, $A_{L}^{\ell}$, $S_{H}^{\ell}$, and $S_{L}^{\ell}$ are infinite sequences that specify the arrival and service times of jobs that arrive after time $t^{\ell}$, analogously to how $A_{H}$, $A_{L}$, $S_{H}$, and $S_{L}$ specify the arrival and service times of the jobs that arrive after time 0. 

Formally defining $\Omega^{\ell}$ will require a few steps. First, initialize $A_{H}^{0} = A_{H}$, $A_{L}^{0} = A_{L}$, $S_{H}^{0} = S_{H}$, $S_{L}^{0} = S_{L}$, and $R^{0} = ()$, where $()$ denotes an empty vector. Second, define $\iota^{\ell+1} \equiv (\iota_{C}^{\ell+1}, \iota_{c}^{\ell+1}, \iota_{H}^{\ell+1}, \iota_{h}^{\ell+1}, \iota_{L}^{\ell+1}, \iota_{l}^{\ell+1})$ as a collection of mutually-exclusive events that characterize the $(\ell+1)$th action: 
\begin{itemize}
	\item $\iota_{C}^{\ell+1}$ is the completion of the accepted type-$L$ job,
	\item $\iota_{c}^{\ell+1}$ is the completion of a job other than the last accepted $L$,
	\item $\iota_{H}^{\ell+1}$ is the acceptance of a type-$H$ job, 
	\item $\iota_{h}^{\ell+1}$ is the rejection of a type-$H$ job, 
	\item $\iota_{L}^{\ell+1}$ is the acceptance of a type-$L$ job, and
	\item $\iota_{l}^{\ell+1}$ is the rejection of a type-$L$ job.
\end{itemize}
Third, define $m_{i}$, for $i \in \{C, c, H, h, L, l\}$, as the transformation function that maps $\Omega^{\ell}$ to $\Omega^{\ell+1}$, under event $\iota_{i}^{\ell+1}$. For example, if $\iota_{H}^{\ell+1}$ holds then $\Omega^{\ell+1} \equiv m_{H}(\Omega^{\ell})$, where 
\begin{align*}
	m_{H}(\Omega^{\ell})  \equiv & \Big( \\
	& \qquad (A_{H}^{\ell}(2), A_{H}^{\ell}(3), \cdots ),\\
	& \qquad (A_{L}^{\ell}(1) - \lambda_{L} A_{H}^{\ell}(1) / \lambda_{H}, A_{L}^{\ell}(2), A_{L}^{\ell}(3), \cdots),\\
	& \qquad (S_{H}^{\ell}(2), S_{H}^{\ell}(3), \cdots),\\
	& \qquad S_{L}^{\ell},\\
	& \qquad \text{Sort}(S_{H}^{\ell}(1), R^{\ell}(1) - \mu A_{H}^{\ell}(1) / (\lambda_{H} N), R^{\ell}(2) - \mu A_{H}^{\ell}(1) / (\lambda_{H} N), \cdots)\\
	& \Big),
\end{align*}
and $\text{Sort}()$ is a function that arranges a vector in ascending order. This transformation is complex because the various exponential clocks run at different rates. Since the type-$H$ arrival process runs at rate $\lambda_{H} N$, the time elapsed between the $\ell$th and $(\ell+1)$th actions is $t^{\ell+1} - t^{\ell} = A_{H}^{\ell}(1) / (\lambda_{H} N)$. And since the server processes run at rate $\mu$, this inter-action time decreases the elements of $R^{\ell}$ by $\mu (t^{\ell+1} - t^{\ell}) = \mu A_{H}^{\ell}(1) / (\lambda_{H} N)$. Likewise, since the type-$L$ arrival process runs at rate $\lambda_{L} N$, the elapsed inter-action time decreases the first element of $A_{L}^{\ell}$ by $\lambda_{L}N (t^{\ell+1} - t^{\ell}) = \lambda_{L}N A_{H}^{\ell}(1) / (\lambda_{H} N) = \lambda_{L} A_{H}^{\ell}(1) /\lambda_{H}$ (note, only the first element of $A_{L}^{\ell}$ is a live exponential clock). Finally, accepting a type-$H$ exhausts the first element of $A_{H}^{\ell}$ and shifts the first element of $S_{H}^{\ell}$ to $R^{\ell+1}$, but it does not influence the sequence of future type-$L$ jobs, so $S_{L}^{\ell+1} = S_{L}^{\ell}$.

Mapping $m_{h}$ is analogous to $m_{H}$, except without the movement in $S_{H}^{\ell}(1)$ (so that $S_{H}^{\ell+1} = S_{H}^{\ell}$), and mappings $m_{L}$ and $m_{l}$ are analogous to mappings $m_{H}$ and $m_{H}$, but with the $H$ and $L$ subscripts interchanged. Finally, mappings $m_{C}$ and $m_{c}$ are one and the same:
\begin{align*}
	m_{C}(\Omega^{\ell}) \equiv m_{c}(\Omega^{\ell}) \equiv & \Big( \\
	& \qquad (A_{H}^{\ell}(1) - \lambda_{H} N R^{\ell}(1) / \mu, A_{H}^{\ell}(2), A_{H}^{\ell}(3), \cdots ),\\
	& \qquad (A_{L}^{\ell}(1) - \lambda_{L} N R^{\ell}(1) / \mu, A_{L}^{\ell}(2), A_{L}^{\ell}(3), \cdots),\\
	& \qquad S_{H}^{\ell},\\
	& \qquad S_{L}^{\ell},\\
	& \qquad (R^{\ell}(2) - R^{\ell}(1), R^{\ell}(3) - R^{\ell}(1), \cdots)\\
	& \Big).
\end{align*}
Residual service time $R^{\ell}(1)$ corresponds to the last accepted $L$ under event $\iota_{C}^{\ell+1}$, and corresponds to any other job under event $\iota_{c}^{\ell+1}$. Note that the $\text{Sort}()$ operation above ensures that the first element of $R^{\ell}$ is the smallest, and hence that the inter-action time satisfies $t^{\ell+1} - t^{\ell} = R^{\ell}(1) / \mu$ (recall that the service time clocks run at rate $\mu$). This elapsed time reduces the other elements of $R^{\ell}$ by $\mu (t^{\ell+1} - t^{\ell}) = R^{\ell}(1)$, reduces the first element of $A_{H}^{\ell}$ by $\lambda_{H}N (t^{\ell+1} - t^{\ell}) = \lambda_{H}NR^{\ell}(1) / \mu$, and reduces the first element of $A_{L}^{\ell}$ by $\lambda_{L}N (t^{\ell+1} - t^{\ell}) = \lambda_{L}NR^{\ell}(1) / \mu$.
 
The definitions of $\Omega^{0}$ and $m_{i}$ inductively define $\Omega^{\ell}$. This explicit definition, however, isn't useful, as it entangles the infinite random variables of $\Omega^{0}$ in a complex fashion. Rather than specify the \emph{value} of $\Omega^{l}$ in terms of $\Omega^{0}$, it will be more useful to specify the \emph{distribution} of $\Omega^{l}$ in terms $\mathcal{H}^{\ell}$, the information set at time $t^{\ell}$ of an external observer privy to the process' arrivals, acceptance decisions, and departures. The lemma below characterizes this distribution in terms of the following objects:
\begin{itemize}
	\item $\mathcal{H}^{\ell}$ is the $\sigma$-algebra generated by the arrivals, acceptance decisions, and departures that realize prior to time $t^{\ell}$, under the PFI algorithm.
	\item $Y^{\ell}$ is the number of idle servers at time $t^{\ell}$.
	\item $r^{\ell}$ is the difference between $t^{\ell}$ and the time of the last $L$ rejection prior to $t^{\ell}$.
	\item $a^{\ell}$ is the difference between $t^{\ell}$ and the time of the last $L$ acceptance prior to $t^{\ell}$.
	\item $s^{\ell}$ is the service time of the last accepted $L$ job.
	\item $\tau_{y}^{\ell}$ is the value of $\tau_{y}$ at time $t^{\ell}$ (i.e., the length of time after $t^{\ell}$ required for the reject-all-$L$-jobs idle-server process to reach $y$).
	\item $\hat{\tau}_{y}^{\ell}$ represents the counterfactual value of $\tau_{y}^{\ell}$ assuming infinite service time for the most recently accepted type-$L$ job (or most recently accepted type-$H$ job, if no type-$L$ jobs are in service).
	\item $\underleftarrow{\tau}_{y}(t)$ is the length of time prior to $t$ at which the idle-server process last equaled $y$. (Note that $\underleftarrow{\tau}_{y}$ pertains to the actual idle-server process, whereas $\tau_{y}$ pertains to the hypothetical reject-all-$L$-jobs idle-server process.)	
	\item $\underleftarrow{\tau}_{y}^{\ell} = \underleftarrow{\tau}_{y}(t^{\ell})$.
	\item $\mathsf{U}^{\ell} \equiv r^{\ell} < \underleftarrow{\tau}_{1}^{\ell}$ is the event that a type-$L$ job was rejected since the last time the idle-server process equaled 1.
	\item $\mathsf{P}^{\ell} \equiv \big\{a^{\ell} < \underleftarrow{\tau}_{\lfloor \sqrt{N}\rfloor}^{\ell} , \ \ a^{\ell}\wedge s^{\ell} < \underleftarrow{\tau}_{\lfloor \sqrt{N}\rfloor-1}(t^{\ell}\wedge(t^{\ell}-a^{\ell}+s^{\ell}))\big\}$ is the event that (i) a type-$L$ job was accepted since the last time the idle-server process equaled $\lfloor \sqrt{N}\rfloor$ and (ii) the idle-server process did not equal $\lfloor \sqrt{N}\rfloor-1$ while this job was in service (i.e., between its acceptance time and the earlier of $t^{\ell}$ and its completion time).
	\item $\mathsf{W}^{\ell} \equiv a^{\ell} < s^{\ell}$ is the event that the last $L$ job accepted prior to time $t^{\ell}$ is still in service at time $t^{\ell}$.
	\item $\mathcal{S}(\Phi)$ is the subset of $\mathbb{R}^{\infty}$ for which $\Omega^{\ell} \in \mathcal{S}(\Phi)$ implies that condition $\Phi$ holds under sample path $\Omega^{\ell}$ (e.g., $\mathcal{S}(\tau_{y}^{\ell} < \tau_{y'}^{\ell})$ is the set of realizations for which $\tau_{y}^{\ell} < \tau_{y'}^{\ell}$).
	\item $\mathcal{S}_{E}(\Phi)$ is the set $S(\Phi)$ if event $E$ holds, and is the set $\mathbb{R}^{\infty}$ otherwise.
	\item $\xi$ is an infinite sequence of independent Exp(1) random variables.
	\item $\mathcal{L}(\xi)$ denotes the law of $\xi$. For example, writing $\Omega^{\ell} \sim \mathcal{L}(\xi)$ indicates that $\Omega^{\ell}$ comprises independent Exp(1) random variables.
	\item $\mathcal{L}(\xi \barBreak \xi \in \mathcal{S})$ denotes the law of $\xi$ conditional on $\xi\in \mathcal{S}$. For example, writing $\Omega^{\ell} \sim \mathcal{L}(\xi \barBreak \xi \in \mathcal{S})$ indicates that $\Pr(\Omega^{\ell} \in \mathcal{C}) = \Pr(\xi \in \mathcal{C} \barBreak \xi \in \mathcal{S})$.
	\item $\mathcal{L}^{\ell}$ denotes the law of $\Omega^{\ell}$ under the PFI algorithm, conditional on $\mathcal{H}^{\ell}$.
\end{itemize}

\begin{lemma}[Markovian reduction]\label{l:nearmemorylessness}
	History $\mathcal{H}^{\ell}$ is conditionally independent of $\Omega^{\ell}$, given $Y^{\ell}$, $\indicator{\mathsf{U}^{\ell}}$, $\indicator{\mathsf{P}^{\ell}}$, and $\indicator{\mathsf{W}^{\ell}}$:
	\begin{align*}
		\mathcal{L}^{\ell} & = \mathcal{L}\big(\xi \barBreak \xi \in \mathcal{S}_{\mathsf{U}^{\ell}}(\tau_{1}^{\ell}<\tau_{\lfloor\sqrt{N}\rfloor}^{\ell})
		\cap
		\mathcal{S}_{\mathsf{P}^{\ell}\cap\mathsf{W}^{\ell}}(\hat{\tau}_{\lfloor\sqrt{N}\rfloor-1}^{\ell}<\hat{\tau}_{0}^{\ell})
		\cap
		\mathcal{S}_{\mathsf{P}^{\ell}\cap\mathsf{W}^{\ell c}}(\tau_{\lfloor\sqrt{N}\rfloor}^{\ell}<\tau_{1}^{\ell}) \big).
	\end{align*}
\end{lemma}

Lemma \ref{l:nearmemorylessness} asserts that $\Omega^{\ell}$ would comprise mutually independent Exp(1) random variables, conditional on $\mathcal{H}^{\ell}$, but for the following restrictions: 
\begin{enumerate}
	\item $\tau_{1}^{\ell} < \tau_{\lfloor \sqrt{N}\rfloor}^{\ell}$ under event $\mathsf{U}^{\ell}$, 
	\item $\hat{\tau}_{\lfloor \sqrt{N}\rfloor-1}^{\ell} < \hat{\tau}_{0}^{\ell}$ under events $\mathsf{P}^{\ell}$ and $\mathsf{W}^{\ell}$, and 
	\item $\tau_{\lfloor \sqrt{N}\rfloor}^{\ell} < \tau_{1}^{\ell}$ under events $\mathsf{P}^{\ell}$ and $\mathsf{W}^{\ell c}$.
\end{enumerate}
The first restriction stems from a rejected $L$ job: The PFI algorithm rejecting a type-$L$ job indicates that $\tau_{1} < \tau_{\lfloor \sqrt{N}\rfloor}$, a condition that remains extant until the number of idle servers next reaches 1. In other words, event $\mathsf{U}^{\ell}$ indicates that the system is on its way to becoming fully \emph{utilized}. The latter two restrictions stem from an accepted $L$ job: the PFI algorithm accepting a type-$L$ job indicates that $\tau_{\lfloor \sqrt{N}\rfloor} < \tau_{1}$ would have held had this job been rejected instead of accepted. The second restriction holds when the job is still in service, in which case the hypothetical restraint is weakened to account for the server occupied by the accepted $L$ job. The third restriction holds when the job in question has completed service, in which case the previously hypothetical restraint becomes an actual one.  Put differently, the pair $(\mathsf{W}^{\ell}, \mathsf{P}^{\ell})$ implies a \emph{weak promise} of future idleness, $\hat{\tau}_{\lfloor \sqrt{N}\rfloor-1}^{\ell} < \hat{\tau}_{0}^{\ell}$, whereas $(\mathsf{W}^{\ell c}, \mathsf{P}^{\ell})$ implies a \emph{strong promise}, $\tau_{\lfloor \sqrt{N}\rfloor} < \tau_{1}$. 

I will use Lemma \ref{l:nearmemorylessness} to divide time into independent epochs, which will facilitate the use of the renewal reward theorem. A new epoch begins immediately before the $\ell$th action if and only if $Y^{\ell} = 1$ and events $\mathsf{U}^{\ell c}$, $\mathsf{P}^{\ell}$, and $\mathsf{W}^{\ell}$ hold. In other words, a new epoch begins whenever (i) the system has one idle server and (ii) it operates under restriction $\hat{\tau}_{\lfloor \sqrt{N}\rfloor-1}^{\ell} < \hat{\tau}_{0}^{\ell}$, and no others. I denote the start and end of a representative epoch as $t^{\text{start}}$ and $t^{\text{end}}$.

\subsection{Proof of Theorem \ref{thm:main}}

Together, the following three results establish that $\mathcal{R}_{\text{fluid}}^{N}-\mathcal{R}_{\text{PFI}}^{N} = O(1)$ as $N \rightarrow \infty$, which proves Theorem \ref{thm:main}.

\begin{lemma}[Performance gap characterization]\label{l:gap}
	The difference between the fluid and PFI rewards is a linear combination of (i) the ratio of the total expected server idleness during an epoch and the expected duration of an epoch, and (ii) the ratio of the expected number of high-type jobs rejected during an epoch and the expected duration of an epoch:
	\begin{align*}
		\mathcal{R}_{\emph{fluid}}^{N} - \mathcal{R}_{\emph{PFI}}^{N} 	& = r_{L}\mu\frac{\E\big(\int_{t=t^{\emph{start}}}^{t^{\emph{end}}} Y(t)dt\big)}{\E(t^{\emph{end}} - t^{\emph{start}})} + (r_{H} - r_{L})\frac{\E\big(\int_{t=t^{\emph{start}}}^{t^{\emph{end}}} \indicator{\tau_{0}(t) = 0} d\mathcal{A}_{H}(\lambda_{H}Nt)\big)}{\E(t^{\emph{end}} - t^{\emph{start}})} . 
	\end{align*}
\end{lemma}

\begin{proposition}[Upper-bounded idleness]\label{l:serverIdleness}
	The ratio of the total expected server idleness during a PFI epoch and the expected duration of a PFI epoch is uniformly bounded in the number of servers: 
	\begin{align*}
		\frac{\E\big(\int_{t=t^{\emph{start}}}^{t^{\emph{end}}} Y(t)dt\big)}{\E(t^{\emph{end}} - t^{\emph{start}})}= O(1) \ \text{as} \ N \rightarrow \infty.			
	\end{align*}
\end{proposition}
\begin{proposition}[Upper-bounded rejection rate]\label{l:littleo}
	The ratio of the expected number of high-type jobs rejected during a PFI epoch and the expected duration of a PFI epoch falls exponentially quickly in the square root of the number of servers: 
	\begin{align*}
		\frac{\E\big(\int_{t=t^{\emph{start}}}^{t^{\emph{end}}} \indicator{Y(t) = 0} d\mathcal{A}_{H}(\lambda_{H}Nt)\big)}{\E(t^{\emph{end}} - t^{\emph{start}})} = \exp(-\Omega(\sqrt{N})) \ \text{as} \ N \rightarrow \infty.
	\end{align*}

\end{proposition}

The lemma follows from a standard renewal reward argument, but the propositions are more difficult to prove. Since $\E(t^{\text{end}} - t^{\text{start}})$ is no less than the expected time between two consecutive actions, which is $\Omega(1/N)$, proving these propositions amounts to showing that their numerators are $O(1/N)$. 

To bound Proposition \ref{l:serverIdleness}'s numerator, I divide the given epoch into four sub-epochs, based on various events, such as a second server becoming idle, or the rejection of an $L$ job. I bound each sub-epoch's contribution to the integral with the maximum value the idle-server process achieves during the sub-epoch, multiplied by the duration of the sub-epoch. I then constrain each product by bounding its corresponding factors in terms of the number of transitions made by the idle-server process over the sub-epoch. It's easier to bound the number of transitions when the idle-server process has negative unconditional drift (i.e., when the system accepts $L$ jobs) than when it has positive unconditional drift (i.e., when the system rejects $L$ jobs). Note, I wrote positive \emph{unconditional} drift, because while the job outflow rate is nominally larger than its inflow rate when $L$ jobs are rejected, the idle server count is prevented from reaching $\lfloor \sqrt{N}\rfloor$ before 1 at this time, which compels the conditional process to drift downward (like under Doob's $h$-transformation). 

To bound Proposition \ref{l:littleo}'s numerator, I leverage two facts: (i) the probability of an upward drifting random walk traveling from $y$ to 0 falls exponentially quickly in $y$, and (ii) an epoch won't comprise a type-$H$ rejection until the reject-all-$L$-jobs idle-server process travels from $\lfloor \sqrt{N}\rfloor - 1$ to 0. (To be clear, there is nothing special about $\sqrt{N}$ here: e.g., we could make the type-$H$ rejection rate $\exp(-\Omega(N))$ by changing the PFI algorithm's type-$L$ acceptance condition from $\tau_{\lfloor \sqrt{N}\rfloor} < \tau_{1}$ to $\tau_{\lfloor \delta N\rfloor} < \tau_{1}$, for some sufficiently small $\delta > 0$.)

\section{Bounding Long-Run Uncertainty Cost}

\subsection{Algorithm}

I will prove Theorem \ref{thm:lastMin} with the Single Short Sensor (SSS) algorithm, which is similar to the PFI algorithm, except with the type-$L$ acceptance criterion changed from $\tau_{\lfloor \sqrt{N}\rfloor}(t) < \tau_{1}(t)$ to $w < \tau_{0}(t) \wedge \tau_{1}(t)$, for $w \equiv \frac{8(7\lambda_H+\mu)(\lambda_H+\mu)(\lambda_H+\lambda_L+\mu)^2}{\lambda_{H}(\mu-\lambda_H)^2(\lambda_H+\lambda_L-\mu)^2} \frac{\log(N)}{N}$. This algorithm belongs to $\Pi_{w}^{N}$, because indicator variables $\indicator{0 < \tau_{0}(t)}$ and $\indicator{w < \tau_{0}(t) \wedge \tau_{1}(t)}$ are measurable with respect to the information available by time $t + w$ (although the variables that underlie them, $\tau_{0}(t)$ and $\tau_{1}$, are not). 

Since it sets $\pi_{H}(t) = \indicator{0 < \tau_{0}(t)}$ and $\pi_{L}(t) = \indicator{w < \tau_{0}(t) \wedge \tau_{1}(t)}$, the SSS algorithm has long-run average reward
\begin{align*}
	\mathcal{R}_{\text{SSS}}^{N} &\equiv \liminf_{T\rightarrow \infty}\E\Big(\frac{1}{T}\int_{t=0}^{T} r_{H}\indicator{0 < \tau_{0}(t)} d\mathcal{A}_{H}(\lambda_{H}Nt) + r_{L} \indicator{w < \tau_{0}(t) \wedge \tau_{1}(t)} d\mathcal{A}_{L}(\lambda_{L}Nt) \Big).
\end{align*}

Finally, the name \emph{Single Short Sensor} arises from the following analogy: Consider the graph of the reject-all-$L$-jobs idle-server process, $y$, over time. The PFI algorithm places \emph{two infinite sensors} on this graph---along the $y=\lfloor\sqrt{N}\rfloor$ and $y=1$ horizontal lines. The idle-server process then evolves until it hits one of them: if the upper sensor fires first, the system accepts the job; if the lower sensor fires first, it rejects it. In contrast, the SSS policy uses a \emph{single short sensor}, running for $w$ hours along the $y=1$ line: if the idle-server process hits this sensor, the system rejects the job; otherwise, it accepts it. (This analogy supposes that identifying $y(t) = 0$ would not require a separate sensor.)

\begin{algorithm}[h!!]
	\renewcommand{\thealgorithm}{}
	\floatname{algorithm}{Algorithm}\caption{Single Short Sensor (SSS)}
	Follow the PFI algorithm, except change Step \ref{State:low}'s condition from $\tau_{\lfloor \sqrt{N}\rfloor} < \tau_{1}$ to $w < \tau_{0} \wedge \tau_{1}$, where $w \equiv\frac{8(7\lambda_H+\mu)(\lambda_H+\mu)(\lambda_H+\lambda_L+\mu)^2}{\lambda_{H}(\mu-\lambda_H)^2(\lambda_H+\lambda_L-\mu)^2} \frac{\log(N)}{N}$.
\end{algorithm}

\subsection{System Dynamics}

The lemma below characterizes the law of the process under the SSS algorithm, in terms of the following objects (and those defined in Section \ref{s:dynamicsPFI}):
\begin{itemize}
	\item $Y_{\text{SSS}}$ is the idle service level under the SSS policy.
	\item $f^{\ell}$ is the difference between $t^{\ell}$ and the time of the first $L$ rejection after $\underleftarrow{\tau}_{1}^{\ell}$.
	\item $\mathsf{P}_{\text{SSS}}^{\ell} \equiv a^{\ell} < w$ is the event that a type-$L$ job was accepted within the last $w$ hours.
	\item $\Omega_{\text{SSS}}^{\ell}$, $\mathcal{H}_{\text{SSS}}^{\ell}$, and $\mathcal{L}_{\text{SSS}}^{\ell}$ are the analogs of $\Omega^{\ell}$, $\mathcal{H}^{\ell}$, and $\mathcal{L}^{\ell}$ under the SSS policy.
\end{itemize}

\begin{lemma}[SSS Markovian reduction]\label{l:nearmemorylessnessSSS}
	History $\mathcal{H}_{\emph{SSS}}^{\ell}$ is conditionally independent of $\Omega_{\emph{SSS}}^{\ell}$, given $Y_{\emph{SSS}}^{\ell}$, $f^{\ell}$, $a^{\ell}$, $\indicator{\mathsf{U}^{\ell}}$, $\indicator{\mathsf{P}_{\emph{SSS}}^{\ell}}$, and $\indicator{\mathsf{W}^{\ell}}$:
	\begin{align*}
		\mathcal{L}_{\emph{SSS}}^{\ell} & = \mathcal{L}\big(\xi \barBreak \xi \in \mathcal{S}_{\mathsf{U}^{\ell}}(\tau_{0}^{\ell} \wedge \tau_{1}^{\ell} < w - f^{\ell})
		\cap
		\mathcal{S}_{\mathsf{P}_{\emph{SSS}}^{\ell}\cap\mathsf{W}^{\ell}}(w - a^{\ell} < \hat{\tau}_{0}^{\ell})
		\cap
		\mathcal{S}_{\mathsf{P}_{\emph{SSS}}^{\ell}\cap\mathsf{W}^{\ell c}}(w - a^{\ell} < \tau_{1}^{\ell}) \big).
	\end{align*}
\end{lemma}

The probability law in Lemma \ref{l:nearmemorylessnessSSS} mirrors that of Lemma \ref{l:nearmemorylessness}. The SSS law, like the PFI law, is shaped by three restrictions: 
\begin{enumerate}
	\item a promise of near-full utilization (event $\mathsf{U}^{\ell}$), which upper-bounds the time until the reject-all-$L$-jobs idle-server process reaches level $1$, 
	\item a weak promise of future idleness (event $\mathsf{P}_{\text{SSS}}^{\ell}\cap\mathsf{W}^{\ell}$), which lower-bounds the time until the reject-all-$L$-jobs idle-server process can reach level $0$ (under the counterfactual in which the last accepted $L$ job has infinite service time), and 
	\item a strong promise of future idleness ($\mathsf{P}_{\text{SSS}}^{\ell}\cap\mathsf{W}^{\ell c}$), which lower-bounds the time until the reject-all-$L$-jobs idle-server process can reach level $1$.
\end{enumerate}

The key distinction between the PFI and SSS laws lies in what we benchmark hitting times $\tau_{1}^{\ell}$ and $\hat{\tau}_{0}^{\ell}$ against. In Lemma \ref{l:nearmemorylessness}, we compare these quantities to hitting times $\tau_{\lfloor\sqrt{N}\rfloor}^{\ell}$ and $\hat{\tau}_{\lfloor\sqrt{N}\rfloor-1}^{\ell}$, and in Lemma \ref{l:nearmemorylessnessSSS}, we compare them to residual lookahead windows $w - f^{\ell}$ and $w - a^{\ell}$. The former, $w - f^{\ell}$, provides an \emph{upper bound} on how long the process can remain above level $1$, whereas the latter, $w - a^{\ell}$, provides a \emph{lower bound} on how long it is guaranteed to remain above level $1$ (or above $0$, if the last accepted $L$ job remains in service).

As before, I will use Lemma \ref{l:nearmemorylessness} to divide time into independent epochs. Under the SSS policy, a new epoch begins before the $\ell$th action if and only if $Y_{\text{SSS}}^{\ell} = 0$. I denote the start and end of a representative SSS epoch as $t^{\text{start}}_{\text{SSS}}$ and $t^{\text{end}}_{\text{SSS}}$.

\subsection{Proof of Theorem \ref{thm:lastMin}}

Together, the following three results establish that $\mathcal{R}_{\text{fluid}}^{N}-\mathcal{R}_{\text{SSS}}^{N} = O(1)$ as $N \rightarrow \infty$, which proves Theorem \ref{thm:lastMin}.

\begin{lemma}[SSS performance gap characterization]\label{l:gapSSS}
	The difference between the fluid and SSS rewards is a linear combination of (i) the ratio of the total expected server idleness during an SSS epoch and the expected duration of an SSS epoch, and (ii) the ratio of the expected number of high-type jobs rejected during an SSS epoch and the expected duration of an SSS epoch:
	\begin{align*}
		\mathcal{R}_{\emph{fluid}}^{N} - \mathcal{R}_{\emph{SSS}}^{N} 	& = r_{L}\mu\frac{\E\big(\int_{t=t_{\emph{SSS}}^{\emph{start}}}^{t_{\emph{SSS}}^{\emph{end}}} Y_{\emph{SSS}}(t)dt\big)}{\E(t_{\emph{SSS}}^{\emph{end}} - t_{\emph{SSS}}^{\emph{start}})} + (r_{H} - r_{L})\frac{\E\big(\int_{t=t_{\emph{SSS}}^{\emph{start}}}^{t_{\emph{SSS}}^{\emph{end}}} \indicator{\tau_{0}(t) = 0} d\mathcal{A}_{H}(\lambda_{H}Nt)\big)}{\E(t_{\emph{SSS}}^{\emph{end}} - t_{\emph{SSS}}^{\emph{start}})} . 
	\end{align*}
\end{lemma}

\begin{proposition}[Upper-bounded SSS idleness]\label{l:serverIdlenessSSS}
	The ratio of the total expected server idleness during an SSS epoch and the expected duration of an SSS epoch is uniformly bounded in the number of servers: 
	\begin{align*}
		\frac{\E\big(\int_{t=t_{\emph{SSS}}^{\emph{start}}}^{t_{\emph{SSS}}^{\emph{end}}} Y_{\emph{SSS}}(t)dt\big)}{\E(t_{\emph{SSS}}^{\emph{end}} - t_{\emph{SSS}}^{\emph{start}})}= O(1) \ \text{as} \ N \rightarrow \infty.			
	\end{align*}
\end{proposition}

\begin{proposition}[Upper-bounded SSS rejection rate]\label{l:littleoSSS}
	The ratio of the expected number of high-type jobs rejected during an SSS epoch and the expected duration of an SSS epoch is uniformly bounded in the number of servers: 
	\begin{align*}
		\frac{\E\big(\int_{t=t_{\emph{SSS}}^{\emph{start}}}^{t_{\emph{SSS}}^{\emph{end}}} \indicator{\tau_{0}(t) = 0} d\mathcal{A}_{H}(\lambda_{H}Nt)\big)}{\E(t_{\emph{SSS}}^{\emph{end}} - t_{\emph{SSS}}^{\emph{start}})} = O(1) \ \text{as} \ N \rightarrow \infty.
	\end{align*}
\end{proposition}

I establish the lemma by recycling the renewal-reward argument underpinning Lemma \ref{l:gap}. I establish the propositions by benchmarking the SSS idle-server process against three auxiliary policies: PFI, Accept-if-Either (AE), and Reject-if-Either (RE), where AE accepts a job if and only if either SSS or PFI does so, and RE rejects a job if and only if either SSS or PFI does so. The key observation is that the SSS idle-server process cannot decouple from the PFI process without either AE or RE also decoupling.

I show that the RE process decouples from PFI only if the reject-all-$L$-jobs idle-server process falls from $\lfloor \sqrt{N}\rfloor$ to 1 within $w$ hours---an event whose probability decays exponentially in $\sqrt{N}$. And when this rare event occurs, its impact is limited, as the idle-server count must return to 1 within $w$ hours, and it can't exceed $N$ in the meantime.

The AE process decouples more frequently, so handling this case requires additional care. Crucially, however, AE actions are measurable with respect to the PFI information set whenever the PFI idle-server level equals one. This compatibility allows me to invoke Lemma \ref{l:nearmemorylessness} at the start of each PFI epoch, even when conditioning jointly on the AE and PFI histories. And this, in turn, enables me to bound the expected idleness and rejection contributions induced by AE–PFI decouplings by analyzing the behavior of the AE idle service level over a representative PFI epoch. 

\section{Conclusion}

There are two types of dynamic allocation problems: those with single-use resources (dynamic stochastic knapsack problems) and those with multi-use resources (reusable resource problems). The simplest reusable resource problem is the overloaded loss network studied in this article. And the simplest single-use problem is the multi-secretary problem, in which a decision maker allocates one unit of a single resource (i.e., a job) to each accepted applicant. 

\cite{Arlotto2019} proved that the expected difference between the latter problem's offline and online values is universally bounded---i.e., that the hiring manager will never make more than some constant number of mistakes, in expectation, regardless of the number of roles to fill. It is natural to wonder whether this bounded regret extends to the reusable resource setting. I show that it does not: even the simplest reusable resource problem yields an unbounded number of admittance mistakes. Accordingly, I establish a fundamental difference between the two families of resource allocation problems: the value of future information tends to a finite constant for stochastic knapsacks, and tends to infinity for reusable resources.

In fact, I show that the offline--online gap is as large as possible, with the offline value bumping up against the hard limit provided by the LP fluid approximation. Since \cite{Xie2025} showed that the fluid value exceeds the online value by $\Theta(\log N)$, the $\Theta(1)$ gap between the fluid and offline values implies an analogous $\Theta(\log N)$ gap between the offline and online values.

The asymptotic equivalence of the offline and fluid benchmarks is useful, as the latter model is \emph{far} easier to work with. Indeed, the fluid model is a simple linear program, whereas the offline problem is a difficult dynamic program. My result maintains that we can disregard the less tractable benchmark.

The asymptotic equivalence of the offline and fluid benchmarks is also striking. It indicates that it's possible to interweave four Poisson processes---$A_{H}$, $A_{L}$, $S_{H}$, and $S_{L}$---in such a way as to ensure that the time-average number of high-type customers rejected and the time-average number of idle servers remain bounded as the Poisson rates go to infinity. Put differently, the result indicates that only $\Theta(1)$ idle servers are required to match demand and supply processes with $\Theta(N)$ variances. (The bounded buffer can handle the unbounded variability because the surplus type-$L$ jobs absorb the excess stochasticity.)

The policy I used to establish the asymptotic equivalence of the offline and fluid benchmarks anticipates the dynamics of the system for an unbounded amount of time into the future. However, when they defined the rules of engagement for queueing models with future information, \citet[p. 2106]{Spencer2014} explicitly excluded models with unbounded lookahead windows, since ``In practice, infinite prediction into the future is certainly too much to ask for." To honor this admonition, I show that switching to a finite-horizon analog of my infinite-lookahead policy sacrifices only $O(1)$ value. Specifically, I show that an $\Omega(\log(N)/N)$ lookahead is sufficient to bound costs.

\section{Acknowledgments}

Itai Gurvich contributed so much to this manuscript that he should probably be an author. He furnished the core idea and initial analysis. In fact, Itai made significant progress towards a proof of Theorem \ref{thm:main} before introducing me to the problem. However, when he gave me the document, I pursued a different path, effectively starting from scratch. In hindsight, this was a mistake, as my proofs, in their early form, were impenetrable to anyone but me. After seeing the mess I had made of his work, Itai graciously withdrew from the project. I offered to list him as a co-author for the work he had already done, but he was too much of a gentleman to accept.

\pagebreak

\section*{Appendix: Proofs}

\begin{proof}[Proof of Lemma \ref{l:cc}]
	Since $\Pi_{\text{on}}^{N} \subset \Pi_{W}^{N} \subset \Pi_{\text{off}}^{N}$, it suffices to show that $\mathcal{C}_{\text{vol}}^{N} \ge 0$. To begin, define $B_{i}^{\pi}(T)$ and $F_{i}^{\pi}(T)$ as the number of jobs of type $i\in\{H,L\}$ that begin service and finish service by time $T$ under generic offline policy $\pi \in \Pi_{\text{off}}^{N}$. Since there are at most $N$ busy servers, $F_{i}^{\pi}(T)$ must be no less than $B_{i}^{\pi}(T) - N$, and hence
	\begin{align*}
		NT \ge \sum_{j = 1}^{F_{H}^{\pi}(T)} S_{H}(j)/\mu + \sum_{j = 1}^{F_{L}^{\pi}(T)} S_{L}(j)/\mu \ge \sum_{j = 1}^{B_{H}^{\pi}(T) - N} S_{H}(j)/\mu + \sum_{j = 1}^{B_{L}^{\pi}(T) - N} S_{L}(j)/\mu . 
	\end{align*}
	The first inequality above holds because the time the servers spend on the finished jobs cannot exceed the total available server time. Further, the law of large numbers indicates that $\lim_{T\rightarrow \infty} \frac{\sum_{j = 1}^{B_{i}^{\pi}(T)-N} S_{i}(j)/\mu}{B_{i}^{\pi}(T)-N} = 1/\mu$ a.s., for $i \in \{L, H\}$. And with the bound above, this implies that
	\begin{align}
		N &\ge \liminf_{T\rightarrow \infty} \sum_{j = 1}^{B_{H}^{\pi}(T) - N} S_{H}(j)/(T\mu) + \sum_{j = 1}^{B_{L}^{\pi}(T) - N} S_{L}(j)/(T\mu) \nonumber\\
		& = \liminf_{T\rightarrow \infty} \frac{B_{H}^{\pi}(T)-N}{T} \cdot \frac{\sum_{j = 1}^{B_{H}^{\pi}(T)-N} S_{H}(j)/\mu}{B_{H}^{\pi}(T)-N} + \frac{B_{L}^{\pi}(T)-N}{T} \cdot \frac{\sum_{j = 1}^{B_{L}^{\pi}(T)-N} S_{L}(j)/\mu}{B_{L}^{\pi}(T)-N} \nonumber\\
		& = \liminf_{T\rightarrow \infty} B_{H}^{\pi}(T)/(T\mu) + B_{L}^{\pi}(T)/(T\mu) \ \ \text{a.s.} \label{eq:constraint1}
	\end{align}
	
	Next, the number of accepted $H$ jobs cannot exceed the number of arriving $H$ jobs, which, by the strong law of large numbers, implies that 
	\begin{align}
		\liminf_{T\rightarrow \infty} B_{H}^{\pi}(T)/T \le \liminf_{T\rightarrow \infty} \mathcal{A}_{H}(\lambda_{H}NT)/T \le \lambda_{H}N \ \ \text{a.s.}	\label{eq:constraint2}
	\end{align}
	Finally, combining constraints \eqref{eq:constraint1} and \eqref{eq:constraint2} yields:
	\begin{align*}
		\mathcal{R}_{\text{off}}^{N} & = \sup_{\pi\in\Pi_{\text{off}}^{N}} \E\Big(\liminf_{T\rightarrow \infty} r_{H}  B_{H}^{\pi}(T)/T + r_{L}B_{L}^{\pi}(T)/T\Big)  \\
		& \le \max_{B_{H}, B_{L} \ge 0} r_{H}B_{H} + r_{L}B_{L} \ \ \text{s.t.} \ \  B_{H}^{\pi} + B_{L}^{\pi} \le \mu N, \ \   B_{H}^{\pi} \le \lambda_{H}N \\
		& = N r_H\lambda_H+ N r_L (\mu - \lambda_H) \\
		& = \mathcal{R}_{\text{fluid}}^{N}.
	\end{align*}
\end{proof}

\begin{proof}[Proof of Lemma \ref{l:nearmemorylessness}]
	This will be a proof by induction. The initial case holds trivially, so we will only have to establish the inductive step. In other words, we must show that the result holds through to the $(\ell+1)$th action when it holds through the $\ell$th action. The inductive hypothesis establishes that $\Omega^{\ell}$ and $\mathcal{H}^{\ell}$ are conditionally independent, given $Y^{\ell}$ and $\mathcal{E}^{\ell} \equiv (\indicator{\mathsf{U}^{\ell}}, \indicator{\mathsf{P}^{\ell}}, \indicator{\mathsf{W}^{\ell}})$. And since $\Omega^{\ell+1}$ is a deterministic function of $\Omega^{\ell}$, this in turn implies that $\Omega^{\ell+1}$ and $\mathcal{H}^{\ell}$ are conditionally independent, given $Y^{\ell}$ and $\mathcal{E}^{\ell}$. Further, since $\mathcal{H}^{\ell}$ and $\iota^{\ell+1}$ characterize $\mathcal{H}^{\ell+1}$, it follows that $\Omega^{\ell+1}$ and $\mathcal{H}^{\ell+1}$ are conditionally independent, given $Y^{\ell}$, $\mathcal{E}^{\ell}$, and $\iota^{\ell+1}$. Hence, it suffices to characterize $\Omega^{\ell+1}$ in terms of $Y^{\ell}$, $\mathcal{E}^{\ell}$, and $\iota^{\ell+1}$. I will condition on $\mathcal{E}^{\ell}$ and $\iota^{\ell+1}$ explicitly, but will condition on $Y^{\ell}$ only implicitly, as this variable is baked into $\Omega^{\ell}$, via the number of elements in $R^{\ell}$. 
	
	The following analysis hinges on the fact that transformation function $m_{i}$, for $i \in \{C, c, H, h, L, l\}$, maps \iid\ processes to \iid\ processes. More specifically, the exponential distribution's memorylessness property ensures that 
	\begin{align}
		\mathcal{L}^{\ell} = \mathcal{L}(\xi) & \Rightarrow \underline{\mathcal{L}}_{i}^{\ell+1} \sim \mathcal{L}(\xi),\label{eq:twoSim}
	\end{align}
	where $\underline{\mathcal{L}}_{i}^{\ell+1}$ is the law of $m_{i}(\Omega^{\ell})$ conditional on event 
	\begin{align}
		\underline{\iota}_{i}^{\ell+1} & \equiv
		\begin{cases}
			\iota_{L}^{\ell+1} \cup \iota_{l}^{\ell+1} & i \in \{L, l\}\\
			\iota_{i}^{\ell+1} & \text{otherwise}.
		\end{cases}	\nonumber
	\end{align}
	
	Line \eqref{eq:twoSim} states that if $\Omega^{\ell}$ comprises independent Exp(1) random variables, then (i) $m_{C}(\Omega^{\ell})$, $m_{c}(\Omega^{\ell})$, $m_{H}(\Omega^{\ell})$, and $m_{h}(\Omega^{\ell})$ comprise independent Exp(1) random variables, conditional on $\iota_{C}^{\ell+1}$, $\iota_{c}^{\ell+1}$, $\iota_{H}^{\ell+1}$, and $\iota_{h}^{\ell+1}$, respectively, and (ii) $m_{L}(\Omega^{\ell})$ and $m_{l}(\Omega^{\ell})$ comprise independent Exp(1) random variables, conditional on $\iota_{L}^{\ell+1} \cup \iota_{l}^{\ell+1}$. Note that whereas events $\iota_{L}^{\ell+1}$ and $\iota_{l}^{\ell+1}$ inform us about the future sample path of the reject-all-$L$-jobs idle-server process, event $\iota_{L}^{\ell+1} \cup \iota_{l}^{\ell+1}$ does not.
	
	I will now extend result \eqref{eq:twoSim} to the case in which $\mathcal{L}^{\ell} = \mathcal{L}(\xi \barBreak \mathcal{S})$, for generic $\mathcal{S} \subset \mathbb{R}^{\infty}$. Doing so will be delicate, as the conditioning statement sacrifices the memorylessness that facilitates the transformation from $\mathcal{L}^{\ell}$ to $\underline{\mathcal{L}}_{i}^{\ell+1}$. The solution will be to identify $c_{i}(\mathcal{S}) \subset \mathbb{R}^{\infty}$ for which events $\Omega^{\ell} \in \mathcal{S}$ and $m_{i}(\Omega^{\ell}) \in c_{i}(\mathcal{S})$ almost surely equal, under $\underline{\iota}_{i}^{\ell+1}$. Doing so will enable us to remove the restriction on $\Omega^{\ell}$, thereby restoring the memoryless structure.
	
	To begin, let $\Pr(\Omega^{\ell} \in \mathcal{S} \barBreak \underline{\iota}_{i}^{\ell+1},\, m_{i}(\Omega^{\ell}) = x)$ denote a \emph{regular conditional probability}---i.e., a Borel-measurable function of $x$ that satisfies the following, for every Borel set $\mathcal{C}\subset \mathbb{R}^{\infty}$:
	\begin{align*}
		\Pr(\Omega^{\ell} \in \mathcal{S}, m_{i}(\Omega^{\ell}) \in \mathcal{C} \barBreak \underline{\iota}_{i}^{\ell+1}) = \int_{\mathcal{C}} \Pr(\Omega^{\ell} \in \mathcal{S} \barBreak \underline{\iota}_{i}^{\ell+1}, m_{i}(\Omega^{\ell}) = x) \Pr(m_{i}(\Omega^{\ell}) \in dx \barBreak \underline{\iota}_{i}^{\ell+1}).
	\end{align*}
	Such a function exists on $\mathbb{R}^{\infty}$ (a standard Borel space) and is unique almost everywhere. Next, define $\mathcal{S} \subset \mathbb{R}^{\infty}$ as \emph{$m_{i}$-determined} if it satisfies the following:
	\begin{align*}
		\Pr(\Omega^{\ell} \in \mathcal{S} \barBreak \underline{\iota}_{i}^{\ell+1},\, m_{i}(\Omega^{\ell}) = x) \in \{0, 1\} \ \ \text{a.s.}
	\end{align*}
	And for $m_{i}$-determined $\mathcal{S}$, define
	\begin{align*}
		c_{i}(\mathcal{S}) \equiv \left\{x \barBreak \Pr(\Omega^{\ell} \in \mathcal{S} \barBreak \underline{\iota}_{i}^{\ell+1},\, m_{i}(\Omega^{\ell}) = x) = 1\right\}.
	\end{align*}
	By construction, $\indicator{\Omega^{\ell} \in \mathcal{S}} = \indicator{m_{i}(\Omega^{\ell}) \in c_{i}(\mathcal{S})}$, a.s., when $\mathcal{S}$ is $m_{i}$-determined. 
	
	Intuitively, $m_{i}$-determined events are those whose occurrence is almost surely resolved by the value of $m_{i}(\Omega^{\ell})$ (given $\underline{\iota}_{i}^{\ell+1}$). To characterize this class of events, note that conditioning on $\underline{\iota}_{i}^{\ell+1}$ ensures that the time between the $\ell$th and $(\ell+1)$th actions is the information stored in $\Omega^{\ell}$ and not in $m_{i}(\Omega^{\ell})$. Accordingly, event $\mathcal{S} \subset \mathbb{R}^{\infty}$ is $m_{i}$-determined if and only if its occurrence does not depend on inter-action time $t^{\ell+1} - t^{\ell}$, except on null sets.
	
	And now if we let $\Pr_{\mathcal{L}(\xi |\xi \in \mathcal{S})}$ and $\Pr$ denote probability under laws $\mathcal{L}^{\ell} = \mathcal{L}(\xi \barBreak \mathcal{S})$ and $\mathcal{L}^{\ell} = \mathcal{L}(\xi)$, respectively, then we get the following for $m_{i}$-determined set $\mathcal{S}$ and Borel set $\mathcal{C}$:
	\begin{align}
		\Pr_{\mathcal{L}(\xi |\xi \in \mathcal{S})} (&m_{i}(\Omega^{\ell}) \in \mathcal{C} \barBreak \underline{\iota}_{i}^{\ell+1}) \nonumber \\
		& = \frac{
			\Pr(m_{i}(\Omega^{\ell}) \in \mathcal{C}, \Omega^{\ell} \in \mathcal{S} \barBreak \underline{\iota}_{i}^{\ell+1})
		}{
			\Pr(\Omega^{\ell} \in \mathcal{S} \barBreak \underline{\iota}_{i}^{\ell+1})
		} \nonumber\\
		& = \frac{
			\Pr(m_{i}(\Omega^{\ell}) \in \mathcal{C} \cap c_{i}(\mathcal{S}) \barBreak \underline{\iota}_{i}^{\ell+1})
		}{
			\Pr(m_{i}(\Omega^{\ell}) \in c_{i}(\mathcal{S}) \barBreak \underline{\iota}_{i}^{\ell+1})
		} \nonumber\\
		& = \frac{
			\Pr(\xi \in \mathcal{C} \cap c_{i}(\mathcal{S}))
		}{
			\Pr(\xi \in c_{i}(\mathcal{S}))
		} \nonumber\\
		& =
		\Pr(\xi \in \mathcal{C} \barBreak \xi \in c_{i}(\mathcal{S})), \label{eq:noActionYet}
	\end{align}
	Note, the first line above follows from identity $\Pr(A | B \cap C) = \Pr(A \cap B | C) / \Pr(B | C)$, the second line from the definition of $c_{i}(\mathcal{S})$, and the third line from \eqref{eq:twoSim}. 
	
	Next, I will now strengthen the conditioning event from $\underline{\iota}_{i}^{\ell+1}$ to $\iota_{i}^{\ell+1}$---i.e., I will accommodate the information implied by the type-$L$ acceptance condition. To this end, note that we have the following, for $m_{i}$-determined set $\mathcal{S}$ and Borel set $\mathcal{C}$:
	\begin{align*}
		\Pr_{\mathcal{L}(\xi |\xi \in \mathcal{S})}(&m_{L}(\Omega^{\ell}) \in \mathcal{C} \barBreak \iota_{L}^{\ell+1}) \\
		& = \Pr_{\mathcal{L}(\xi |\xi \in \mathcal{S})}\big(m_{L}(\Omega^{\ell}) \in \mathcal{C} \barBreak \Omega^{\ell} \in \mathcal{S}(\tau_{\lfloor \sqrt{N}\rfloor}^{\ell} < \tau_{1}^{\ell}) \cap \underline{\iota}_{L}^{\ell+1}\big) \\
		& = \frac{\Pr_{\mathcal{L}(\xi |\xi \in \mathcal{S})}(m_{L}(\Omega^{\ell}) \in \mathcal{C}, \Omega^{\ell} \in \mathcal{S}(\tau_{\lfloor \sqrt{N}\rfloor}^{\ell} < \tau_{1}^{\ell})\barBreak \underline{\iota}_{L}^{\ell+1})}{\Pr_{\mathcal{L}(\xi |\xi \in \mathcal{S})}(\Omega^{\ell} \in \mathcal{S}(\tau_{\lfloor \sqrt{N}\rfloor}^{\ell} < \tau_{1}^{\ell}) \barBreak \underline{\iota}_{L}^{\ell+1})} \\
		& = \frac{\Pr_{\mathcal{L}(\xi |\xi \in \mathcal{S})}\big(m_{L}(\Omega^{\ell}) \in \mathcal{C} \cap c_{L}(\mathcal{S}(\tau_{\lfloor \sqrt{N}\rfloor}^{\ell} < \tau_{1}^{\ell}))\barBreak \underline{\iota}_{L}^{\ell+1}\big)}{\Pr_{\mathcal{L}(\xi |\xi \in \mathcal{S})}\big(m_{L}(\Omega^{\ell}) \in c_{L}(\mathcal{S}(\tau_{\lfloor \sqrt{N}\rfloor}^{\ell} < \tau_{1}^{\ell})) \barBreak \underline{\iota}_{L}^{\ell+1}\big)} \\
		& = \frac{\Pr\big(\xi \in \mathcal{C} \cap c_{L}(\mathcal{S}(\tau_{\lfloor \sqrt{N}\rfloor}^{\ell} < \tau_{1}^{\ell}))\barBreak \xi \in c_{L}(\mathcal{S})\big)}{\Pr\big(\xi \in c_{L}(\mathcal{S}(\tau_{\lfloor \sqrt{N}\rfloor}^{\ell} < \tau_{1}^{\ell})) \barBreak \xi \in c_{L}(\mathcal{S})\big)}  \\
		& = \Pr\big(\xi \in \mathcal{C} \barBreak \xi \in c_{L}(\mathcal{S}) \cap c_{L}(\mathcal{S}(\tau_{\lfloor \sqrt{N}\rfloor}^{\ell} < \tau_{1}^{\ell}))\big) \\		
		& = \Pr\big(\xi \in \mathcal{C} \barBreak \xi \in c_{L}(\mathcal{S} \cap \mathcal{S}(\tau_{\lfloor \sqrt{N}\rfloor}^{\ell} < \tau_{1}^{\ell}))\big).		
	\end{align*}
	
	The first line above holds because event $\iota_{L}^{\ell+1}$ is equivalent to events $\underline{\iota}_{L}^{\ell+1}$ and $\tau_{\lfloor \sqrt{N}\rfloor}^{\ell} < \tau_{1}^{\ell}$, the second line by identity $\Pr(A | B \cap C) = \Pr(A \cap B | C) / \Pr(B | C)$, the third line by the definition of $c_{L}(\mathcal{S}(\tau_{\lfloor \sqrt{N}\rfloor}^{\ell} < \tau_{1}^{\ell}))$ (which exists because event $\tau_{\lfloor \sqrt{N}\rfloor}^{\ell} < \tau_{1}^{\ell}$ is $m_{L}$-determined), the fourth by \eqref{eq:noActionYet}, the fifth by identity $\Pr(A | B \cap C) = \Pr(A \cap B | C) / \Pr(B | C)$, and the sixth line holds because if $\mathcal{S}_{1}$ and $\mathcal{S}_{2}$ are $m_{i}$-determined then $\mathcal{S}_{1} \cap \mathcal{S}_{2}$ is $m_{i}$-determined, with $c_{i}(\mathcal{S}_{1} \cap \mathcal{S}_{2}) = c_{i}(\mathcal{S}_{1}) \cap c_{i}(\mathcal{S}_{2})$ (modulo null sets).
	
	And an analogous argument establishes an analogous result for the $\iota_{l}^{\ell+1}$ case: 
	\begin{align*}
		\Pr_{\mathcal{L}(\xi |\xi \in \mathcal{S})}(&m_{l}(\Omega^{\ell}) \in \mathcal{C} \barBreak \iota_{l}^{\ell+1}) = \Pr\big(\xi \in \mathcal{C} \barBreak \xi \in c_{l}(\mathcal{S} \cap \mathcal{S}(\tau_{1}^{\ell} < \tau_{\lfloor \sqrt{N}\rfloor}^{\ell}))\big),		
	\end{align*}
	when $\mathcal{S}$ is $m_{i}$-determined and $\mathcal{C}$ is Borel.
	
	Finally, combining the preceding three results yields the following, for $m_{i}$-determined $\mathcal{S}$:
	\begin{align}
		\mathcal{L}^{\ell} = \mathcal{L}(\xi |\xi \in \mathcal{S}) \Rightarrow \mathcal{L}_{i}^{\ell+1} = 
		\begin{cases}
			\mathcal{L}\big(\xi \barBreak \xi \in c_{L}(\mathcal{S} \cap \mathcal{S}(\tau_{\lfloor \sqrt{N}\rfloor}^{\ell} < \tau_{1}^{\ell}))\big) & i = L,\\
			\mathcal{L}\big(\xi \barBreak \xi \in c_{l}(\mathcal{S} \cap \mathcal{S}(\tau_{1}^{\ell} < \tau_{\lfloor \sqrt{N}\rfloor}^{\ell}))\big) & i = l,\\
			\mathcal{L}\big(\xi \barBreak \xi \in c_{i}(\mathcal{S})\big) & \text{otherwise},
		\end{cases} \label{eq:usesCiTerms}
	\end{align}
	where $\mathcal{L}_{i}^{\ell+1}$ denotes the law of $m_{i}(\Omega^{\ell})$ conditional on $\iota_{i}^{\ell+1}$. 
	
	Now, to conclude the proof, I will use \eqref{eq:usesCiTerms} to show that setting $\mathcal{L}^{\ell} = \mathcal{L}(\xi |\xi \in \mathcal{S}^{\ell})$ yields $\mathcal{L}_{i}^{\ell+1} = \mathcal{L}(\xi |\xi \in \mathcal{S}^{\ell+1})$, under event $\iota_{i}^{\ell+1}$, where 
	\begin{align*}
		\mathcal{S}^{\ell} \equiv \mathcal{S}_{\mathsf{U}^{\ell}}(\tau_{1}^{\ell}<\tau_{\lfloor\sqrt{N}\rfloor}^{\ell}) \cap \mathcal{S}_{\mathsf{P}^{\ell}\cap\mathsf{W}^{\ell}}(\hat{\tau}_{\lfloor\sqrt{N}\rfloor-1}^{\ell}<\hat{\tau}_{0}^{\ell}) \cap \mathcal{S}_{\mathsf{P}^{\ell}\cap\mathsf{W}^{\ell c}}(\tau_{\lfloor\sqrt{N}\rfloor}^{\ell}<\tau_{1}^{\ell})	
	\end{align*}
	is the set that constrains $\Omega^{\ell}$, by the inductive hypothesis. This set is $m_{i}$-determined, since events $\tau_{1}^{\ell}<\tau_{\lfloor\sqrt{N}\rfloor}^{\ell}$, $\hat{\tau}_{\lfloor\sqrt{N}\rfloor-1}^{\ell}<\hat{\tau}_{0}^{\ell}$, and $\tau_{\lfloor\sqrt{N}\rfloor}^{\ell}<\tau_{1}^{\ell}$ are independent of inter-action time $t^{\ell+1} - t^{\ell}$. Put differently, I must show that if $\iota_{i}^{\ell+1}$ holds then
	\begin{align*}
		\mathcal{S}^{\ell + 1} = 
		\begin{cases}
			c_{L}(\mathcal{S}^{\ell} \cap \mathcal{S}(\tau_{\lfloor \sqrt{N}\rfloor}^{\ell} < \tau_{1}^{\ell})) & i = L,\\
			c_{l}(\mathcal{S}^{\ell} \cap \mathcal{S}(\tau_{1}^{\ell} < \tau_{\lfloor \sqrt{N}\rfloor}^{\ell})) & i = l,\\
			c_{i}(\mathcal{S}^{\ell}) & \text{otherwise}.
		\end{cases}
	\end{align*}
	
	I will begin with the $i = L$ case. First, since $\tau_{\lfloor \sqrt{N}\rfloor}^{\ell} < \tau_{1}^{\ell}$ implies $\hat{\tau}_{\lfloor\sqrt{N}\rfloor-1}^{\ell}<\hat{\tau}_{0}^{\ell}$, it follows that $\mathcal{S}^{\ell} \cap \mathcal{S}(\tau_{\lfloor \sqrt{N}\rfloor}^{\ell} < \tau_{1}^{\ell}) = \mathcal{S}(\tau_{\lfloor \sqrt{N}\rfloor}^{\ell} < \tau_{1}^{\ell})$. And restricting $\Omega^{\ell}$ to satisfy $\mathcal{S}(\tau_{\lfloor \sqrt{N}\rfloor}^{\ell} < \tau_{1}^{\ell})$ is equivalent to restricting $m_{L}(\Omega^{\ell})$ to satisfy $\hat{\tau}_{\lfloor\sqrt{N}\rfloor-1}^{\ell+1}<\hat{\tau}_{0}^{\ell+1}$, given $\iota_{L}^{\ell+1}$. Consequently, $c_{L}(\mathcal{S}^{\ell} \cap \mathcal{S}(\tau_{\lfloor \sqrt{N}\rfloor}^{\ell} < \tau_{1}^{\ell})) = \mathcal{S}(\hat{\tau}_{\lfloor\sqrt{N}\rfloor-1}^{\ell+1}<\hat{\tau}_{0}^{\ell+1})$, given $\iota_{L}^{\ell+1}$. And this indeed equals $\mathcal{S}^{\ell + 1}$, since $\iota_{L}^{\ell+1}$ implies $\mathsf{P}^{\ell+1}\cap\mathsf{W}^{\ell+1}$ (unless $Y^{\ell + 1} \in \{\lfloor\sqrt{N}\rfloor-1, \lfloor\sqrt{N}\rfloor\}$, in which case the constraint is redundant).
	
	Next, I will consider the $i = l$ case. First, $\mathcal{S}^{\ell} \cap \mathcal{S}(\tau_{1}^{\ell} < \tau_{\lfloor \sqrt{N}\rfloor}^{\ell})$ can take two possible values: $\mathcal{S}(\hat{\tau}_{\lfloor\sqrt{N}\rfloor-1}^{\ell}<\hat{\tau}_{0}^{\ell}) \cap \mathcal{S}(\tau_{1}^{\ell} < \tau_{\lfloor \sqrt{N}\rfloor}^{\ell})$ if $\mathsf{P}^{\ell}\cap\mathsf{W}^{\ell}$ holds and $\mathcal{S}(\tau_{1}^{\ell} < \tau_{\lfloor \sqrt{N}\rfloor}^{\ell})$ otherwise. Further, rejecting the $L$ job that arrives in period $\ell + 1$ mimics the reject-all-$L$-jobs policy that underlies random variables $\tau_{y}^{\ell}$ and $\hat{\tau}_{y}^{\ell}$. Accordingly, $m_{l}(\Omega^{\ell})$ is subject to the same restrictions as $\Omega^{\ell}$, given $\iota_{l}^{\ell+1}$. Indeed, all that changes in this case is the superscripts: $c_{l}(\mathcal{S}(\hat{\tau}_{\lfloor\sqrt{N}\rfloor-1}^{\ell}<\hat{\tau}_{0}^{\ell}) \cap \mathcal{S}(\tau_{1}^{\ell} < \tau_{\lfloor \sqrt{N}\rfloor}^{\ell})) = \mathcal{S}(\hat{\tau}_{\lfloor\sqrt{N}\rfloor-1}^{\ell+1}<\hat{\tau}_{0}^{\ell+1}) \cap \mathcal{S}(\tau_{1}^{\ell+1} < \tau_{\lfloor \sqrt{N}\rfloor}^{\ell+1})$ and $c_{l}(\mathcal{S}(\tau_{1}^{\ell}<\tau_{\lfloor\sqrt{N}\rfloor}^{\ell}))  = \mathcal{S}(\tau_{1}^{\ell+1}<\tau_{\lfloor\sqrt{N}\rfloor}^{\ell+1})$. And these values indeed equal $\mathcal{S}^{\ell + 1}$, since $\iota_{l}^{\ell+1}$ and $\mathsf{P}^{\ell}\cap\mathsf{W}^{\ell}$ imply $\mathsf{P}^{\ell+1}\cap\mathsf{W}^{\ell+1}$, and $\iota_{l}^{\ell+1}$ implies $\mathsf{U}^{\ell+1}$ (unless $Y^{\ell+1} = 1$, in which case the $\tau_{1}^{\ell+1}<\tau_{\lfloor\sqrt{N}\rfloor}^{\ell+1}$ constraint is redundant). 
	
	Next, I will consider the $i = C$ case. As before, the system dynamics mimic those under the reject-all-$L$-jobs policy, so that whatever conditions hold in period $\ell$ also holds in period $\ell+1$, just with an updated superscript: e.g., $c_{C}(\mathcal{S}(\tau_{1}^{\ell}<\tau_{\lfloor\sqrt{N}\rfloor}^{\ell})) = \mathcal{S}(\tau_{1}^{\ell+1}<\tau_{\lfloor\sqrt{N}\rfloor}^{\ell+1})$. However, there is one exception, if $\mathcal{S}^{\ell} = \mathcal{S}(\hat{\tau}_{\lfloor\sqrt{N}\rfloor-1}^{\ell}<\hat{\tau}_{0}^{\ell})$ then $c_{C}(\mathcal{S}^{\ell}) = \mathcal{S}(\tau_{\lfloor\sqrt{N}\rfloor}^{\ell+1} < \tau_{1}^{\ell+1})$, since $\hat{\tau}_{\lfloor\sqrt{N}\rfloor-1}^{\ell+1}<\hat{\tau}_{0}^{\ell+1}$ and $\tau_{\lfloor\sqrt{N}\rfloor}^{\ell+1} < \tau_{1}^{\ell+1}$ are equivalent under $\iota_{C}^{\ell+1}$. And this tracks with $\mathcal{S}^{\ell+1}$, since $\iota_{C}^{\ell+1}$ implies $\mathsf{W}^{\ell+1c}$, $\mathsf{U}^{\ell+1} = \mathsf{U}^{\ell}$, and $\mathsf{P}^{\ell+1} = \mathsf{P}^{\ell}$.
	
	Next, the $i \in \{H, c\}$ cases are the same, except without the $\mathcal{S}^{\ell} = \mathcal{S}(\hat{\tau}_{\lfloor\sqrt{N}\rfloor-1}^{\ell}<\hat{\tau}_{0}^{\ell})$ exception. Indeed, if $\iota_{H}^{\ell+1}$ or $\iota_{c}^{\ell+1}$ hold then the system mimics that under the reject-all-$L$-jobs policy, and the conditions are updated by incrementing the superscripts from $\ell$ to $\ell+1$. Further, in these cases $\mathsf{U}^{\ell+1} = \mathsf{U}^{\ell+1}$, $\mathsf{P}^{\ell+1} = \mathsf{P}^{\ell+1}$, and $\mathsf{W}^{\ell+1} = \mathsf{W}^{\ell+1}$, unless the idle-server process reaches an upper or lower boundary, in which case the corresponding restrictions are moot.
	
	Finally, the $i = h$ case holds only if $Y^{\ell} = 0$, in which case it's trivial to verify that the lemma holds. 
\end{proof}

\begin{proof}[Proof of Corollary \ref{cor:shortRun}] 
	This follows directly from Theorem \ref{thm:main}, Lemma \ref{l:cc}, and bound \eqref{eq:OmegaGap2}
\end{proof}

\begin{proof}[Proof of Corollary \ref{cor:shortRun2}] 
	This follows directly from Theorem \ref{thm:lastMin} and Corollary \ref{cor:shortRun}
\end{proof}

\begin{proof}[Proof of Lemma \ref{l:gap}]
	The law of large numbers and the renewal reward theorem quantify the long-run average reward from type-$H$ jobs:
	\begin{align}
		\lim_{T\rightarrow \infty} &\frac{1}{T}\int_{t=0}^{T} \indicator{0 < \tau_{0}(t)} d\mathcal{A}_{H}(\lambda_{H}Nt)\nonumber\\
		& = \lim_{T\rightarrow \infty} \frac{1}{T}\int_{t=0}^{T} d\mathcal{A}_{H}(\lambda_{H}Nt) -  \frac{1}{T}\int_{t=0}^{T} \indicator{\tau_{0}(t) = 0} d\mathcal{A}_{H}(\lambda_{H}Nt) \nonumber\\
		& = N\lambda_{H} - \frac{\E\big(\int_{t=t^{\text{start}}}^{t^{\text{end}}} \indicator{Y(t) = 0} d\mathcal{A}_{H}(\lambda_{H}Nt)\big)}{\E(t^{\text{end}} - t^{\text{start}})} \ \ \text{a.s.} \label{eq:firstOfGap}
	\end{align}
	
	Quantifying the long-run average reward from type-$L$ jobs will require more work. To begin, respectively define $B_i(T)$ and $F_i(T)$ as the number of jobs of type $i \in \{H,L\}$ jobs that begin and finish service by time $T$, under the PFI policy. These variables must satisfy 
	\begin{align}
		F_i(T) \le B_i(T) \le F_i(T) + N. \label{eq:inequalityDouble1}
	\end{align}
	Next, define $\mathcal{S}_i$ as the service-process analog of $\mathcal{A}_i$: 
	\begin{align*}
		\mathcal{S}_i(t) \equiv \sum_{j=1}^{\infty} \indicator{S_i(1) + \cdots + S_i(j) \le t}.
	\end{align*}
	Further, define $X_i(t)$ as the number of servers working on type-$i$ jobs at time $t$, so that $\mathcal{S}_i(\mu\int_{t=0}^{T} X_i(t) dt)$ represents the number of type-$i$ jobs that would be completed by time $T$ if all server effort were applied sequentially to a single stream of jobs. Divvying this effort across $N$ parallel servers will leave at most $N$ partially completed jobs at time $T$, which implies: 
	\begin{align}
		\mathcal{S}_i\Big(\mu\int_{t=0}^{T} X_i(t) dt\Big) - N \le F_i(T) \le \mathcal{S}_i\Big(\mu\int_{t=0}^{T} X_i(t) dt\Big).	\label{eq:inequalityDouble2}
	\end{align}
	In addition, the renewal reward theorem establishes that $\lim_{t \rightarrow \infty} \mathcal{S}_i(t)/t = 1$, a.s., and hence that 
	\begin{align*}
		\lim_{T \rightarrow \infty} \frac{\mathcal{S}_i\big(\mu\int_{t=0}^{T} X_i(t) dt\big)}{\mu\int_{t=0}^{T} X_i(t) dt} = 1, \ \  \text{a.s.}
	\end{align*}
	Combining the preceding results yields
	\begin{align*}
		\liminf_{T \rightarrow \infty} B_i(T)/T & \ge \liminf_{T \rightarrow \infty} F_i(T)/T \\
		& \ge \liminf_{T \rightarrow \infty} \Big(\mathcal{S}_i\Big(\mu\int_{t=0}^{T} X_i(t) dt\Big) - N\Big) / T\\
		& = \liminf_{T \rightarrow \infty} (\mu/T)\int_{t=0}^{T} X_i(t) dt \frac{\mathcal{S}_i\Big(\mu\int_{t=0}^{T} X_i(t) dt\Big) - N}{\mu\int_{t=0}^{T} X_i(t) dt}\\
		& = \liminf_{T \rightarrow \infty} (\mu/T)\int_{t=0}^{T} X_i(t) dt  \ \ \text{a.s.} 
	\end{align*}
	Further, applying the other inequalities provided in \eqref{eq:inequalityDouble1} and \eqref{eq:inequalityDouble2} analogously yields
	\begin{align*}
		\limsup_{T \rightarrow \infty} B_i(T)/T & \le \limsup_{T \rightarrow \infty} (\mu/T)\int_{t=0}^{T} X_i(t) dt  \ \ \text{a.s.} 
	\end{align*}
	Combining these results with the renewal reward theorem establishes the following:
	\begin{align*}
		\lim_{T \rightarrow \infty} B_{H}(T)/T + B_{L}(T)/T & = \lim_{T \rightarrow \infty} (\mu/T)\int_{t=0}^{T} (X_{H}(t) + X_{L}(t)) dt  \ \ \text{a.s.} \\
		& = \lim_{T \rightarrow \infty} (\mu/T)\int_{t=0}^{T} (N - Y(t)) dt \\
		& = \lim_{T\rightarrow \infty} \mu\Big(N - \frac{1}{T}\int_{t=0}^{T} Y(t)dt\Big) \\
		& = N\mu - \mu\frac{\E\big(\int_{t=t^{\text{start}}}^{t^{\text{end}}} Y(t)dt\big)}{\E(t^{\text{end}} - t^{\text{start}})} \ \ \text{a.s.}
	\end{align*}
	And combining this with \eqref{eq:firstOfGap} yields 
	\begin{align}
		\lim_{T\rightarrow \infty}&\frac{1}{T}\int_{t=0}^{T} \indicator{\tau_{\lfloor \sqrt{N}\rfloor}(t) < \tau_{1}(t)} d\mathcal{A}_{L}(\lambda_{L}Nt)\nonumber \\
		& = \lim_{T \rightarrow \infty}  B_{L}(T)/T\nonumber\\
		& =  N\mu - \mu \frac{\E\big(\int_{t=t^{\text{start}}}^{t^{\text{end}}} Y(t)dt\big)}{\E(t^{\text{end}} - t^{\text{start}})} - \lim_{T\rightarrow \infty} B_{H}(T)/T \ \ \text{a.s.} \nonumber\\		
		& =  N(\mu - \lambda_{H}) -  \mu\frac{\E\big(\int_{t=t^{\text{start}}}^{t^{\text{end}}} Y(t)dt\big)}{\E(t^{\text{end}} - t^{\text{start}})} +  \frac{\E\big(\int_{t=t^{\text{start}}}^{t^{\text{end}}} \indicator{\tau_{0}(t) = 0} d\mathcal{A}_{H}(\lambda_{H}Nt)\big)}{\E(t^{\text{end}} - t^{\text{start}})} \ \ \text{a.s.} \label{eq:boundOnL}
	\end{align}
	
	Further, the following implies that $B_{L}(T)/T$ is uniformly integrable:
	\begin{align*}
		\sup_{T\ge 1} \E\big((B_{L}(T)/T)^{2}\big) 
		& = \sup_{T\ge 1} \E\Big(\Big(\frac{1}{T}\int_{t=0}^{T}\indicator{\tau_{\lfloor \sqrt{N}\rfloor}(t) < \tau_{1}(t)} d\mathcal{A}_{L}(\lambda_{L}Nt)\Big)^{2}\Big) \\
		& \le \sup_{T \ge 1} \E\Big(\Big(\frac{1}{T}\int_{t=0}^{T} d\mathcal{A}_{L}(\lambda_{L}Nt)\Big)^{2}\Big)\\
		& = \sup_{T \ge 1} \E\big(\mathcal{A}_{L}(\lambda_{L}NT)^{2}\big)/T^{2}\\
		& = \sup_{T \ge 1}\lambda_{L}N/T + \lambda_{L}^{2}N^{2} \\
		& < \infty.
	\end{align*}
	And an analogous argument establishes the uniform integrability of $B_{H}(T)/T$. Since $B_{L}(T)/T$ and $B_{H}(T)/T$ both converge to a limit, almost surely, this uniform integrability enables us to commute the limit and expectation in \eqref{eq:defOfRFI}, which with \eqref{eq:firstOfGap} and \eqref{eq:boundOnL} yields the result.
\end{proof}

\begin{proof}[Proof of Proposition \ref{l:serverIdleness}] 
	The ratio's denominator is $\Omega(1/N)$, as $N \rightarrow \infty$. To see this, note that the expected duration of an epoch is no less than the expected time until its first action, and hence
	\begin{align}
		\E(t^{\text{end}} - t^{\text{start}}) \ge 1/(\lambda_{H} N + \lambda_{L} N + \mu N). \label{eq:denom}
	\end{align} 
	Accordingly, it will suffice to show that 
	\begin{align}
		\E_{\mathcal{L}^{\ell(\text{start})} = \mathcal{L}(\xi \barBreak \xi \in  \mathcal{S}(\hat{\tau}_{\lfloor \sqrt{N}\rfloor-1} < \hat{\tau}_{0}))}\Big(\int_{t=t^{\text{start}}}^{t^{\text{end}}} Y(t)dt \Big) = O(1/N). \label{eq:takeAwayMessageFromThisProp}
	\end{align}
	
	Establishing \eqref{eq:takeAwayMessageFromThisProp} will take several steps. First, take $N$ large enough so that $\lfloor \sqrt{N}\rfloor < N (\mu - \lambda_{H})/(2\mu)$, in which case the total service completion rate is at least $N (\lambda_{H} + \mu) / 2$ when $Y < \lfloor \sqrt{N}\rfloor$.
	
	Second, define the following times:
	\begin{itemize}
		\item $t^{Y=2}$: the first time after $t^{\text{start}}$ that the system has more than one idle server,
		\item $t^{\text{reject}}$: the first time after $t^{Y=2}$ that the system rejects an $L$ job or the idle-server process reaches zero, and
		\item $t^{Y=1}$: the first time at or after $t^{\text{reject}}$ that the idle-server process reaches one.
	\end{itemize} 
	Further, define corresponding action indices $\ell(Y=2)$, $\ell(\text{reject})$, and $\ell(Y=1)$, so that $\mathcal{L}^{\ell(Y=2)}$, $\mathcal{L}^{\ell(\text{reject})}$, and $\mathcal{L}^{\ell(Y=1)}$ represent the probability laws at times $t^{Y=2}$, $t^{\text{reject}}$, and $t^{Y=1}$. 
	
	Third, Lemma \ref{l:nearmemorylessness} establishes that $\mathcal{L}^{\ell(\text{start})} = \mathcal{L}(\xi \barBreak \xi \in  \mathcal{S}(\hat{\tau}_{\lfloor \sqrt{N}\rfloor-1} < \hat{\tau}_{0}))$. This probability law is cumbersome, so I will use the identity $\E(A \barBreak B) \le \E(A) / \Pr(B)$ to reduce the analysis to the simpler case in which the elements of $\Omega^{\ell(\text{start})}$ are independent Exp(1) random variables:
	\begin{align}
		\E_{\mathcal{L}^{\ell(\text{start})} = \mathcal{L}(\xi \barBreak \xi \in  \mathcal{S}(\hat{\tau}_{\lfloor \sqrt{N}\rfloor-1} < \hat{\tau}_{0}))}&\Big(\int_{t=t^{\text{start}}}^{t^{\text{end}}} Y(t)dt \Big) \nonumber\\
		& \le \frac{\E_{\mathcal{L}^{\ell(\text{start})} = \mathcal{L}(\xi)}\big(\int_{t=t^{\text{start}}}^{t^{\text{end}}} Y(t)dt\big)}{\Pr_{\mathcal{L}^{\ell(\text{start})} = \mathcal{L}(\xi)}\big(\hat{\tau}_{\lfloor \sqrt{N}\rfloor-1}^{\ell(\text{start})} < \hat{\tau}_{0}^{\ell(\text{start})}\big)}. \label{eq:conditionalDistFraction}
	\end{align}
	To lower bound the denominator in \eqref{eq:conditionalDistFraction}, note that the probability of event $\hat{\tau}_{\lfloor \sqrt{N}\rfloor-1}^{\ell(\text{start})} < \hat{\tau}_{0}^{\ell(\text{start})}$ under law $\mathcal{L}(\xi)$ is at least as large as the probability of a random walk never reaching 0, conditional on it starting at 1, incrementing at rate $N(\lambda_{H} + \mu)/2$, and decrementing at rate $N\lambda_{H}$:
	\begin{align}
		\Pr_{\mathcal{L}^{\ell(\text{start})} = \mathcal{L}(\xi)}\big(\hat{\tau}_{\lfloor \sqrt{N}\rfloor-1}^{\ell(\text{start})} < \hat{\tau}_{0}^{\ell(\text{start})}\big) \ge \frac{\mu - \lambda_{H}}{\lambda_{H} + \mu}. \label{eq:boundDenom}
	\end{align}

	Fourth, the numerator in \eqref{eq:conditionalDistFraction} satisfies:
	\begin{align}
		\E_{\mathcal{L}^{\ell(\text{start})} =\mathcal{L}(\xi)}\Big(\int_{t=t^{\text{start}}}^{t^{\text{end}}} Y(t)dt\Big)
		& \le \E_{\mathcal{L}^{\ell(\text{start})} =\mathcal{L}(\xi)}(t^{Y=2} - t^{\text{start}}) \nonumber\\
		& \quad + \E_{\mathcal{L}^{\ell(\text{start})} =\mathcal{L}(\xi)}\big((t^{\text{reject}} - t^{Y=2})\mathcal{Y}_{\leftarrow}\big)\nonumber\\
		& \quad + \E_{\mathcal{L}^{\ell(\text{start})} =\mathcal{L}(\xi)}\big((t^{Y=1} - t^{\text{reject}}) \mathcal{Y}_{\rightarrow}\big)\nonumber\\
		& \quad + \E_{\mathcal{L}^{\ell(\text{start})} =\mathcal{L}(\xi)}\Big(\int_{t=t^{Y=1} \wedge t^{\text{end}}}^{t^{\text{end}}}Y(t)dt\Big), \label{eq:threeTerms}\\
		\wq \mathcal{Y}_{\leftarrow} & \equiv \max_{t \in [t^{Y=2}, t^{\text{reject}}]} Y(t) \nonumber\\
		\aq \mathcal{Y}_{\rightarrow} & \equiv \max_{t \in [t^{\text{reject}}, t^{Y=1}]} Y(t). \nonumber
	\end{align}
	
	Fifth, I will create an $O(1/N)$ bound for the first expectation in \eqref{eq:threeTerms}. When $\mathcal{L}^{\ell(\text{start})} = \mathcal{L}(\xi)$, the expected value of $t^{Y=2} - t^{\text{start}}$ is no more than the expected length of time for a random walk to reach 2, given that it starts at 1, increments at rate $(N-1)\mu$, and decrements at rate $N(\lambda_{H} + \lambda_{L})$, which is $1/(N(\mu - \lambda_{H} - \lambda_{L}) - \mu)$.
	
	Sixth, I will create an $O(1/N)$ bound for the second expectation in \eqref{eq:threeTerms}. To begin, note that $\mathcal{L}^{\ell(\text{start})} = \mathcal{L}(\xi)$ implies $\mathcal{L}^{\ell(Y = 2)} = \mathcal{L}(\xi)$, since the system trivially rejects $L$ jobs when $Y = 1$. Thus, for our purposes, $\Omega^{\ell(Y = 2)}$ will comprise independent Exp(1) random variables.
	
	Next, observe that $t^{\text{reject}} - t^{Y=2}$ equals the number of times the idle-server process transitions between $t^{Y=2}$ and $t^{\text{reject}}$ multiplied by the average inter-transition time. Further, the idle-server process' inter-transition times are no more than the type-$H$ inter-arrival times. Consequently, if $\mathcal{L}^{\ell(\text{start})} = \mathcal{L}(\xi)$ (and thus $\mathcal{L}^{\ell(Y = 2)} = \mathcal{L}(\xi)$) then $t^{\text{reject}} - t^{Y=2}$ is no larger than a Gamma$(\chi, N\lambda_{H})$ random variable, where $N\lambda_{H}$ is the rate of type-$H$ arrival process and $\chi$ is the number of times the idle-server process jumps between times $t^{Y=2}$ and $t^{\text{reject}}$. This fact holds conditional on the sequence of transitions---and hence conditional on the specific value of $\mathcal{Y}_{\leftarrow}$---because $\min_{i} x_{i}$ and $\argmin_{i} x_{i}$ are independent when $\{x_{i}\}_{i}$ are independent exponential random variables. With this, the law of iterated expectation yields
	\begin{align}
		\E_{\mathcal{L}^{\ell(\text{start})} = \mathcal{L}(\xi)}\big(t^{\text{reject}} &- t^{Y=2} \barBreak \mathcal{Y}_{\leftarrow}\big) \nonumber\\
		& = \E_{\mathcal{L}^{\ell(\text{start})} = \mathcal{L}(\xi)}\big(\E(t^{\text{reject}} - t^{Y=2} \barBreak \chi) \barBreak \mathcal{Y}_{\leftarrow} \big) \nonumber\\
		& \le \E_{\mathcal{L}^{\ell(\text{start})} = \mathcal{L}(\xi)}\big(\chi /(N\lambda_{H}) \barBreak \mathcal{Y}_{\leftarrow}\big) .\label{eq:Nlamb}
	\end{align}
	Further, since $Y(t^{Y=2}) = 2$, by definition, we must also have:
	\begin{align}
		\mathcal{Y}_{\leftarrow} \le 2 + \chi.  \label{eq:YboundSideQuest}
	\end{align}
	Combining \eqref{eq:Nlamb} and \eqref{eq:YboundSideQuest} bounds the second expectation of \eqref{eq:threeTerms} in terms of the first two moments of $\chi$:
	\begin{align}
		\E_{\mathcal{L}^{\ell(\text{start})} = \mathcal{L}(\xi)}\big(&(t^{\text{reject}} - t^{Y=2}) \mathcal{Y}_{\leftarrow}\big) \nonumber\\
		& = \E_{\mathcal{L}^{\ell(\text{start})} = \mathcal{L}(\xi)}\big(\E\big(t^{\text{reject}} - t^{Y=2} \barBreak \mathcal{Y}_{\leftarrow}\big) \mathcal{Y}_{\leftarrow}\big) \nonumber\\
		& \le \E_{\mathcal{L}^{\ell(\text{start})} = \mathcal{L}(\xi)}\big(\E\big(\chi/(N\lambda_{H}) \barBreak \mathcal{Y}_{\leftarrow}\big) \mathcal{Y}_{\leftarrow}\big) \nonumber\\
		& \le \E_{\mathcal{L}^{\ell(\text{start})} = \mathcal{L}(\xi)}\big(\chi \mathcal{Y}_{\leftarrow}\big) /(N\lambda_{H})\nonumber\\
		& \le \E_{\mathcal{L}^{\ell(\text{start})} = \mathcal{L}(\xi)}(\chi (2+\chi))/(N\lambda_{H}). \label{eq:secondMoment}
	\end{align}
	
	Unfortunately, $\chi$ is difficult to directly control, since it depends on complex random stopping time $t^{\text{reject}}$. To avoid this random stopping time, define $\hat{Y}$ as what the idle-server process would be if the system accepted all jobs until it ran out of idle servers. Since $\hat{Y}$ is a continuation of $Y$, we must have
	\begin{align*}
		\chi & \le \hat{\chi},
	\end{align*}
	where $\hat{\chi}$ is the number of times the $\hat{Y}$ process transitions before hitting zero.
	
	I will now benchmark the $\hat{Y}$ process against the simpler random walk $\breve{Y}$, which starts at 2, increments at rate $N \mu$, and decrements at rate $N(\lambda_{H} + \lambda_{L})$. Further, we can couple these processes so that $\hat{Y} \le \breve{Y}$, and hence
	\begin{align*}
		\hat{\chi} & \le \breve{\chi},
	\end{align*}
	where $\breve{\chi}$ denotes the number of times the $\breve{Y}$ process transitions before hitting zero.	
	
	Since $\breve{Y}$ starts at two and ends at zero, at least $x/2 - 1$ of its first $x$ transitions must be upward jumps. But this process' jump-down rate exceeds its jump-up rate, so the probability of getting at least $x/2 - 1$ upward jumps in the first $x$ transitions falls exponentially quickly in $x$, by the Binomial Chernoff bound. And with the previous results, this implies that there exists $C_{1}, C_{2}>0$ for which 
	\begin{align}
		\Pr_{\mathcal{L}^{\ell(\text{start})} = \mathcal{L}(\xi)}(\chi \ge x) & \le \Pr_{\mathcal{L}^{\ell(\text{start})} = \mathcal{L}(\xi)}(\hat{\chi} \ge x) \nonumber \\
		& \le \Pr_{\mathcal{L}^{\ell(\text{start})} = \mathcal{L}(\xi)}(\breve{\chi} \ge x)  \nonumber \\
		& \le C_{1}\exp(-C_{2}x).  \label{eq:c1Bound}
	\end{align}
	This universal tail bound implies that $\E_{\mathcal{L}^{\ell(\text{start})} = \mathcal{L}(\xi)}(\chi (2+\chi)) = O(1)$ as $N \rightarrow \infty$, which with \eqref{eq:secondMoment} yields the desired bound for the second expectation in \eqref{eq:threeTerms}:
	\begin{align*}
		\E_{\mathcal{L}^{\ell(\text{start})} = \mathcal{L}(\xi)}&\big((t^{\text{reject}} - t^{Y=2}) \mathcal{Y}_{\leftarrow}\big) \\
		& \le \E_{\mathcal{L}^{\ell(\text{start})} = \mathcal{L}(\xi)}(\chi (2+\chi))/(N\lambda_{H}) \\
		& = O(1)/(N\lambda_{H}) \\	
		& = O(1/N) .	
	\end{align*}
	
	Seventh, I will control the tail of $\mathcal{Y}_{\leftarrow}$ with \eqref{eq:YboundSideQuest} and \eqref{eq:c1Bound}, which imply the following, for sufficiently large $y$:
	\begin{align}
		\Pr_{\mathcal{L}^{\ell(\text{start})} = \mathcal{L}(\xi)}(\mathcal{Y}_{\leftarrow} \ge y ) &   \le \Pr_{\mathcal{L}^{\ell(\text{start})} = \mathcal{L}(\xi)}\big(\chi \ge y - 2\big) \nonumber\\
		&  \le C_{1}\exp(-C_{2}(y-2)). \label{eq:extTailBound}
	\end{align}
	Note that this exponential tail bound implies that $\mathcal{Y}_{\leftarrow}$ has a bounded second moment:
	\begin{align}
		\E_{\mathcal{L}^{\ell(\text{start})} = \mathcal{L}(\xi)}(\mathcal{Y}_{\leftarrow}^{2}) = O(1). \label{eq:extSecondMoment}
	\end{align}
	
	Eighth, I will create an $O(1/N)$ bound for the third expectation in \eqref{eq:threeTerms}. Since $Y^{\ell(\text{reject})} = 0$ implies $(t^{Y=1} - t^{\text{reject}}) \mathcal{Y}_{\rightarrow} = 0$, we can restrict attention to the $Y^{\ell(\text{reject})} > 0$ case. Lemma \ref{l:nearmemorylessness} establishes that there are two possible values for $\mathcal{L}^{\ell(\text{reject})}$ in this case: if the system accepted \emph{any} $L$ jobs between times $t^{Y = 2}$ and $t^{\text{reject}}$ then $\mathcal{L}^{\ell(\text{reject})} = \mathcal{L}_{\text{any}} \equiv \mathcal{L}(\xi \barBreak \xi \in  \mathcal{S}(\tau_{1} < \tau_{\lfloor \sqrt{N}\rfloor}) \cap \mathcal{S}(\hat{\tau}_{\lfloor \sqrt{N}\rfloor-1} < \hat{\tau}_{0}))$, and otherwise $\mathcal{L}^{\ell(\text{reject})} =\mathcal{L}_{\text{none}} \equiv \mathcal{L}(\xi \barBreak \xi \in  \mathcal{S}(\tau_{1} < \tau_{\lfloor \sqrt{N}\rfloor}))$. 
	
	To avoid analyzing two separate laws, I will now use line \eqref{eq:boundDenom} and identity $\E(A \mid B) \le \E(A) / \Pr(B)$ to bound the expectation under $\mathcal{L}_{\text{any}}$ in terms of the expectation under $\mathcal{L}_{\text{none}}$ (the second inequality below holds because $\hat{\tau}_{\lfloor \sqrt{N}\rfloor-1}^{\ell(Y = 1)} < \hat{\tau}_{0}^{\ell(Y = 1)}$ implies $\hat{\tau}_{\lfloor \sqrt{N}\rfloor-1}^{\ell(\text{reject})} < \hat{\tau}_{0}^{\ell(\text{reject})}$):
	\begin{align*}
		\E_{\mathcal{L}^{\ell(\text{reject})} = \mathcal{L}_{\text{any}}}&\big((t^{Y=1} - t^{\text{reject}}) \mathcal{Y}_{\rightarrow}\big) \nonumber \\
		& \le \frac{\E_{\mathcal{L}^{\ell(\text{reject})} = \mathcal{L}_{\text{none}}}\big((t^{Y=1} - t^{\text{reject}}) \mathcal{Y}_{\rightarrow}\big)}{\Pr_{\mathcal{L}^{\ell(\text{reject})}=\mathcal{L}_{\text{none}}}(\hat{\tau}_{\lfloor \sqrt{N}\rfloor-1}^{\ell(\text{reject})} < \hat{\tau}_{0}^{\ell(\text{reject})})} \nonumber \\
		& \le \frac{\E_{\mathcal{L}^{\ell(\text{reject})} = \mathcal{L}_{\text{none}}}\big((t^{Y=1} - t^{\text{reject}}) \mathcal{Y}_{\rightarrow}\big)}{\Pr_{\mathcal{L}^{\ell(Y = 1)}=\mathcal{L}(\xi)}(\hat{\tau}_{\lfloor \sqrt{N}\rfloor-1}^{\ell(Y = 1)} < \hat{\tau}_{0}^{\ell(Y = 1)})} \nonumber \\
		& = \frac{\E_{\mathcal{L}^{\ell(\text{reject})} = \mathcal{L}_{\text{none}}}\big((t^{Y=1} - t^{\text{reject}}) \mathcal{Y}_{\rightarrow}\big)}{\Pr_{\mathcal{L}^{\ell(\text{start})} = \mathcal{L}(\xi)}\big(\hat{\tau}_{\lfloor \sqrt{N}\rfloor-1}^{\ell(\text{start})} < \hat{\tau}_{0}^{\ell(\text{start})}\big) } \nonumber \\
		& \le \frac{\mu+\lambda_{H}}{\mu-\lambda_{H}}\E_{\mathcal{L}^{\ell(\text{reject})} = \mathcal{L}_{\text{none}}}\big((t^{Y=1} - t^{\text{reject}}) \mathcal{Y}_{\rightarrow}\big). 
	\end{align*} 
	Note, the result above holds under \emph{all} feasible values of $Y^{\ell(\text{reject})}$, which laws $\mathcal{L}_{\text{none}}$ and $\mathcal{L}_{\text{any}}$ implicitly condition on. Indeed, whenever I condition on $\mathcal{L}^{\ell(\text{reject})} = \mathcal{L}_{\text{any}}$ or $\mathcal{L}^{\ell(\text{reject})} = \mathcal{L}_{\text{none}}$, I also implicitly condition on the value of $Y^{\ell(\text{reject})}$. For example, the preceding result implies the following:
	\begin{align}
		\E_{\mathcal{L}^{\ell(\text{start})} =\mathcal{L}(\xi)} &\big((t^{Y=1} - t^{\text{reject}}) \mathcal{Y}_{\rightarrow} \barBreak Y^{\ell(\text{reject})}\big) \nonumber \\
		& \le \max_{\mathcal{L} \in \{\mathcal{L}_{\text{any}}, \mathcal{L}_{\text{none}}\}} \E_{\mathcal{L}^{\ell(\text{reject})} = \mathcal{L}}\big((t^{Y=1} - t^{\text{reject}}) \mathcal{Y}_{\rightarrow} \barBreak Y^{\ell(\text{reject})}\big) \nonumber \\
		& \le \frac{\mu+\lambda_{H}}{\mu-\lambda_{H}}  \E_{\mathcal{L}^{\ell(\text{reject})} = \mathcal{L}_{\text{none}}}\big((t^{Y=1} - t^{\text{reject}}) \mathcal{Y}_{\rightarrow} \barBreak Y^{\ell(\text{reject})}\big). \label{eq:switchToL2}
	\end{align}
	However, to streamline the notation, I will henceforth suppress the $Y^{\ell(\text{reject})}$ conditioning.
	
	I will now bound the expectation in \eqref{eq:switchToL2} with the insight underpinning line \eqref{eq:Nlamb}: that the idle-server process' inter-transition times are no more than rate-$\lambda_{H}$ exponential random variables, regardless of the idle-server process trajectory. As in \eqref{eq:Nlamb}, this fact enables us to replace an expected timespan with an expected jump count:
	\begin{align}
		\E_{\mathcal{L}^{\ell(\text{reject})} = \mathcal{L}_{\text{none}}}&\big((t^{Y=1} - t^{\text{reject}}) \mathcal{Y}_{\rightarrow}\big) \nonumber\\
		& \le \sum_{i \in \{\text{before}, \text{after}\}}\E_{\mathcal{L}^{\ell(\text{reject})} = \mathcal{L}_{\text{none}}}\big(\chi_{i}\mathcal{Y}_{\rightarrow}\big)/(N\lambda_{H}) \nonumber\\
		& = \sum_{i \in \{\text{before}, \text{after}\}}\E_{\mathcal{L}^{\ell(\text{reject})} = \mathcal{L}_{\text{none}}}\big(\E(\chi_{i} \barBreak \mathcal{Y}_{\rightarrow})\mathcal{Y}_{\rightarrow}\big)/(N\lambda_{H}), \label{eq:littlem}
	\end{align}
	where $\chi_{\text{before}}$ and $\chi_{\text{after}}$ denote the number of times the idle-server process transitions before and after reaching $\mathcal{Y}_{\rightarrow}$, respectively. 
	
	Now to bound $\E(\chi_{\text{before}} \barBreak \mathcal{Y}_{\rightarrow})$ in \eqref{eq:littlem}, note that the distribution of the post-$t^{\text{reject}}$, pre-$t^{Y=1}$ idle-server process conditional on $\mathcal{Y}_{\rightarrow} = y$ is equivalent to the distribution of this process conditional on $\tau_{y}^{\ell(\text{reject})} < \tau_{1}^{\ell(\text{reject})}$ and $\tau_{1}^{\ell(Y = y)} < \tau_{y+1}^{\ell(Y = y)}$. Also, note that conditioning on $\mathcal{Y}_{\rightarrow} = y$ makes $\chi_{\text{before}}$ independent of restriction $\tau_{1}^{\ell(Y = y)} < \tau_{y+1}^{\ell(Y = y)}$. Hence, given $\mathcal{Y}_{\rightarrow} = y$, the distribution of $\chi_{\text{before}}$ equals the distribution of the number of jumps made by the idle-server process before hitting $y$, conditional on it not hitting 1 along the way. Further, it's simple to show that dropping the ``not hitting 1 along the way" condition would not decrease the expected number of transitions required to hit level $y$. Likewise reducing the idle-server process' jump-up rate from $\mu(N - Y)$ to $N (\lambda_{H} + \mu) / 2$ would also not decrease the expected number of transitions required to hit level $y$ (keep in mind that $\mathcal{Y}_{\rightarrow} < \lfloor \sqrt{N}\rfloor$). Consequently, we have
	\begin{align}
		\E(\chi_{\text{before}} \barBreak \mathcal{Y}_{\rightarrow}) \le \frac{\mu + 3\lambda_{H}}{\mu - \lambda_{H}}\mathcal{Y}_{\rightarrow}, \label{eq:hitCalY}
	\end{align}
	where the expression on the right is the expected number of transitions required for a random walk to hit $\mathcal{Y}_{\rightarrow}$, conditional on it starting at 0, incrementing at rate $N (\lambda_{H} + \mu) / 2$, and decrementing at rate $N \lambda_{H}$.
	
	I will bound $\E(\chi_{\text{after}} \barBreak \mathcal{Y}_{\rightarrow})$ in \eqref{eq:littlem} in a similar fashion. First, recall that the distribution of the post-$t^{\text{reject}}$, pre-$t^{Y=1}$ idle-server process conditional on $\mathcal{Y}_{\rightarrow} = y$ is equivalent to the distribution of this process conditional on $\tau_{y}^{\ell(\text{reject})} < \tau_{1}^{\ell(\text{reject})}$ and $\tau_{1}^{\ell(Y = y)} < \tau_{y+1}^{\ell(Y = y)}$. And, as before, conditioning on $\mathcal{Y}_{\rightarrow} = y$ makes $\chi_{\text{after}}$ independent of restriction $\tau_{y}^{\ell(\text{reject})} < \tau_{1}^{\ell(\text{reject})}$. Hence, given $\mathcal{Y}_{\rightarrow} = y$, the distribution of $\chi_{\text{after}}$ equals the distribution of the number of jumps made by the idle-server process before hitting 1, conditional on it not hitting $y + 1$ along the way. The probability of $\chi_{\text{after}}$ exceeding $x$, therefore, is no larger than the unconditional probability of the idle-server process hitting 1 in $x$ steps divided by the unconditional probability of hitting 1 before $y + 1$. To bound the denominator, note that the unconditional probability of hitting 1 before $y + 1$ is no less than the unconditional probability of at least $y$ type-$H$ jobs arriving before the next service completion, which is no less than $\big(\frac{\lambda_{H}}{\lambda_{H} + \mu}\big)^{y}$. And to bound the numerator, note that the unconditional probability of the idle-server process hitting 1 in $x$ steps is no larger than the probability that at least half of the next $x$ transitions are downward hops, which is less than $K_{1}\exp(-K_{2}x)$, for some $K_{1},K_{2} > 0$, by the binomial Chernoff. With this, we find the following:
	\begin{align*}
		\Pr(\chi_{\text{after}} > x \barBreak \mathcal{Y}_{\rightarrow} = y) \le \frac{K_{1}\exp(-K_{2}x)}{\big(\frac{\lambda_{H}}{\lambda_{H} + \mu}\big)^{y}} = K_{1}\exp\Big(-K_{2}x - \log\Big(\frac{\lambda_{H}}{\lambda_{H} + \mu}\Big)y\Big),
	\end{align*}
	and hence
	\begin{align}
		\Pr\Big(\chi_{\text{after}} > x - \log\Big(\frac{\lambda_{H}}{\lambda_{H} + \mu}\Big)y/K_{2} \barBreak \mathcal{Y}_{\rightarrow} = y\Big) \le K_{1}\exp(-K_{2}x). \label{eq:seeThisOne}
	\end{align}
	And, with this, we get the desired conditional expectation bound:
	\begin{align*}
		\E(\chi_{\text{after}} \barBreak \mathcal{Y}_{\rightarrow}) & =  \int_{x = 0}^{\infty}
		\Pr(\chi_{\text{after}} > x \barBreak \mathcal{Y}_{\rightarrow})
		dx \\
		& \le \int_{x = 0}^{-\log\big(\frac{\lambda_{H}}{\lambda_{H} + \mu}\big)\mathcal{Y}_{\rightarrow}/K_{2}} dx + \int_{x = -\log\big(\frac{\lambda_{H}}{\lambda_{H} + \mu}\big)\mathcal{Y}_{\rightarrow}/K_{2}}^{\infty}  \Pr(\chi_{\text{after}} > x \barBreak \mathcal{Y}_{\rightarrow}) dx \\
		& = -\log\Big(\frac{\lambda_{H}}{\lambda_{H} + \mu}\Big)\mathcal{Y}_{\rightarrow}/K_{2} + \int_{x = 0}^{\infty}  \Pr(\chi_{\text{after}} > x - \log\Big(\frac{\lambda_{H}}{\lambda_{H} + \mu}\Big)\mathcal{Y}_{\rightarrow}/K_{2} \barBreak \mathcal{Y}_{\rightarrow}) dx \\
		& \le -\log\Big(\frac{\lambda_{H}}{\lambda_{H} + \mu}\Big)\mathcal{Y}_{\rightarrow}/K_{2} + \int_{x = 0}^{\infty}  K_{1}\exp(-K_{2}x) dx \\
		& = \Big(K_{1}-\log\Big(\frac{\lambda_{H}}{\lambda_{H} + \mu}\Big)\mathcal{Y}_{\rightarrow}\Big)/K_{2} .
	\end{align*}
	
	Combining the preceding result with \eqref{eq:littlem} and \eqref{eq:hitCalY} establishes the following, for some universal constant $M > 0$:
	\begin{align}
		\E_{\mathcal{L}^{\ell(\text{reject})} = \mathcal{L}_{\text{none}}}&\big((t^{Y=1} - t^{\text{reject}}) \mathcal{Y}_{\rightarrow}\big) \nonumber \\
		& \le \E_{\mathcal{L}^{\ell(\text{reject})} = \mathcal{L}_{\text{none}}}\Big(\frac{\mu + 3\lambda_{H}}{\mu - \lambda_{H}}\mathcal{Y}_{\rightarrow}^{2}\Big)/(N\lambda_{H}) \nonumber \\
		& \qquad + \E_{\mathcal{L}^{\ell(\text{reject})} = \mathcal{L}_{\text{none}}}\Big(\Big(K_{1}-\log\Big(\frac{\lambda_{H}}{\lambda_{H} + \mu}\Big)\mathcal{Y}_{\rightarrow}\Big)\mathcal{Y}_{\rightarrow}/K_{2}\Big)/(N\lambda_{H}) \nonumber \\
		& \le M\E_{\mathcal{L}^{\ell(\text{reject})} = \mathcal{L}_{\text{none}}}(\mathcal{Y}_{\rightarrow}^{2})/N. \label{eq:quadraticForm}
	\end{align}
	
	To bound the expectation in \eqref{eq:quadraticForm}, I will now show that the tail of the $\mathcal{Y}_{\rightarrow}$ probability distribution decays exponentially. To this end, define $t^{Y = y}$ as the first time after $t^{\text{reject}}$ that the idle-server process reaches some given $y$, define $t^{\text{return}}$ as the first time after $t^{Y = y}$ that this process returns to the level $Y^{\ell(\text{reject})}$, and define corresponding action indices $\ell(Y = y)$ and $\ell(\text{return})$. With this, we have:
	\begin{align}
		& \Pr_{\mathcal{L}^{\ell(\text{reject})} = \mathcal{L}_{\text{none}}}(\mathcal{Y}_{\rightarrow} \ge y) \nonumber \\
		& \qquad =
		\Pr_{\mathcal{L}^{\ell(\text{reject})} = \mathcal{L}_{\text{none}}}\big(\tau_{y}^{\ell(\text{reject})} < \tau_{1}^{\ell(\text{reject})}\big) \nonumber \\
		& \qquad =  \Pr_{\mathcal{L}^{\ell(\text{reject})} = \mathcal{L}(\xi)}\big(\tau_{y}^{\ell(\text{reject})} < \tau_{1}^{\ell(\text{reject})} \barBreak \tau_{1}^{\ell(\text{reject})} < \tau_{\lfloor \sqrt{N}\rfloor}^{\ell(\text{reject})}\big) \nonumber \\
		& \qquad = \frac{\Pr_{\mathcal{L}^{\ell(\text{reject})} = \mathcal{L}(\xi)}\big(\tau_{y}^{\ell(\text{reject})} < \tau_{1}^{\ell(\text{reject})}  ,\ \tau_{Y^{\ell(\text{reject})}}^{\ell(Y = y)} < \tau_{\lfloor \sqrt{N}\rfloor}^{\ell(Y = y)},\ \tau_{1}^{\ell(\text{return})} < \tau_{\lfloor \sqrt{N}\rfloor}^{\ell(\text{return})}\big)}{\Pr_{\mathcal{L}^{\ell(\text{reject})} = \mathcal{L}(\xi)}\big(\tau_{1}^{\ell(\text{reject})} < \tau_{\lfloor \sqrt{N}\rfloor}^{\ell(\text{reject})}\big)} \nonumber \\
		& \qquad = \frac{\prod_{E \in \left\{\big(\tau_{y}^{\ell(\text{reject})} < \tau_{1}^{\ell(\text{reject})}\big), \big(\tau_{Y^{\ell(\text{reject})}}^{\ell(Y = y)} < \tau_{\lfloor \sqrt{N}\rfloor}^{\ell(Y = y)}\big), \big(\tau_{1}^{\ell(\text{return})} < \tau_{\lfloor \sqrt{N}\rfloor}^{\ell(\text{return})}\big)\right\}} \Pr_{\mathcal{L}^{\ell(\text{reject})} = \mathcal{L}(\xi)}(E)}{\Pr_{\mathcal{L}^{\ell(\text{reject})} = \mathcal{L}(\xi)}\big(\tau_{1}^{\ell(\text{reject})} < \tau_{\lfloor \sqrt{N}\rfloor}^{\ell(\text{reject})}\big)} \nonumber \\
		& \qquad \le \Pr_{\mathcal{L}^{\ell(\text{reject})} = \mathcal{L}(\xi)}\big(\tau_{Y^{\ell(\text{reject})}}^{\ell(Y = y)} < \tau_{\lfloor \sqrt{N}\rfloor}^{\ell(Y = y)}\big) \nonumber \\
		& \qquad \le \Big(\frac{2\lambda_{H}}{\lambda_{H} + \mu}\Big)^{y - Y^{\ell(\text{reject})}}. \label{eq:controlrightarrowY}
	\end{align}
	
	This display is complex, so I will explain it one line at a time. The first inequality asserts that $\mathcal{Y}_{\rightarrow}$ cannot exceed $y$ unless the reject-all-$L$-jobs idle-server process reaches $y$ before 1 (keep in mind that the actual idle-server process tracks the reject-all-$L$-jobs idle-server between times $t^{\text{reject}}$ and $t^{Y=1}$). The second line makes the $\mathcal{L}^{\ell(\text{reject})} = \mathcal{L}_{\text{none}}$ initialization explicit: law $\mathcal{L}_{\text{none}}$ is equivalent to law $\mathcal{L}(\xi)$ plus condition $\tau_{1}^{\ell(\text{reject})} < \tau_{\lfloor \sqrt{N}\rfloor}^{\ell(\text{reject})}$. The third line holds by Bayes rule: $\tau_{y}^{\ell(\text{reject})} < \tau_{1}^{\ell(\text{reject})} < \tau_{\lfloor \sqrt{N}\rfloor}^{\ell(\text{reject})}$ holds if and only if the reject-all-$L$-jobs idle-server process travels from $Y^{\ell(\text{reject})}$ to $y$ without touching 1, then travels from $y$ to $Y^{\ell(\text{reject})}$ without touching $\lfloor \sqrt{N}\rfloor$, and then travels from $Y^{\ell(\text{reject})}$ to 1 without touching $\lfloor \sqrt{N}\rfloor$. The fourth line holds because events $\tau_{y}^{\ell(\text{reject})} < \tau_{1}^{\ell(\text{reject})}$, $\tau_{Y^{\ell(\text{reject})}}^{\ell(Y = y)} < \tau_{\lfloor \sqrt{N}\rfloor}^{\ell(Y = y)}$, and $\tau_{1}^{\ell(\text{return})} < \tau_{\lfloor \sqrt{N}\rfloor}^{\ell(\text{return})}$ are independent when $\mathcal{L}^{\ell(\text{reject})} = \mathcal{L}(\xi)$. The penultimate line holds because probability $\Pr_{\mathcal{L}^{\ell} = \mathcal{L}(\xi)}(\tau_{1}^{\ell} < \tau_{\lfloor \sqrt{N}\rfloor}^{\ell})$ does not vary with index $\ell$. And the final line holds because the probability of the reject-all-$L$-jobs idle-server process traveling from $y$ to $Y^{\ell(\text{reject})}$ before $\lfloor \sqrt{N}\rfloor$ is no larger than the probability of a random walk ever reaching $Y^{\ell(\text{reject})}$, given that it increments at rate $N(\lambda_{H} + \mu)/2$, decrements at rate $N\lambda_{H}$, and starts at $y$, which is $\big(\frac{2\lambda_{H}}{\lambda_{H} + \mu}\big)^{y - Y^{\ell(\text{reject})}}$. 
	
	Line \eqref{eq:controlrightarrowY}'s tail bound implies the following second-moment bound, for some universal constant $Q > 0$:
	\begin{align*}
		\E_{\mathcal{L}^{\ell(\text{reject})} = \mathcal{L}_{\text{none}}}(\mathcal{Y}_{\rightarrow}^{2}) \le Q (Y^{\ell(\text{reject})})^{2}.
	\end{align*} 
	When parsing this expression, keep in mind that the expectations we've been working with recently implicitly condition on $Y^{\ell(\text{reject})}$. Accordingly, we can combine the bound above with \eqref{eq:switchToL2} and \eqref{eq:quadraticForm} to establish that
	\begin{align*}
		\E_{\mathcal{L}^{\ell(\text{start})} =\mathcal{L}(\xi)} \big((t^{Y=1} - t^{\text{reject}}) \mathcal{Y}_{\rightarrow} \barBreak Y^{\ell(\text{reject})}\big) \le MQ \frac{\mu+\lambda_{H}}{\mu-\lambda_{H}} (Y^{\ell(\text{reject})})^{2}/N.
	\end{align*}
	Finally, combining this with \eqref{eq:extSecondMoment} and the law of iterated expectations yields the desired $O(1/N)$ convergence rate:	
	\begin{align*}
		\E_{\mathcal{L}^{\ell(\text{start})} =\mathcal{L}(\xi)} \big((t^{Y=1} - t^{\text{reject}}) \mathcal{Y}_{\rightarrow}\big) & = \E_{\mathcal{L}^{\ell(\text{start})} =\mathcal{L}(\xi)}\Big(\E_{\mathcal{L}^{\ell(\text{start})} =\mathcal{L}(\xi)} \big((t^{Y=1} - t^{\text{reject}}) \mathcal{Y}_{\rightarrow} \barBreak Y^{\ell(\text{reject})}\big)\Big) \\
		& \le MQ \frac{\mu+\lambda_{H}}{\mu-\lambda_{H}} \E_{\mathcal{L}^{\ell(\text{start})} =\mathcal{L}(\xi)}\big((Y^{\ell(\text{reject})})^{2}\big)/N \\
		& \le MQ \frac{\mu+\lambda_{H}}{\mu-\lambda_{H}} \E_{\mathcal{L}^{\ell(\text{start})} =\mathcal{L}(\xi)}(\mathcal{Y}_{\leftarrow}^{2})/N \\
		& = MQ \frac{\mu+\lambda_{H}}{\mu-\lambda_{H}} O(1)/N \\
		& = O(1/N).		
	\end{align*}
	
	Eighth, I will conclude the proof by accounting for the fourth expectation in \eqref{eq:threeTerms}. Note that the integral in this expectation is zero unless $t^{Y=1} < t^{\text{end}}$, an event that will not hold unless $\max_{t \in [t^{Y=2}, t^{Y = 1}]} Y(t) \ge \lfloor \sqrt{N}\rfloor-1$. We will now upper-bound the probability of this event. 
	
	To begin, note that the argument underpinning \eqref{eq:switchToL2} implies that 
	\begin{align*}
		\Pr_{\mathcal{L}^{\ell(\text{start})} =\mathcal{L}(\xi)}(\mathcal{Y}_{\rightarrow} \ge y \barBreak Y^{\ell(\text{reject})}) \le \frac{\mu+\lambda_{H}}{\mu-\lambda_{H}}  \Pr_{\mathcal{L}^{\ell(\text{reject})} = \mathcal{L}_{\text{none}}}(\mathcal{Y}_{\rightarrow} \ge y \barBreak Y^{\ell(\text{reject})}).
	\end{align*}
	Combining this with \eqref{eq:extTailBound} and \eqref{eq:controlrightarrowY} yields the following exponential tail bound:
	\begin{align*}
		\Pr_{\mathcal{L}^{\ell(\text{start})} =\mathcal{L}(\xi)}\big(\mathcal{Y}_{\rightarrow} \ge y \big) & \le \Pr_{\mathcal{L}^{\ell(\text{start})} =\mathcal{L}(\xi)}\big(Y^{\ell(\text{reject})} \ge y/2 \ \cup\ \mathcal{Y}_{\rightarrow} - Y^{\ell(\text{reject})}\ge y/2  \big) \nonumber \\
		& \le \Pr_{\mathcal{L}^{\ell(\text{start})} =\mathcal{L}(\xi)}\big(Y^{\ell(\text{reject})} \ge y/2\big) +  \Pr_{\mathcal{L}^{\ell(\text{start})} =\mathcal{L}(\xi)}\big(\mathcal{Y}_{\rightarrow} - Y^{\ell(\text{reject})}\ge y/2  \big) \nonumber \\
		& \le \Pr_{\mathcal{L}^{\ell(\text{start})} =\mathcal{L}(\xi)}\big(\mathcal{Y}_{\leftarrow} \ge y/2\big) + \frac{\mu+\lambda_{H}}{\mu-\lambda_{H}}  \Pr_{\mathcal{L}^{\ell(\text{reject})} = \mathcal{L}_{\text{none}}}\big(\mathcal{Y}_{\rightarrow}\ge y/2 + Y^{\ell(\text{reject})}\big) \nonumber \\
		& \le C_{1}\exp(-C_{2}(y/2-2)) + \frac{\mu+\lambda_{H}}{\mu-\lambda_{H}}\Big(\frac{2\lambda_{H}}{\lambda_{H} + \mu}\Big)^{y/2}.
	\end{align*}
	And combining this with \eqref{eq:extTailBound} yields the following:
	\begin{align}
		\Pr_{\mathcal{L}^{\ell(\text{start})} =\mathcal{L}(\xi)}&\Big(\max_{t \in [t^{Y=2}, t^{Y = 1}]} Y(t) \ge y\Big)\nonumber \\
		& = \Pr_{\mathcal{L}^{\ell(\text{start})} =\mathcal{L}(\xi)}(\mathcal{Y}_{\leftarrow} \ge y \ \cup\ \mathcal{Y}_{\rightarrow} \ge y) \nonumber\\
		& \le \Pr_{\mathcal{L}^{\ell(\text{start})} =\mathcal{L}(\xi)}(\mathcal{Y}_{\leftarrow} \ge y) + \Pr_{\mathcal{L}^{\ell(\text{start})} =\mathcal{L}(\xi)}(\mathcal{Y}_{\rightarrow} \ge y) \nonumber\\
		& \le C_{1}\exp(-C_{2}(y-2)) + C_{1}\exp(-C_{2}(y/2-2)) + \frac{\mu+\lambda_{H}}{\mu-\lambda_{H}}\Big(\frac{2\lambda_{H}}{\lambda_{H} + \mu}\Big)^{y/2}\nonumber\\
		& = \exp(-\Omega(y)). \label{eq:ohhhExpBound} 
	\end{align}
	Accordingly, the probability that $t^{Y=1} < t^{\text{end}}$---and hence the probability that the integral in the fourth expectation of \eqref{eq:threeTerms} is non-zero---is no larger than $\exp(-\Omega(\sqrt{N}))$. 
	
	Further if $t^{Y=1} < t^{\text{end}}$, then $\mathcal{L}^{\ell(Y = 1)} =\mathcal{L}(\xi)$, by Lemma \ref{l:nearmemorylessness}. And with \eqref{eq:ohhhExpBound} this implies the following:
	\begin{align*}
		\E_{\mathcal{L}^{\ell(\text{start})} =\mathcal{L}(\xi)} &\Big(\int_{t=t^{Y=1} \wedge t^{\text{end}}}^{t^{\text{end}}}Y(t)dt\Big) \\
		& = \Pr_{\mathcal{L}^{\ell(\text{start})} =\mathcal{L}(\xi)}(t^{Y=1} < t^{\text{end}}) \E_{\mathcal{L}^{\ell(\text{start})} =\mathcal{L}(\xi)}\Big(\int_{t=t^{Y=1} \wedge t^{\text{end}}}^{t^{\text{end}}}Y(t)dt \barBreak t^{Y=1} < t^{\text{end}}\Big) \\
		& \le \Pr_{\mathcal{L}^{\ell(\text{start})} =\mathcal{L}(\xi)}\Big(\max_{t \in [t^{Y=2}, t^{Y = 1})} Y(t) \ge \lfloor \sqrt{N}\rfloor-1\Big) \E_{\mathcal{L}^{\ell(Y = 1)} =\mathcal{L}(\xi)}\Big(\int_{t=t^{Y=1}}^{t^{\text{end}}}Y(t)dt\Big) \\
		& = \exp(-\Omega(\sqrt{N}))	\E_{\mathcal{L}^{\ell(\text{start})} =\mathcal{L}(\xi)}\Big(\int_{t=t^{\text{start}}}^{t^{\text{end}}} Y(t)dt\Big).
	\end{align*}
	Note, we can replace $\E_{\mathcal{L}^{\ell(Y = 1)} =\mathcal{L}(\xi)}\Big(\int_{t=t^{Y=1}}^{t^{\text{end}}}Y(t)dt\Big)$ with $\E_{\mathcal{L}^{\ell(\text{start})} =\mathcal{L}(\xi)}\Big(\int_{t=t^{\text{start}}}^{t^{\text{end}}} Y(t)dt\Big)$ in the last line above, because the post-$t^{Y=1}$, pre-$t^{\text{end}}$ idle-server process conditional on $\mathcal{L}^{\ell(Y = 1)} =\mathcal{L}(\xi)$ follows the same law as the post-$t^{\text{start}}$, pre-$t^{\text{end}}$ idle-server process conditional on $\mathcal{L}^{\ell(\text{start})} =\mathcal{L}(\xi)$.
	
	Combining the bound above with \eqref{eq:threeTerms} yields the following:
	\begin{align}
		\E_{\mathcal{L}^{\ell(\text{start})} =\mathcal{L}(\xi)}\Big(\int_{t=t^{\text{start}}}^{t^{\text{end}}} Y(t)dt\Big)
		& = O(1/N) + \E_{\mathcal{L}^{\ell(\text{start})} =\mathcal{L}(\xi)}\Big(\int_{t=t^{Y=1} \wedge t^{\text{end}}}^{t^{\text{end}}}Y(t)dt\Big) \nonumber\\
		& = O(1/N) + \exp(-\Omega(\sqrt{N})) 	\E_{\mathcal{L}^{\ell(\text{start})} =\mathcal{L}(\xi)}\Big(\int_{t=t^{\text{start}}}^{t^{\text{end}}} Y(t)dt\Big) \nonumber\\
		& = \frac{O(1/N)}{1 - \exp(-\Omega(\sqrt{N}))} \nonumber \\
		& = O(1/N), \label{eq:saveForWayLater}
	\end{align}
	where the $O(1/N)$ term in the first line in the display above represents the first three expectations on the right-hand side of \eqref{eq:threeTerms}. Finally, combining \eqref{eq:saveForWayLater} with \eqref{eq:conditionalDistFraction} and \eqref{eq:boundDenom} establishes \eqref{eq:takeAwayMessageFromThisProp}, which concludes the proof.
	
	However, before I end, I will derive a few miscellaneous results, which I will use in future proofs. First, working through the Chernoff bound, we find that the constant $C_{2}$ in line \eqref{eq:extTailBound} can be set to $\frac{(\lambda_{H} + \lambda_{L} - \mu)^{2}}{2(\lambda_{H} + \lambda_{L} + \mu)^{2}}$. And combining this with the logic of \eqref{eq:conditionalDistFraction} and \eqref{eq:boundDenom}, and we find that there exists $C > 0$ such that
	\begin{align}
		\Pr_{\mathcal{L}^{\ell(\text{start})} = \mathcal{L}(\xi \barBreak \xi \in  \mathcal{S}(\hat{\tau}_{\lfloor \sqrt{N}\rfloor-1} < \hat{\tau}_{0}))}(\mathcal{Y}_{\leftarrow} \ge y ) &  \le C\exp\Big(-\frac{(\lambda_{H} + \lambda_{L} - \mu)^{2}}{2(\lambda_{H} + \lambda_{L} + \mu)^{2}}y\Big). \label{eq:createConstantK}
	\end{align}
	Next, combining \eqref{eq:conditionalDistFraction}, \eqref{eq:boundDenom}, and \eqref{eq:ohhhExpBound} in a similar fashion yields
	\begin{align}
		\Pr_{\mathcal{L}^{\ell(\text{start})} = \mathcal{L}(\xi \barBreak \xi \in  \mathcal{S}(\hat{\tau}_{\lfloor \sqrt{N}\rfloor-1} < \hat{\tau}_{0}))}\Big(\max_{t \in [t^{Y = 2}, t^{Y=1}]} Y(t) > y\Big) = \exp(-\Omega(y)). \label{eq:FinaExpBoundForFuture}
	\end{align}
	And since $\max_{t \in [t^{Y = 2}, t^{Y=1}]} Y(t) < \sqrt{N}\rfloor-1$ implies $t^{\text{end}} = t^{Y=1}$, the result above yields the following:
	\begin{align}
		\Pr\Big(&\max_{t \in [t^{\text{start}}, t^{\text{end}}]} Y(t) > y\Big) \nonumber\\
		& \le \Pr\Big(\max_{t \in [t^{\text{start}}, t^{\text{end}}]} Y(t) > \max_{t \in [t^{Y = 2}, t^{Y=1}]} Y(t)\Big) + \Pr\Big(\max_{t \in [t^{Y = 2}, t^{Y=1}]} Y(t) > y\Big) \nonumber\\
		& \le \Pr\Big(\max_{t \in [t^{Y = 2}, t^{Y=1}]} Y(t) > \lfloor \sqrt{N}\rfloor-2 \Big) + \Pr\Big(\max_{t \in [t^{Y = 2}, t^{Y=1}]} Y(t) > y\Big) \nonumber\\
		& = \exp(-\Omega(\lfloor \sqrt{N}\rfloor \wedge y)), \label{eq:FinaExpBoundForFuture2}
	\end{align}
	where the probabilities are taken with respect to the $\mathcal{L}^{\ell(\text{start})} = \mathcal{L}(\xi \barBreak \xi \in  \mathcal{S}(\hat{\tau}_{\lfloor \sqrt{N}\rfloor-1} < \hat{\tau}_{0}))$ law.
\end{proof}

\begin{proof}[Proof of Proposition \ref{l:littleo}]
	Line \eqref{eq:denom} establishes that the ratio's denominator is $\Omega(1/N)$. Accordingly, it will suffice to show that the ratio's numerator is $\exp(-\Omega(\sqrt{N}))$.  
	
	Begin as before, by taking $N$ large enough so that the total service completion rate is at least $N (\lambda_{H} + \mu) / 2$ when $Y < \lfloor \sqrt{N}\rfloor$.
	
	Next, divide the epoch into sub-epochs. The first sub-epoch starts at the beginning of the epoch, and a new sub-epoch begins whenever the idle-server process travels from $\lfloor \sqrt{N}\rfloor - 1$ to zero. The first sub-epoch cannot reject any type-$H$ jobs, by design, because it starts with condition $\hat{\tau}_{\lfloor \sqrt{N}\rfloor-1} < \hat{\tau}_{0}$. Further, this first sub-epoch concludes the epoch unless $\max_{t \in [t^{Y=2}, t^{Y = 1}]} Y(t) \ge \lfloor \sqrt{N}\rfloor-1$, where $t^{Y=2}$ and $t^{Y = 1}$ are defined in the proof of Proposition \ref{l:serverIdleness}. Line \eqref{eq:FinaExpBoundForFuture} of that proposition's proof asserts that the probability of this event is $\exp(-\Omega(\sqrt{N}))$, as $N \rightarrow \infty$.
	
	Next, Lemma \ref{l:nearmemorylessness} establishes that the system follows law $\mathcal{L}(\xi)$ at the beginning of all but the first sub-epoch. Accordingly, the system enters a renewal process at the start of the second sub-epoch, with the system renewing with each subsequent sub-epoch, until the end of the epoch. 
	
	I would like to apply Wald's equation to this renewal process, to assert that the expected number of $H$ jobs rejected after the start of the second sub-epoch equals the expected number of $H$ jobs rejected in the second sub-epoch divided by the probability of the second sub-epoch concluding the epoch. Unfortunately, I can't do so directly, because the number of $H$ jobs rejected in the second sub-epoch influences the probability of this sub-epoch concluding the epoch. 
	
	To create a renewal process that's amenable to Wald's equation, suppose the service system rejected all $L$ jobs between the start of the second sub-epoch and the next time the idle-server process equaled $\lfloor \sqrt{N}\rfloor-1$, at which time it reverted back to the PFI algorithm. This change does not influence the number of type-$H$ jobs rejected during the second sub-epoch because this sub-epoch cannot reject $H$ jobs after it accepts an $L$ job. This change also does not increase the probability of the second sub-epoch concluding the epoch because both policies follow the same probability law after the idle-server process reaches $\lfloor \sqrt{N}\rfloor-1$, and only the PFI policy can terminate the epoch before the idle-server process reaches this level. Accordingly, modifying the policy so that all but the first sub-epoch behave like this modified second sub-epoch would yield a renewal process that rejects no fewer $H$ jobs per epoch, in expectation, than the PFI algorithm. Further, this alternative renewal process is compatible with Wald's equation, because the number of type-$H$ jobs rejected in one of its sub-epochs does not depend on whether the sub-epoch concludes the epoch---the former depending on the system dynamics before $Y$ reaches $\lfloor \sqrt{N}\rfloor-1$ and the latter depending on the system dynamics after $Y$ reaches $\lfloor \sqrt{N}\rfloor-1$. 
	
	The number of type-$H$ jobs rejected in the second sub-epoch under the alternative policy is not more than the number of transitions until a given random walk last touches zero, given that it starts at zero, increments at rate $N(\lambda_{H} + \mu)/2$, and decrements at rate $N\lambda_{H}$. The expectation of this last hitting time is $\frac{8 \lambda_H (\lambda_H + \mu)}{(\mu - \lambda_H)^2}$.
	
	Next, I will lower-bound the probability of the second sub-epoch concluding the epoch under the alternative policy. Following the alternative policy ensures that the idle-server process reaches $\lfloor \sqrt{N}\rfloor-1$ during the sub-epoch. As the idle server level falls from $\lfloor \sqrt{N}\rfloor-1$ to $\lfloor \sqrt{N}\rfloor-2$, three things can happen: (i) the idle-server process can reach 1 before returning to $\lfloor \sqrt{N}\rfloor-1$ without accepting any type-$L$ jobs along the way, (ii) the idle-server process can reach 1 before returning to $\lfloor \sqrt{N}\rfloor-1$ with at least one $L$ job accepted along the way, and (iii) the idle-server process can return to $\lfloor \sqrt{N}\rfloor-1$ before reaching 1. The probability of the sub-epoch not concluding the epoch is no more than the probability of event (i), conditional on event (iii) not happening.
	
	The unconditional probability of event (i) is the probability of the idle-server process traveling from $\lfloor \sqrt{N}\rfloor-2$ to 1 before $\lfloor \sqrt{N}\rfloor-1$, given that the system accepts only type-$H$ jobs. This probability is no more than the probability of a random walk reaching 1 before reaching $\lfloor \sqrt{N}\rfloor-1$, given that it starts at $\lfloor \sqrt{N}\rfloor-2$, increments at rate $N(\lambda_{H} + \mu)/2$, and decrements at rate $N\lambda_{H}$. And this latter probability is $\frac{(\frac{\lambda_{H} + \mu}{2 \lambda_{H}})^{2}-1}{(\frac{\lambda_{H} + \mu}{2 \lambda_{H}})^{\lfloor \sqrt{N}\rfloor-1} - 1} \le \frac{(\frac{\lambda_{H} + \mu}{2 \lambda_{H}})^{2}}{(\frac{\lambda_{H} + \mu}{2 \lambda_{H}})^{\lfloor \sqrt{N}\rfloor-1}} = (\frac{\lambda_{H} + \mu}{2 \lambda_{H}})^{3-\lfloor \sqrt{N}\rfloor}$. 
	
	The probability of event (iii) is no larger than the probability of the idle-server process reaching $\lfloor \sqrt{N}\rfloor-1$ before reaching $1$, given that it starts at $\lfloor \sqrt{N}\rfloor-2$ and the system accepts all jobs. This probability is no more than the probability of a random walk ever reaching $\lfloor \sqrt{N}\rfloor-1$, given that it starts at $\lfloor \sqrt{N}\rfloor-2$, increments at rate $\mu N$, and decrements at rate $N(\lambda_{H}+\lambda_{L})$. This latter probability is $\frac{\mu}{\lambda_{H} + \lambda_{L}}$. Thus, the probability of event (iii) \emph{not} occurring is at least $\frac{\lambda_{H} + \lambda_{L} - \mu}{\lambda_{H} + \lambda_{L}}$.
	
	When combined, the previous two results indicate that the probability of the second sub-epoch not concluding the epoch is no more than 
	\begin{align*}
		\frac{\big(\frac{\lambda_{H} + \mu}{2 \lambda_{H}}\big)^{3-\lfloor \sqrt{N}\rfloor}}{\frac{\lambda_{H} + \lambda_{L} - \mu}{\lambda_{H} + \lambda_{L}}}.
	\end{align*}
	And since all subsequent sub-epochs resemble this second sub-epoch, the expected number of sub-epochs after the start of the second sub-epoch is therefore no more than
	\begin{align*}
		\frac{1}{1 - \frac{\big(\frac{\lambda_{H} + \mu}{2 \lambda_{H}}\big)^{3-\lfloor \sqrt{N}\rfloor}}{\frac{\lambda_{H} + \lambda_{L} - \mu}{\lambda_{H} + \lambda_{L}}}}.
	\end{align*}
	Further, since each sub-epoch rejects no more than $\frac{8 \lambda_H (\lambda_H + \mu)}{(\mu - \lambda_H)^2}$ type-$H$ jobs, in expectation, the total expected number of type-$H$ jobs rejected after the start of the second sub-epoch---conditional on there being a second sub-epoch---is no more than 
	\begin{align*}
		\frac{\frac{8 \lambda_H (\lambda_H + \mu)}{(\mu - \lambda_H)^2}}{1 - \frac{\big(\frac{\lambda_{H} + \mu}{2 \lambda_{H}}\big)^{3-\lfloor \sqrt{N}\rfloor}}{\frac{\lambda_{H} + \lambda_{L} - \mu}{\lambda_{H} + \lambda_{L}}}} = O(1).
	\end{align*}
	And since only $\exp(-\Omega(\sqrt{N}))$ epochs comprise a second sub-epoch, this implies the result.
\end{proof}

\begin{proof}[Proof of Lemma \ref{l:nearmemorylessnessSSS}] 
	The proof of Lemma \ref{l:nearmemorylessnessSSS} parallels the inductive argument used for Lemma \ref{l:nearmemorylessness}. The only substantive difference is the inclusion of the continuous state variables $f^{\ell}$ and $a^{\ell}$, which record the time elapsed since the most recent $L$-rejection and $L$-acceptance, respectively. Under the SSS policy, a type-$L$ rejection implies that the reject-all-$L$ idle-server process would reach 0 or 1 within the next $w$ time units. The residual window $w - f^{\ell}$ tracks the remaining time on this countdown clock. Conversely, accepting an $L$-arrival implies that the reject-all-$L$ process would \emph{not} have hit 0 or 1 within $w$ time units. In this case, the residual window $w - a^{\ell}$ denotes the remaining time the process is guaranteed to remain above level 1 (or the remaining time it is guaranteed to remain above level 0 while the last accepted $L$-job is still in service). Including $f^{\ell}$ and $a^{\ell}$ in the state ensures that the augmented state variables fully encode all information implied by past acceptance and rejection decisions that affect the conditional law of future arrivals and service completions. Consequently, the conditional distribution of $m_i(\Omega^{\ell})$ depends only on this augmented state, and the proof follows that of Lemma \ref{l:nearmemorylessness}.
\end{proof}

\begin{proof}[Proof of Lemma \ref{l:gapSSS}]
	The proof of Lemma \ref{l:gap} applies \emph{mutatis mutandis}.
\end{proof}

\begin{proof}[Proof of Proposition \ref{l:serverIdlenessSSS}] 
	I will prove this result by establishing the following: 
	\begin{align}
		\E\Big(\int_{t=t_{\text{SSS}}^{\text{start}}}^{t_{\text{SSS}}^{\text{end}}} Y_{\text{SSS}}(t)dt\Big) & = \frac{O(1/N)}{1 - \Pr(t_{\text{PFI}}^{\text{end}} < t_{\text{PFI} \rightarrow \text{AE}}^{\text{end}})} \label{eq:thisResultEstablishThat0}\\
		\aq \E(t_{\text{SSS}}^{\text{end}} - t_{\text{SSS}}^{\text{start}}) & =\frac{\Omega(1/N)}{1 - \Pr(t_{\text{PFI}}^{\text{end}} < t_{\text{PFI} \rightarrow \text{AE}}^{\text{end}})}, \label{eq:thisResultEstablishThat}
	\end{align}
	where times $t_{\text{PFI} \rightarrow \text{AE}}^{\text{end}}$ and $t_{\text{PFI}}^{\text{end}}$ will be defined in the sequel. 
	
	I will begin by proving \eqref{eq:thisResultEstablishThat0}. To this end, I must introduce a few new objects. First, define the Accept if Either (AE) policy as that which accepts a job if either the SSS or PFI policies do so, and rejects it otherwise. Second, define $Y_{\text{AE}}$ as the idle-server process when the system controller switches from SSS to AE at time $t_{\text{SSS}}^{\text{start}}$. Third, define $t_{\text{SSS} \ne \text{AE}}$ as the first time that processes $Y_{\text{SSS}}$ and $Y_{\text{AE}}$ decouple. And fourth, define $t_{\text{SSS} \ne \text{AE}}^{Y = 1}$ as the first time after $t_{\text{SSS} \ne \text{AE}}$ that $Y_{\text{SSS}} = 1$. Fifth, define $t_{\text{AE}}^{\text{end}}$ as the first time that $Y_{\text{AE}}$ reaches zero after $t_{\text{SSS}}^{\text{start}}$. With this, note the following:
	\begin{align*}
		\int_{t=t_{\text{SSS}}^{\text{start}}}^{t_{\text{SSS}}^{\text{end}}} & Y_{\text{SSS}}(t)dt \nonumber\\
		& = \int_{t=t_{\text{SSS}}^{\text{start}}}^{t_{\text{SSS} \ne \text{AE}} \wedge t_{\text{SSS}}^{\text{end}}} Y_{\text{AE}}(t)dt  + \indicator{t_{\text{SSS} \ne \text{AE}}< t_{\text{SSS}}^{\text{end}}}\Big(\int_{t=t_{\text{SSS} \ne \text{AE}}}^{t_{\text{SSS} \ne \text{AE}}^{Y = 1}} Y_{\text{SSS}}(t)dt + \int_{t_{\text{SSS} \ne \text{AE}}^{Y = 1}}^{t_{\text{SSS}}^{\text{end}}} Y_{\text{SSS}}(t)dt\Big) \nonumber\\
		& \le \int_{t=t_{\text{SSS}}^{\text{start}}}^{t_{\text{AE}}^{\text{end}}} Y_{\text{AE}}(t)dt  + \indicator{t_{\text{SSS} \ne \text{AE}}< t_{\text{SSS}}^{\text{end}}}\Big(Nw + \int_{t_{\text{SSS} \ne \text{AE}}^{Y = 1}}^{t_{\text{SSS}}^{\text{end}}} Y_{\text{SSS}}(t)dt\Big). 
	\end{align*}
	The second line above holds because (i) $t_{\text{SSS} \ne \text{AE}} \wedge t_{\text{SSS}}^{\text{end}} \le t_{\text{AE}}^{\text{end}}$ and (ii) $\int_{t=t_{\text{SSS} \ne \text{AE}}}^{t_{\text{SSS} \ne \text{AE}}^{Y = 1}} Y_{\text{SSS}}(t)dt \le N w$. Inequality (i) holds because $t_{\text{AE}}^{\text{end}} < t_{\text{SSS} \ne \text{AE}}$ implies $Y_{\text{SSS}}(t_{\text{AE}}^{\text{end}}) = Y_{\text{AE}}(t_{\text{AE}}^{\text{end}}) = 0$, and thus $t_{\text{SSS}}^{\text{end}} \le t_{\text{AE}}^{\text{end}}$. Inequality (ii) holds because (iia) $Y_{\text{SSS}} \le N$ and (iib) $\tau_{\lfloor \sqrt{N}\rfloor} <  \tau_{1} < w$ must hold at time $t_{\text{SSS} \ne \text{AE}}$, which guarantees that $Y_{\text{SSS}}$ will return to 1 within $w$ hours.
	
	Applying identity $\E(\indicator{A}B) \le \E(B \barBreak A)$ to result above yields
	\begin{align*}
		\E\Big(\int_{t=t_{\text{SSS}}^{\text{start}}}^{t_{\text{SSS}}^{\text{end}}} Y_{\text{SSS}}(t)dt\Big) & \le \E\Big(\int_{t=t_{\text{SSS}}^{\text{start}}}^{t_{\text{AE}}^{\text{end}}} Y_{\text{AE}}(t)dt\Big)  + \Pr(t_{\text{SSS} \ne \text{AE}}< t_{\text{SSS}}^{\text{end}})Nw \nonumber\\
		& \qquad + \Pr(t_{\text{SSS} \ne \text{AE}}< t_{\text{SSS}}^{\text{end}})\E\Big(\int_{t_{\text{SSS} \ne \text{AE}}^{Y = 1}}^{t_{\text{SSS}}^{\text{end}}} Y_{\text{SSS}}(t)dt \barBreak t_{\text{SSS} \ne \text{AE}}< t_{\text{SSS}}^{\text{end}}\Big). 
	\end{align*}
	
	Unfortunately, addressing the $t_{\text{SSS} \ne \text{AE}}< t_{\text{SSS}}^{\text{end}}$ condition in the last term above will require care, as this event seems to depend on the history of the AE process, and we have never characterized the probability law of $Y_{\text{SSS}}$ conditional on $\mathcal{H}_{\text{AE}}$, the information set available to an external observer who see the arrivals, departures, and acceptance decisions associated with the $Y_{\text{AE}}$ process. Fortunately, conditioning on $t_{\text{AE}}< t_{\text{SSS}}^{\text{end}}$ does not invalidate Lemma \ref{l:nearmemorylessnessSSS}, as the pre-$t_{\text{SSS} \ne \text{AE}}^{Y = 1}$ history of the AE process, $\mathcal{H}_{\text{AE}}(t_{\text{SSS} \ne \text{AE}}^{Y = 1})$, is measurable with respect to the information set generated by the pre-$t_{\text{SSS} \ne \text{AE}}^{Y = 1}$ history of the SSS process, $\sigma(\mathcal{H}_{\text{SSS}}(t_{1}))$. To establish this measurability, note that an SSS acceptance implies an AE acceptance, so that an AE action is incrementally informative only when it coincides with an SSS rejection. Now suppose SSS rejects an $L$ job at time $t_{1}$. In this case, the corresponding AE action will indicate whether event $\tau_{1}(t_{1}) < \tau_{\lfloor \sqrt{N}\rfloor}(t_{1})$ holds (an event that is generally \emph{not} measurable with respect to $\mathcal{H}_{\text{SSS}}(t_{1})$). Next, define $t_{2}$ as the first time after $t_{1}$ that $Y_{\text{SSS}} = 1$. Since SSS rejected an $L$ job at $t_{1}$, process $Y_{\text{SSS}}$ will follow the reject-all-$L$-jobs idle-server process between times $t_{1}$ and $t_{2}$; further, this segment of the $Y_{\text{SSS}}$ process starts at time $t_{1}$ and terminates at level 1, and thus indicates whether event $\tau_{1}(t_{1}) < \tau_{\lfloor \sqrt{N}\rfloor}(t_{1})$ holds. Accordingly, the AE action at time $t_{1}$ is measurable with respect to the $\sigma$-algebra generated by $Y_{\text{SSS}}$, before time $t_{2}$. And in this way, we find that SSS future process $\Omega_{\text{SSS}}(t_{\text{SSS} \ne \text{AE}}^{Y = 1})$, is conditionally independent of AE history $\mathcal{H}_{\text{AE}}(t_{\text{SSS} \ne \text{AE}}^{Y = 1})$, given SSS history $\mathcal{H}_{\text{SSS}}(t_{\text{SSS} \ne \text{AE}}^{Y = 1})$. And conditioning only on $\mathcal{H}_{\text{SSS}}(t_{\text{SSS} \ne \text{AE}}^{Y = 1})$ ensures that Lemma \ref{l:nearmemorylessnessSSS} still holds.
	
	More specifically, Lemma \ref{l:nearmemorylessnessSSS} establishes that $\Omega_{\text{SSS}}(t_{\text{SSS} \ne \text{AE}}^{Y = 1}) \sim \mathcal{L}(\xi)$. The lemma also establishes that $\Omega_{\text{SSS}}(t_{\text{SSS}}^{\text{start}}) \sim \mathcal{L}(\xi)$. However, there is a slight discrepancy between the time-$t_{\text{SSS}}^{\text{start}}$ and the time-$t_{\text{SSS} \ne \text{AE}}^{Y = 1}$ laws, as the former corresponds to zero idle servers and the latter corresponds to one idle server. However, the law at $t_{\text{SSS} \ne \text{AE}}^{Y = 1}$ is exactly the same as the law at $t_{\text{SSS}}^{Y = 1}$, defined as the first time after $t_{\text{SSS}}^{\text{start}}$ that $Y_{\text{SSS}} = 1$. And with this, we have
	\begin{align*}
		\E\Big(\int_{t=t_{\text{SSS} \ne \text{AE}}^{Y = 1}}^{t_{\text{SSS}}^{\text{end}}} & Y_{\text{SSS}}(t)dt \barBreak t_{\text{SSS} \ne \text{AE}}< t_{\text{SSS}}^{\text{end}}\Big) \\
		& = \E\Big(\int_{t=t_{\text{SSS}}^{Y = 1}}^{t_{\text{SSS}}^{\text{end}}} Y_{\text{SSS}}(t)dt \barBreak t_{\text{SSS}}^{Y = 1} < t_{\text{SSS}}^{\text{end}}\Big) \nonumber\\
		& = \E\Big(\indicator{t_{\text{SSS}}^{Y = 1} < t_{\text{SSS}}^{\text{end}}}\int_{t=t_{\text{SSS}}^{Y = 1}}^{t_{\text{SSS}}^{\text{end}}} Y_{\text{SSS}}(t)dt\Big) \Big/ \Pr(t_{\text{SSS}}^{Y = 1} < t_{\text{SSS}}^{\text{end}}) \\
		& = \E\Big(\int_{t=t_{\text{SSS}}^{\text{start}}}^{t_{\text{SSS}}^{\text{end}}} Y_{\text{SSS}}(t)dt\Big) / \Pr(t_{\text{SSS}}^{Y = 1} < t_{\text{SSS}}^{\text{end}}).
	\end{align*}
	
	Next, we will suppose the following three results:
	\begin{align}
		\Pr(t_{\text{SSS}}^{Y = 1} < t_{\text{SSS}}^{\text{end}}) & = \mu/(\lambda_{H} + \lambda_{L} + \mu),\nonumber\\
		\E\Big(\int_{t=t_{\text{SSS}}^{\text{start}}}^{t_{\text{AE}}^{\text{end}}} Y_{\text{AE}}(t)dt\Big) & = \frac{O(1/N)}{1 - \Pr(t_{\text{PFI}}^{\text{end}} < t_{\text{PFI} \rightarrow \text{AE}}^{\text{end}})}, \label{eq:threePartDisplay}\\ 
		\aq \Pr(t_{\text{SSS} \ne \text{AE}} < t_{\text{SSS}}^{\text{end}}) & = \exp(-\Omega(\sqrt{N})). \nonumber
	\end{align}
	Combining these assumed results with the two preceding displays yields the following:
	\begin{align*}
		\E\Big(\int_{t=t_{\text{SSS}}^{\text{start}}}^{t_{\text{SSS}}^{\text{end}}} Y_{\text{SSS}}(t)dt\Big) & \le \frac{\E\Big(\int_{t=t_{\text{SSS}}^{\text{start}}}^{t_{\text{AE}}^{\text{end}}} Y_{\text{AE}}(t)dt\Big) +  Nw\Pr(t_{\text{SSS} \ne \text{AE}} < t_{\text{SSS}}^{\text{end}})}{1 - \Pr(t_{\text{SSS} \ne \text{AE}} < t_{\text{SSS}}^{\text{end}}) \big/\Pr(t_{\text{SSS}}^{Y = 1} < t_{\text{SSS}}^{\text{end}})} \\
		& \le \frac{\frac{O(1/N)}{1 - \Pr(t_{\text{PFI}}^{\text{end}} < t_{\text{PFI} \rightarrow \text{AE}}^{\text{end}})}  + N w \exp(-\Omega(\sqrt{N}))}{1 - \frac{\exp(-\Omega(\sqrt{N}))}{\mu/(\lambda_{H} + \lambda_{L} + \mu)}} \\
		& = \frac{O(1/N)}{1 - \Pr(t_{\text{PFI}}^{\text{end}} < t_{\text{PFI} \rightarrow \text{AE}}^{\text{end}})}.
	\end{align*}
	
	Accordingly, proving the three results in display \eqref{eq:threePartDisplay} would establish line \eqref{eq:thisResultEstablishThat0}. The first line in display \eqref{eq:threePartDisplay} holds because $t_{\text{SSS}}^{Y = 1} < t_{\text{SSS}}^{\text{end}}$ is equivalent to the event that the first action after $t_{\text{SSS}}^{\text{start}}$ is a departure. Establishing the second line in \eqref{eq:threePartDisplay} will be more difficult. For this, note that
	\begin{align}
		\E\Big(\int_{t=t_{\text{SSS}}^{\text{start}}}^{t_{\text{AE}}^{\text{end}}} Y_{\text{AE}}(t)dt\Big) & = \E\Big(\indicator{t_{\text{SSS}}^{Y=1} < t_{\text{AE}}^{\text{end}}}\int_{t=t_{\text{SSS}}^{Y=1}}^{t_{\text{AE}}^{\text{end}}} Y_{\text{AE}}(t)dt\Big) \nonumber\\
		& \le \E\Big(\int_{t=t_{\text{SSS}}^{Y=1}}^{t_{\text{AE}}^{\text{end}}} Y_{\text{AE}}(t)dt \barBreak t_{\text{SSS}}^{Y=1} < t_{\text{AE}}^{\text{end}}\Big), \label{eq:dumbYequal1Conditioning}
	\end{align}
	Now, to lighten the notation, I will henceforth implicitly take it as given that $t_{\text{SSS}}^{Y=1} < t_{\text{AE}}^{\text{end}}$, writing $\E(\cdot)$ instead of $\E(\cdot \barBreak t_{\text{SSS}}^{Y=1} < t_{\text{AE}}^{\text{end}})$.
	
	To bound the expectation in \eqref{eq:dumbYequal1Conditioning}, I will benchmark $Y_{\text{AE}}$ to $Y_{\text{PFI}}$, the idle-server process when the system controller switches from SSS to PFI at time $t_{\text{SSS}}^{\text{start}}$. Performing this comparison will require a few new objects. First, define $t_{\text{PFI}}^{\text{start}}$ as the first time after $t_{\text{SSS}}^{\text{start}}$ that the $Y_{\text{PFI}}$ process meets the PFI-epoch initial conditions (i.e., $Y_{\text{PFI}} = 1$, $\mathsf{U}_{\text{PFI}}^{c}$, $\mathsf{P}_{\text{PFI}}$, and $\mathsf{W}_{\text{PFI}}$), and define $t_{\text{PFI}}^{\text{end}}$ as the end time of the PFI epoch that began at $t_{\text{PFI}}^{\text{start}}$. Second, define $Y_{\text{PFI} \rightarrow \text{AE}}$ as the idle-server process when the system controller switches from SSS to PFI at time $t_{\text{SSS}}^{\text{start}}$, and then switches from PFI to AE at time $t_{\text{PFI}}^{\text{start}}$. And third, define $t_{\text{PFI} \rightarrow \text{AE}}^{\text{end}}$ as the first time after $t_{\text{PFI}}^{\text{start}}$ that $Y_{\text{PFI} \rightarrow \text{AE}}$ reaches zero. Now, with this, we have the following:
	\begin{align}
		\E\Big(&\int_{t=t_{\text{SSS}}^{Y=1}}^{t_{\text{AE}}^{\text{end}}} Y_{\text{AE}}(t)dt\Big) \nonumber\\
		& \le \E\Big(\int_{t=t_{\text{SSS}}^{Y=1}}^{t_{\text{PFI}}^{\text{start}}} Y_{\text{PFI}}(t)dt \Big) + \E\Big(\indicator{t_{\text{PFI}}^{\text{start}} < t_{\text{AE}}^{\text{end}}}\int_{t = t_{\text{PFI}}^{\text{start}}}^{t_{\text{AE}}^{\text{end}}} Y_{\text{PFI}}(t)dt\Big) \nonumber\\
		& \le \E\Big(\int_{t=t_{\text{SSS}}^{Y=1}}^{t_{\text{PFI}}^{\text{start}}} Y_{\text{PFI}}(t)dt \Big) + \E\Big(\int_{t = t_{\text{PFI}}^{\text{start}}}^{t_{\text{AE}}^{\text{end}}} Y_{\text{PFI}}(t)dt \barBreak t_{\text{PFI}}^{\text{start}} < t_{\text{AE}}^{\text{end}}\Big)  \nonumber \\
		& = \E\Big(\int_{t=t_{\text{SSS}}^{Y=1}}^{t_{\text{PFI}}^{\text{start}}} Y_{\text{PFI}}(t)dt \Big) + \E\Big(\int_{t = t_{\text{PFI}}^{\text{start}}}^{t_{\text{PFI} \rightarrow \text{AE}}^{\text{end}}} Y_{\text{PFI}}(t)dt\Big).  \label{eq:EboundedbyEEHuh}
	\end{align}
	The first line above holds because $Y_{\text{AE}} \le Y_{\text{PFI}}$ holds through time $t_{\text{AE}}^{\text{end}}$, and the last line above holds because $\Omega_{\text{PFI}}(t_{\text{PFI}}^{\text{start}})$ is conditionally independent of $\mathcal{H}_{\text{AE}}(t_{\text{PFI}}^{\text{start}})$, given $\mathcal{H}_{\text{PFI}}(t_{\text{PFI}}^{\text{start}})$ when $t_{\text{PFI}}^{\text{start}} < t_{\text{AE}}^{\text{end}}$. This conditional independence holds for the same reason that $\Omega_{\text{SSS}}(t_{\text{SSS} \ne \text{AE}}^{Y = 1})$ is conditionally independent of $\mathcal{H}_{\text{AE}}(t_{\text{SSS} \ne \text{AE}}^{Y = 1})$, given $\mathcal{H}_{\text{SSS}}(t_{\text{SSS} \ne \text{AE}}^{Y = 1})$. The upshot is that conditioning on the pre-$t_{\text{PFI}}^{\text{start}}$ histories of $Y_{\text{AE}}$ and $Y_{\text{PFI}}$ is equivalent to conditioning only on the pre-$t_{\text{PFI}}^{\text{start}}$ history of $Y_{\text{PFI}}$, and hence the unconditional time-$t_{\text{PFI}}^{\text{start}}$ law of $Y_{\text{PFI} \rightarrow \text{AE}}$ equals the time-$t_{\text{PFI}}^{\text{start}}$ law of $Y_{\text{AE}}$ conditional on $t_{\text{PFI}}^{\text{start}} < t_{\text{AE}}^{\text{end}}$. Indeed, switching from $Y_{\text{AE}}$ to $Y_{\text{PFI} \rightarrow \text{AE}}$ is basically just shorthand for dropping this condition.
	
	I will now bound the two expectations in \eqref{eq:EboundedbyEEHuh}. Line \eqref{eq:saveForWayLater}, from the proof of Proposition \ref{l:serverIdleness}, yields the following bound for the first expectation:
	\begin{align}
		\E\Big(\int_{t=t_{\text{SSS}}^{Y=1}}^{t_{\text{PFI}}^{\text{start}}} Y_{\text{PFI}}(t)dt\Big) = O(1/N). \label{eq:1overNforYstarting1}
	\end{align}
	
	To bound the second expectation in \eqref{eq:EboundedbyEEHuh}, I will present four related displays. First:
	\begin{align*}
		\E\Big(&\int_{t = t_{\text{PFI}}^{\text{start}}}^{t_{\text{PFI} \rightarrow \text{AE}}^{\text{end}}} Y_{\text{PFI}}(t)dt\Big) \nonumber\\
		& = \big(1 - \Pr(t_{\text{PFI}}^{\text{end}} < t_{\text{PFI} \rightarrow \text{AE}}^{\text{end}})\big)\E\Big(\int_{t = t_{\text{PFI}}^{\text{start}}}^{t_{\text{PFI} \rightarrow \text{AE}}^{\text{end}}} Y_{\text{PFI}}(t)dt \barBreak t_{\text{PFI} \rightarrow \text{AE}}^{\text{end}} \le t_{\text{PFI}}^{\text{end}}\Big) \nonumber\\
		& \qquad + \Pr(t_{\text{PFI}}^{\text{end}} < t_{\text{PFI} \rightarrow \text{AE}}^{\text{end}})\E\Big(\int_{t = t_{\text{PFI}}^{\text{start}}}^{t_{\text{PFI}}^{\text{end}}} Y_{\text{PFI}}(t)dt \barBreak t_{\text{PFI}}^{\text{end}} < t_{\text{PFI} \rightarrow \text{AE}}^{\text{end}}\Big) \nonumber\\
		& \qquad + \Pr(t_{\text{PFI}}^{\text{end}} < t_{\text{PFI} \rightarrow \text{AE}}^{\text{end}})\E\Big(\int_{t = t_{\text{PFI}}^{\text{end}}}^{t_{\text{PFI} \rightarrow \text{AE}}^{\text{end}}} Y_{\text{PFI}}(t)dt \barBreak t_{\text{PFI}}^{\text{end}} < t_{\text{PFI} \rightarrow \text{AE}}^{\text{end}}\Big). \nonumber
	\end{align*}
	Second:
	\begin{align*}
		\big(1 - \Pr(t_{\text{PFI}}^{\text{end}} &< t_{\text{PFI} \rightarrow \text{AE}}^{\text{end}})\big) \E\Big(\int_{t = t_{\text{PFI}}^{\text{start}}}^{t_{\text{PFI} \rightarrow \text{AE}}^{\text{end}}} Y_{\text{PFI}}(t)dt \barBreak t_{\text{PFI} \rightarrow \text{AE}}^{\text{end}} \le t_{\text{PFI}}^{\text{end}}\Big) \\
		& \le \Pr(t_{\text{PFI} \rightarrow \text{AE}}^{\text{end}} \le t_{\text{PFI}}^{\text{end}})\E\Big(\int_{t = t_{\text{PFI}}^{\text{start}}}^{ t_{\text{PFI}}^{\text{end}}} Y_{\text{PFI}}(t)dt \barBreak t_{\text{PFI} \rightarrow \text{AE}}^{\text{end}} \le t_{\text{PFI}}^{\text{end}}\Big) \\
		& \le\E\Big(\int_{t = t_{\text{PFI}}^{\text{start}}}^{ t_{\text{PFI}}^{\text{end}}} Y_{\text{PFI}}(t)dt\Big).
	\end{align*}
	Third:
	\begin{align*}
		\Pr(t_{\text{PFI}}^{\text{end}} < t_{\text{PFI} \rightarrow \text{AE}}^{\text{end}})\E\Big(\int_{t = t_{\text{PFI}}^{\text{start}}}^{t_{\text{PFI}}^{\text{end}}} Y_{\text{PFI}}(t)dt \barBreak t_{\text{PFI}}^{\text{end}} < t_{\text{PFI} \rightarrow \text{AE}}^{\text{end}}\Big) \le \E\Big(\int_{t = t_{\text{PFI}}^{\text{start}}}^{t_{\text{PFI}}^{\text{end}}} Y_{\text{PFI}}(t)dt\Big).
	\end{align*}
	And fourth:
	\begin{align}
		\E\Big(\int_{t = t_{\text{PFI}}^{\text{end}}}^{t_{\text{PFI} \rightarrow \text{AE}}^{\text{end}}}& Y_{\text{PFI}}(t)dt \barBreak t_{\text{PFI}}^{\text{end}} < t_{\text{PFI} \rightarrow \text{AE}}^{\text{end}}\Big) \nonumber\\
		& = \E\Big(\int_{t = t_{\text{PFI}}^{\text{end}}}^{t_{\text{PFI} \Rightarrow \text{AE}}^{\text{end}}} Y_{\text{PFI}}(t)dt\Big) \nonumber\\
		& = \E\Big(\int_{t = t_{\text{PFI}}^{\text{start}}}^{t_{\text{PFI} \rightarrow \text{AE}}^{\text{end}}} Y_{\text{PFI}}(t)dt\Big), \label{eq:lateAddOne}
	\end{align}
	where $t_{\text{PFI} \Rightarrow \text{AE}}^{\text{end}}$ is the first time after $t_{\text{PFI}}^{\text{end}}$ that $Y_{\text{PFI} \Rightarrow \text{AE}}$ hits zero, and $Y_{\text{PFI} \Rightarrow \text{AE}}$ is the idle server level when the system controller switches from SSS to PFI at time $t_{\text{SSS}}^{\text{start}}$, and then switches from PFI to AE at time $t_{\text{PFI}}^{\text{end}}$ (in contrast, $Y_{\text{PFI} \rightarrow \text{AE}}$ switches from PFI to AE at time $t_{\text{PFI}}^{\text{start}}$). The first line in the last display above holds for the same reason that the last line in display \eqref{eq:EboundedbyEEHuh} holds: the future of $Y_{\text{PFI}}$ (i.e., $\Omega_{\text{PFI}}(t_{\text{PFI}}^{\text{end}})$), is conditionally independent of the history of $Y_{\text{PFI} \rightarrow \text{AE}}$ (i.e., $\mathcal{H}_{\text{PFI} \rightarrow \text{AE}}(t_{\text{PFI}}^{\text{end}})$), given the history of $Y_{\text{PFI}}$ (i.e., $\mathcal{H}_{\text{PFI}}(t_{\text{PFI}}^{\text{end}})$). The second line in the last display above holds because the probability law of $Y_{\text{PFI}}$ over interval $[t_{\text{PFI}}^{\text{end}}, t_{\text{PFI} \Rightarrow \text{AE}}^{\text{end}}]$ equals that over interval $[t_{\text{PFI}}^{\text{start}}, t_{\text{PFI} \rightarrow \text{AE}}^{\text{end}}]$, since the end of one epoch marks the start of another.
	
	Now, with a bit of algebra, the four displays above yield the following:
	\begin{align*}
		\E\Big(\int_{t = t_{\text{PFI}}^{\text{start}}}^{t_{\text{PFI} \rightarrow \text{AE}}^{\text{end}}} Y_{\text{PFI}}(t)dt\Big) & \le \frac{2\E\Big(\int_{t = t_{\text{PFI}}^{\text{start}}}^{t_{\text{PFI}}^{\text{end}}} Y_{\text{PFI}}(t)dt\Big)}{1 - \Pr(t_{\text{PFI}}^{\text{end}} < t_{\text{PFI} \rightarrow \text{AE}}^{\text{end}})} \\
		& = \frac{O(1/N)}{1 - \Pr(t_{\text{PFI}}^{\text{end}} < t_{\text{PFI} \rightarrow \text{AE}}^{\text{end}})},
	\end{align*}
	where the last line follows from line \eqref{eq:takeAwayMessageFromThisProp}, from the proof of Proposition \ref{l:serverIdleness}. Finally, combining this result with \eqref{eq:dumbYequal1Conditioning}--\eqref{eq:1overNforYstarting1} yields the second line of display \eqref{eq:threePartDisplay}. 
	
	I will now establish the third line of display \eqref{eq:threePartDisplay}. To begin, note that for $Y_{\text{AE}}$ and $Y_{\text{SSS}}$ to diverge, SSS must reject an $L$ job that PFI accepts. If the converse happens, and SSS accepts a job that PFI rejects, then the SSS epoch will terminate before PFI next accepts a job (since $Y_{\text{SSS}}$ will go to zero no later than $Y_{\text{PFI}}$ goes to one). Accordingly, $Y_{\text{AE}}$ and $Y_{\text{SSS}}$ will not decouple within a given SSS epoch unless SSS rejects a job that PFI accepts before SSS accepts a job that PFI rejects. Equivalently, the processes will not decouple within a given SSS epoch unless $Y_{\text{PFI}}$ decouples from $Y_{\text{AE}}$ later than it decouples from $Y_{\text{RE}}$, where $Y_{\text{RE}}$ is the Reject if Either (RE) analog of $Y_{\text{AE}}$, and the RE policy as that which rejects a job if either the SSS or PFI policies do so, and accepts it otherwise. Hence, we have
	\begin{align}
		\Pr(t_{\text{SSS} \ne \text{AE}} < t_{\text{SSS}}^{\text{end}})
		& \le \Pr(t_{\text{PFI} \ne \text{RE}} < t_{\text{PFI} \ne \text{AE}}), \label{eq:fourTs}
	\end{align}
	where $t_{\text{PFI} \ne \text{AE}}$ and $t_{\text{PFI} \ne \text{RE}}$ denote the times that $Y_{\text{PFI}}$ separates from and $Y_{\text{AE}}$ and $Y_{\text{RE}}$, respectively. 
	
	Next, define $\chi$ and $\psi$ as count processes that respectively record the number of actions and the number of PFI epochs that have transpired under the PFI policy since $t_{\text{PFI}}^{\text{start}}$. And define $t_{n}^{\chi}$ and $t_{n}^{\psi}$ as the times that $\chi$ and $\psi$ first reach $n$. Further, define events $\mathcal{E}_{1} = t_{\text{PFI} \ne \text{RE}} \le t_{\exp(N^{1/3})}^{\chi}$, $\mathcal{E}_{2} = t_{\exp(N^{1/3})}^{\chi} \le t_{\exp(N^{1/4})}^{\psi}$, and $\mathcal{E}_{3} = t_{\exp(N^{1/4})}^{\psi} \le t_{\text{PFI} \ne \text{AE}}$. Finally, suppose that we knew that 
	\begin{align}
		\Pr(\mathcal{E}_{1}) & = \exp(-\Omega(\sqrt{N})),\nonumber\\ 
		\Pr(\mathcal{E}_{2}) & = \exp(-\Omega(\exp(N^{1/3}))), \label{eq:lastThreeBunch}\\ 
		\Pr(\mathcal{E}_{3}) & = \exp(-\exp(\Omega(N^{1/4}))). \nonumber
	\end{align}
	Since $\mathcal{E}_{1} \cap \mathcal{E}_{2} \cap \mathcal{E}_{3}$ implies $t_{\text{PFI} \ne \text{AE}} < t_{\exp(N^{1/4})}^{\psi} < t_{\exp(N^{1/3})}^{\chi} < t_{\text{PFI} \ne \text{RE}}$, display \eqref{eq:lastThreeBunch} implies the following:
	\begin{align*}
		\Pr(t_{\text{PFI} \ne \text{RE}} \le t_{\text{PFI} \ne \text{AE}}) & \le \Pr((\mathcal{E}_{1} \cap \mathcal{E}_{2} \cap \mathcal{E}_{3})^{c})\nonumber \\
		& \le \Pr(\mathcal{E}_{1}^{c}) + \Pr(\mathcal{E}_{2}^{c}) + \Pr(\mathcal{E}_{3}^{c}) \\
		& \le \exp(-\Omega(\sqrt{N})) + \exp(-\Omega(\exp(N^{1/3}))) + \exp(-\exp(\Omega(N^{1/4}))  \\
		& = \exp(-\Omega(\sqrt{N}).
	\end{align*} 
	And combining this with \eqref{eq:fourTs} establishes the third result in display \eqref{eq:threePartDisplay}. Hence, establishing the three results in display \eqref{eq:lastThreeBunch} would complete the proof of line \eqref{eq:thisResultEstablishThat0}.
	
	To establish the first result in display \eqref{eq:lastThreeBunch}, note that $Y_{\text{RE}}$ and $Y_{\text{PFI}}$ won't separate unless an $L$ arrival coincides with event $\tau_{\lfloor \sqrt{N}\rfloor} < \tau_{1} < w$. The probability of this event, under the PFI law specified in Lemma \ref{l:nearmemorylessness}, is $\exp(-\Omega(\sqrt{N}))$ uniformly over $L$ arrival times (this follows from a slight modification of the argument underpinning Proposition \ref{l:littleo}). And with Boole's inequality, this implies that the probability of $Y_{\text{RE}}$ and $Y_{\text{PFI}}$ separating before $\exp(N^{1/3})$ type-$L$ jobs arrive is $\exp(N^{1/3})\cdot \exp(-\Omega(\sqrt{N})) = \exp(-\Omega(\sqrt{N}))$. And therefore the probability of $Y_{\text{RE}}$ and $Y_{\text{PFI}}$ separating before $\exp(N^{1/3})$ actions of any type is also $\exp(-\Omega(\sqrt{N}))$.
	
	To establish the second result in display \eqref{eq:lastThreeBunch}, note that a small modifications to the proof of Proposition \ref{l:serverIdleness} establish that the number of actions that a given PFI epoch comprises, $\chi_{\text{epoch}}$, satisfies an exponential tail bound: $\Pr(\chi_{\text{epoch}} > x) = \exp(-\Omega(x))$ as $x \rightarrow \infty$, where the $\Omega(x)$ constant is universal across $N$ (e.g., see \eqref{eq:c1Bound}, \eqref{eq:seeThisOne}, and \eqref{eq:FinaExpBoundForFuture2}). Accordingly, a Chernoff bound establishes that the probability of the first $\exp(N^{1/4})$ PFI epochs comprising more than $\exp(N^{1/3})$ actions is $\exp(-\Omega(\exp(N^{1/3})))$ as $N\rightarrow \infty$.
	
	Finally, to establish the third result in display \eqref{eq:lastThreeBunch}, note that $Y_{\text{AE}}$ and $Y_{\text{PFI}}$ will separate if an $L$ arrival coincides with the following three events: (i) $Y_{\text{PFI}} > 1$, (ii) $\tau_{1} < \tau_{\lfloor \sqrt{N}\rfloor}$, and (iii) the time until the next action exceeds $w$. I will now lower-bound the joint probability of these events uniformly over $L$ arrival times, under the PFI law. First, conditioning on the current value of $Y_{\text{PFI}}$ makes event (iii) independent of the current history (including the previous instances of event (iii)), because the time between actions does not influence whether PFI accepts a job. Second, the probability of event (iii), given events (i) and (ii), is at least as large as the probability of an $\text{Exp}(N (\lambda_{H} + \lambda_{L} + \mu))$ random variable exceeding $w = O(\log(N) / N)$, which is $\exp(-N (\lambda_{H} + \lambda_{L} + \mu)) O(\log(N) / N)\big) = N^{-O(1)}$. Third, every PFI epoch has an $\Omega(1)$ chance of comprising an $L$ job that arrives under events (i) and (ii), and thus every PFI epoch has an $N^{-O(1)}$ chance of separating $Y_{\text{AE}}$ and $Y_{\text{PFI}}$.  And since each of these has an $N^{-O(1)}$ chance of separating $Y_{\text{AE}}$ and $Y_{\text{PFI}}$, the probability of these processes not separating within the first $\exp(N^{1/4})$ PFI epochs is $(1 - N^{-O(1)})^{\exp(N^{1/4})} = \exp(-\exp(\Omega(N^{1/4})))$.
	
	Thus concludes the proof of line \eqref{eq:thisResultEstablishThat0}. I will now quickly establish line \eqref{eq:thisResultEstablishThat}. To begin, note the following: 
	\begin{align*}
		\E(t_{\text{SSS}}^{\text{end}} - t_{\text{SSS}}^{\text{start}}) & \ge \E(t_{\text{AE}}^{\text{end}} - t_{\text{SSS}}^{\text{start}}) \\
		& \ge \Pr(t_{\text{PFI}}^{\text{start}} < t_{\text{AE}}^{\text{end}})\E\big(t_{\text{AE}}^{\text{end}} - t_{\text{PFI}}^{\text{start}} \barBreak t_{\text{PFI}}^{\text{start}} < t_{\text{AE}}^{\text{end}}\big) \\
		& = \Pr(t_{\text{PFI}}^{\text{start}} < t_{\text{AE}}^{\text{end}})\E(t_{\text{PFI} \rightarrow \text{AE}}^{\text{end}} - t_{\text{PFI}}^{\text{start}}),
	\end{align*}
	where the last line holds for the same reason that the last line of \eqref{eq:EboundedbyEEHuh} holds. Next, note that
	\begin{align*}
		\E(&t_{\text{PFI} \rightarrow \text{AE}}^{\text{end}} - t_{\text{PFI}}^{\text{start}}) \\
		& \ge \E\big(\indicator{t_{\text{PFI}}^{\text{end}} < t_{\text{PFI} \rightarrow \text{AE}}^{\text{end}}}(t_{\text{PFI}}^{\text{end}} -  t_{\text{PFI}}^{\text{start}}) \big) +  \Pr(t_{\text{PFI}}^{\text{end}} < t_{\text{PFI} \rightarrow \text{AE}}^{\text{end}})\E\big(t_{\text{PFI} \rightarrow \text{AE}}^{\text{end}} -  t_{\text{PFI}}^{\text{end}} \barBreak t_{\text{PFI}}^{\text{end}} < t_{\text{PFI} \rightarrow \text{AE}}^{\text{end}}\big) \\
		& = \E\big(\indicator{t_{\text{PFI}}^{\text{end}} < t_{\text{PFI} \rightarrow \text{AE}}^{\text{end}}}(t_{\text{PFI}}^{\text{end}} -  t_{\text{PFI}}^{\text{start}}) \big) +  \Pr(t_{\text{PFI}}^{\text{end}} < t_{\text{PFI} \rightarrow \text{AE}}^{\text{end}})\E\big(t_{\text{PFI} \rightarrow \text{AE}}^{\text{end}} -  t_{\text{PFI}}^{\text{start}}\big),
	\end{align*}
	where the last line in the display above holds for the same reason that \eqref{eq:lateAddOne} holds. The preceding result implies:
	\begin{align*}
		\E(&t_{\text{PFI} \rightarrow \text{AE}}^{\text{end}} - t_{\text{PFI}}^{\text{start}}) \ge \frac{\E\big(\indicator{t_{\text{PFI}}^{\text{end}} < t_{\text{PFI} \rightarrow \text{AE}}^{\text{end}}}(t_{\text{PFI}}^{\text{end}} -  t_{\text{PFI}}^{\text{start}}) \big)}{1 - \Pr(t_{\text{PFI}}^{\text{end}} < t_{\text{PFI} \rightarrow \text{AE}}^{\text{end}})} . 
	\end{align*}
	Finally, let $\mathcal{E}$ denote the event that the first two actions after $t_{\text{PFI}}^{\text{end}}$ are a departure, followed by a type-$H$ arrival. Note that $\mathcal{E}$ implies $t_{\text{PFI}}^{\text{end}} < t_{\text{PFI} \rightarrow \text{AE}}^{\text{end}}$, and thus
	\begin{align*}
		\E\big(&\indicator{t_{\text{PFI}}^{\text{end}} < t_{\text{PFI} \rightarrow \text{AE}}^{\text{end}}}(t_{\text{PFI}}^{\text{end}} -  t_{\text{PFI}}^{\text{start}}) \big) \\
		& \ge
		\Pr(\mathcal{E})\E\big(\indicator{t_{\text{PFI}}^{\text{end}} < t_{\text{PFI} \rightarrow \text{AE}}^{\text{end}}}(t_{\text{PFI}}^{\text{end}} -  t_{\text{PFI}}^{\text{start}}) \barBreak \mathcal{E}\big) \\
		& = \Pr(\mathcal{E})\E\big(t_{\text{PFI}}^{\text{end}} -  t_{\text{PFI}}^{\text{start}} \barBreak \mathcal{E}\big) \\
		& \ge 2\Omega(1)/(N\lambda_{H} + N\lambda_{L} + N\mu).
	\end{align*}
	The last line above holds because $\Pr(\mathcal{E}) = \Omega(1)$ and the PFI epoch spans two actions under $\mathcal{E}$. Finally, combining the preceding results implies line \eqref{eq:thisResultEstablishThat}. 
\end{proof}

\begin{proof}[Proof of Proposition \ref{l:littleoSSS}]
	The ratio's numerator is $\lambda_{H}/(\lambda_{H} + \lambda_{L} + \mu)$, because the corresponding integral equals 1 if the next action is a type-$H$ arrival and equals zero otherwise. Accordingly, it will suffice to establish that 
	\begin{align*}
		\E(t_{\text{SSS}}^{\text{end}} - t_{\text{SSS}}^{\text{start}}) = \Omega(1).	
	\end{align*}
	This fact would follow from line \eqref{eq:thisResultEstablishThat} of the proof of Proposition \eqref{l:serverIdlenessSSS}, if we could establish that 
	\begin{align*}
		\Pr(t_{\text{PFI} \rightarrow \text{AE}}^{\text{end}} \le t_{\text{PFI}}^{\text{end}}) = O(1/N).	
	\end{align*}
	To derive this result, note that
	\begin{align*}
		\Pr(&t_{\text{PFI} \rightarrow \text{AE}}^{\text{end}} \le t_{\text{PFI}}^{\text{end}}) \\
		& \le \Pr\big(t_{\text{PFI}}^{Y = \lfloor \sqrt{N}\rfloor -1} < t_{\text{PFI}}^{\text{end}}\big) +  
		\Pr\big(t_{\text{PFI}}^{\text{end}} < t_{\text{PFI}}^{Y = \lfloor \sqrt{N}\rfloor -1}\big)\Pr\big(t_{\text{PFI} \ne \text{PFI} \rightarrow \text{AE}} \le t_{\text{PFI}}^{\text{end}} \barBreak t_{\text{PFI}}^{\text{end}} < t_{\text{PFI}}^{Y = \lfloor \sqrt{N}\rfloor - 1}\big) \\
		& \le \Pr\big(t_{\text{PFI}}^{Y = \lfloor \sqrt{N}\rfloor - 1} < t_{\text{PFI}}^{\text{end}}\big) + \Pr\big(t_{\text{PFI} \ne \text{PFI} \rightarrow \text{AE}} \le t_{\text{PFI}}^{\text{end}},\ t_{\text{PFI}}^{\text{end}} < t_{\text{PFI}}^{Y = \lfloor \sqrt{N}\rfloor - 1}\big) / \Pr\big(t_{\text{PFI}}^{\text{end}} < t_{\text{PFI}}^{Y = \lfloor \sqrt{N}\rfloor - 1}\big) \\
		& \le \exp(-\Omega(\sqrt{N})) + 2\Pr\big(t_{\text{PFI} \ne \text{PFI} \rightarrow \text{AE}} \le t_{\text{PFI}}^{\text{end}} < t_{\text{PFI}}^{Y = \lfloor \sqrt{N}\rfloor - 1}\big),
	\end{align*}
	where $t_{\text{PFI} \ne \text{PFI} \rightarrow \text{AE}}$ is the first time that $Y_{\text{PFI}}$ and $Y_{\text{PFI} \rightarrow \text{AE}}$ decouple, and $t_{\text{PFI}}^{Y = \lfloor \sqrt{N}\rfloor - 1}$ is the first time after $t_{\text{PFI}}^{\text{start}}$ that $Y_{\text{PFI}} = \lfloor \sqrt{N}\rfloor - 1$. The first line above holds because $Y_{\text{PFI}}$ can't reach zero without first reaching $\lfloor \sqrt{N}\rfloor - 1$, and the last line above follows from line \eqref{eq:FinaExpBoundForFuture2} of the proof of Proposition \ref{l:serverIdleness}, which establishes that $\Pr(t_{\text{PFI}}^{Y = \lfloor \sqrt{N}\rfloor - 1} < t_{\text{PFI}}^{\text{end}}) = \exp(-\Omega(\sqrt{N}))$, and thus that $\Pr(t_{\text{PFI}}^{\text{end}} < t_{\text{PFI}}^{Y = \lfloor \sqrt{N}\rfloor - 1}) \ge 1/2$.
	
	The result above establishes that it will suffice to show that 
	\begin{align*}
		\Pr\big(t_{\text{PFI} \ne \text{PFI} \rightarrow \text{AE}} \le t_{\text{PFI}}^{\text{end}} < t_{\text{PFI}}^{Y = \lfloor \sqrt{N}\rfloor - 1}\big) = O(1/N).
	\end{align*}
	To derive this result, define $t_{\text{PFI}}^{\text{reject}}$ as the time of the first PFI rejection after $t_{\text{PFI}}^{\text{start}}$ (or, more technically, define it as the moment immediately before this rejection time), and define $t_{\text{PFI}}^{Y = 1}$ as the first time after $t_{\text{PFI}}^{\text{reject}}$ that $Y_{\text{PFI}} = 1$. I will first show that event $t_{\text{PFI} \ne \text{PFI} \rightarrow \text{AE}} \le t_{\text{PFI}}^{\text{end}} < t_{\text{PFI}}^{Y = \lfloor \sqrt{N}\rfloor - 1}$ will not hold unless AE rejects the job that PFI accepts at time $t_{\text{PFI}}^{\text{reject}}$. To see this, note that if AE rejects this job then both $Y_{\text{PFI} \rightarrow \text{AE}}$ and $Y_{\text{PFI}}$ will track the reject-all-$L$-jobs idle-server process until time $t_{\text{PFI}}^{Y = 1}$; further, condition $t_{\text{PFI}}^{\text{end}} < t_{\text{PFI}}^{Y = \lfloor \sqrt{N}\rfloor - 1}$ ensures that restriction $\hat{\tau}_{\lfloor\sqrt{N}\rfloor-1}<\hat{\tau}_{0}$ will still be binding at time $t_{\text{PFI}}^{Y = 1}$, which in turn implies that $t_{\text{PFI}}^{Y = 1} = t_{\text{PFI}}^{\text{end}}$. Hence, AE rejecting the job at time $t_{\text{PFI}}^{\text{reject}}$ ensures that $Y_{\text{PFI} \rightarrow \text{AE}}$ and $Y_{\text{PFI}}$ will remain coupled through $t_{\text{PFI}}^{\text{end}}$. 
	
	Next, since PFI acceptances imply AE acceptances, the pre-$t_{\text{PFI}}^{\text{reject}}$ history of the $Y_{\text{PFI} \rightarrow \text{AE}}$ process, $\mathcal{H}_{\text{PFI} \rightarrow \text{AE}}(t_{\text{PFI}}^{\text{reject}})$, is measurable with respect to the information set generated by the pre-$t_{\text{PFI}}^{\text{reject}}$ history of the $Y_{\text{PFI}}$ process, $\sigma(\mathcal{H}_{\text{PFI}}(t_{\text{PFI}}^{\text{reject}}))$. Hence, Lemma \ref{l:nearmemorylessness} still applies at time $t_{\text{PFI}}^{\text{reject}}$, even when we condition on both the $Y_{\text{PFI}}$ and $Y_{\text{PFI} \rightarrow \text{AE}}$ histories. Thus, if we define $t_{\text{PFI}}^{\text{reject}+}$ as the moment right after $t_{\text{PFI}}^{\text{reject}}$ (i.e., the moment right after the PFI rejection), then the distribution of the future random variables, $\Omega$, conditional on $\mathcal{H}_{\text{PFI} \rightarrow \text{AE}}(t_{\text{PFI}}^{\text{reject}})$ and $\mathcal{H}_{\text{PFI}}(t_{\text{PFI}}^{\text{reject}+})$ is $\mathcal{L}\big(\xi \barBreak \xi \in \mathcal{S}(\tau_{1}<\tau_{\lfloor\sqrt{N}\rfloor}) \cap \mathcal{S}(\hat{\tau}_{\lfloor\sqrt{N}\rfloor-1}<\hat{\tau}_{0}) \big)$. Let $\Pr_{\mathcal{L}_{y}(t_{\text{PFI}}^{\text{reject}})}$ denote probability with respect to this law, when $Y_{\text{PFI}}(t_{\text{PFI}}^{\text{reject}}) = y$. (Note that this law conditions on the PFI action at time $t_{\text{PFI}}^{\text{reject}}$, but not on the AE action at this time.)
	
	Next, line \eqref{eq:createConstantK}, from the proof of Proposition \ref{l:serverIdleness} establishes that there exists $C_{1}, C_{2} > 0$ such that
	\begin{align*}
		\Pr(Y_{\text{PFI}}(t_{\text{PFI}}^{\text{reject}}) > y) \le	
		C_{1}\exp(-C_{2}y).
	\end{align*}
	And since $t_{\text{PFI} \ne \text{PFI} \rightarrow \text{AE}} \le t_{\text{PFI}}^{\text{end}} < t_{\text{PFI}}^{Y = \lfloor \sqrt{N}\rfloor - 1}$ requires an AE acceptance at time $t_{\text{PFI}}^{\text{reject}}$, which in turn requires $w < \tau_{1}(t_{\text{PFI}}^{\text{reject}})$, we must therefore have
	\begin{align*}
		\Pr\big(&t_{\text{PFI} \ne \text{PFI} \rightarrow \text{AE}} \le t_{\text{PFI}}^{\text{end}} < t_{\text{PFI}}^{Y = \lfloor \sqrt{N}\rfloor - 1}\big) \\
		& \le \sum_{y = 2}^{\lfloor \sqrt{N}\rfloor - 2} \Pr(Y_{\text{PFI}}(t_{\text{PFI}}^{\text{reject}}) = y) \Pr_{\mathcal{L}_{y}(t_{\text{PFI}}^{\text{reject}})}(w < \tau_{1}(t_{\text{PFI}}^{\text{reject}})) \\
		& \le \Pr(Y_{\text{PFI}}(t_{\text{PFI}}^{\text{reject}}) > \log(N)/C_{2}) + \sum_{y = 1}^{\lfloor\log(N)/C_{2} \rfloor} \Pr(Y_{\text{PFI}}(t_{\text{PFI}}^{\text{reject}}) > y) \Pr_{\mathcal{L}_{y}(t_{\text{PFI}}^{\text{reject}})}(w < \tau_{1}(t_{\text{PFI}}^{\text{reject}})) \\
		& = C_{1}\exp(-C_{2}\log(N)/C_{2}) + \sum_{y = 1}^{\lfloor\log(N)/C_{2} \rfloor} C_{1}\exp(-C_{2}y) \Pr_{\mathcal{L}_{y}(t_{\text{PFI}}^{\text{reject}})}(w < \tau_{1}(t_{\text{PFI}}^{\text{reject}})) \\
		& \le  C_{1}/N + C_{1}(\exp(-C_{2}) - 1)^{-1}\max_{y \le \lfloor\log(N)/C_{2} \rfloor}\Pr_{\mathcal{L}_{y}(t_{\text{PFI}}^{\text{reject}})}(w < \tau_{1}(t_{\text{PFI}}^{\text{reject}})) .
	\end{align*}
	
	To conclude the proof, I must choose $w$ large enough to establish that $\Pr_{\mathcal{L}_{y}(t_{\text{PFI}}^{\text{reject}})}(w < \tau_{1}(t_{\text{PFI}}^{\text{reject}})) = O(1/N)$, uniformly over $y \le \lfloor\log(N)/C_{2} \rfloor$. To this end, first note that 
	\begin{align*}
		\Pr_{\mathcal{L}_{y}(t_{\text{PFI}}^{\text{reject}})}(w < \tau_{1}(t_{\text{PFI}}^{\text{reject}})) = \Theta(1) \Pr_{\hat{\mathcal{L}}_{y}(t_{\text{PFI}}^{\text{reject}})}(w < \tau_{1}(t_{\text{PFI}}^{\text{reject}})),
	\end{align*}
	where $\Pr_{\hat{\mathcal{L}}_{y}(t_{\text{PFI}}^{\text{reject}})}$ denotes probability with respect to law $\mathcal{L}(\xi \barBreak \xi \in \mathcal{S}(\tau_{1}<\tau_{\lfloor\sqrt{N}\rfloor}))$, evaluated with $Y_{\text{PFI}}(t_{\text{PFI}}^{\text{reject}}) = y$. This result follows closely from lines \eqref{eq:conditionalDistFraction} and \eqref{eq:boundDenom} of the proof of Proposition \ref{l:serverIdleness}. 
	
	Next, define $\chi(\tau)$ as the number of times that the reject-all-$L$-jobs idle-server process transitions between times $t_{\text{PFI}}^{\text{reject}}$ and $t_{\text{PFI}}^{\text{reject}} + \tau$. And with this, we have the following, for any given constant $C_{3} > 0$:
	\begin{align}
		\Pr_{\hat{\mathcal{L}}_{y}(t_{\text{PFI}}^{\text{reject}})}&(w < \tau_{1}) \nonumber\\
		& = \Pr_{\hat{\mathcal{L}}_{y}(t_{\text{PFI}}^{\text{reject}})}\big(\chi(w) < C_{3}\log(N) ,\, w < \tau_{1}\big) + \Pr_{\hat{\mathcal{L}}_{y}(t_{\text{PFI}}^{\text{reject}})}\big(\chi(w) \ge C_{3}\log(N),\, w < \tau_{1}\big)\nonumber\\
		& \le \Pr_{\hat{\mathcal{L}}_{y}(t_{\text{PFI}}^{\text{reject}})}(\chi(w) < C_{3}\log(N)) + \Pr_{\hat{\mathcal{L}}_{y}(t_{\text{PFI}}^{\text{reject}})}(\chi(\tau_{1}) \ge C_{3}\log(N)). \label{eq:xtautau}
	\end{align}
	
	I will now create $O(1/N)$ bounds for the two probabilities in \eqref{eq:xtautau}. To bound the first probability, suppose that $N$ is large enough so that the aggregate service rate is at least $N (\lambda_{H} + \mu) / 2$ when $Y < \lfloor \sqrt{N}\rfloor$. In this case, the total transition rate is at least $N (3\lambda_{H} + \mu) / 2 \ge 2\lambda_{H}N$ when $Y < \lfloor \sqrt{N}\rfloor$. And, thus, if we set $w \ge \frac{C_{3}\log(N)}{\lambda_{H}N}$ then the probability of fewer than $C_{3}\log(N)$ transitions in time window $w$ is no more than the probability of a Poisson with mean $2\lambda_{H}Nw \ge 2C_{3}\log(N)$ being less than $C_{3}\log(N)$, which is $O(1/N)$ when $C_{3} \ge 4$, by the Poisson Chernoff bound.
	
	To bound the second probability in \eqref{eq:xtautau}, note that Doob's $h$-transformation establishes that conditioning on $\tau_{1}<\tau_{\lfloor\sqrt{N}\rfloor}$ makes the probability of a jump in the reject-all-$L$-jobs idle-server process being an upward jump converge uniformly to $\lambda_{H}/(\lambda_{H} + \mu)$, for $Y < N^{1/3}$. And since $Y_{\text{PFI}}(t_{\text{PFI}}^{\text{reject}}) = y \le \lfloor\log(N)/C_{2} \rfloor \ll N^{1/3}$, the probability of $\chi(\tau_{1}) \ge C_{3} \log(N)$ under law $\mathcal{L}(\xi \barBreak \xi \in \mathcal{S}(\tau_{1}<\tau_{\lfloor\sqrt{N}\rfloor}))$ (and sufficiently large $N$), is no more than twice the probability of a random walk requiring more than $\lfloor C_{3}\log(N) \rfloor$ transitions to reach 1, conditional on it starting at $y$, incrementing with probability $\lambda_{H}/(\lambda_{H} + \mu)$ and decrementing with probability $\mu/(\lambda_{H} + \mu)$. This latter probability is no larger than the probability that at least $(\lfloor C_{3}\log(N) \rfloor-\lfloor\log(N)/C_{2} \rfloor)/2$ out of the next $\lfloor C_{3}\log(N) \rfloor$ transitions of the reject-all-$L$-jobs idle-server process are upward jumps. And now if we set $N$ sufficiently large and let $C_{3} \ge \frac{2(\lambda_{H} + \mu)}{C_{2}(\mu - \lambda_{H})}$ then this probability is no larger than the probability that at least a $\frac{3\lambda_{H} + \mu}{4\lambda_{H} + 4 \mu}$ fraction of the next $\lfloor C_{3}\log(N) \rfloor$ reject-all-$L$-jobs idle-server process transitions are upward jumps. This, in turn, equals the probability of a $\text{Binomial}(\lfloor C_{3}\log(N) \rfloor,\ \lambda_{H}/(\lambda_{H} + \mu))$ random variable exceeding $\lfloor C_{3}\log(N) \rfloor\frac{3\lambda_{H} + \mu}{4\lambda_{H} + 4 \mu}$, which is $O(1/N)$ when $C_{3} \ge \frac{4(7\lambda_{H} + \mu)(\lambda_{H} + \mu)}{(\mu - \lambda_{H})^{2}}$, by the multiplicative Chernoff bound for binomials. 
	
	Finally, line \eqref{eq:createConstantK} establishes that we can set $C_{2} \equiv \frac{(\lambda_{H} + \lambda_{L} - \mu)^{2}}{2(\lambda_{H} + \lambda_{L} + \mu)^{2}}$. Hence, the argument above holds when we set $w = \frac{C_{3}\log(N)}{\lambda_{H}N}$ and $C_{3} \equiv\frac{8(7\lambda_H+\mu)(\lambda_H+\mu)(\lambda_H+\lambda_L+\mu)^2}{(\mu-\lambda_H)^2(\lambda_H+\lambda_L-\mu)^2}$.
\end{proof}

\bibliographystyle{ormsv080}
\bibliography{library.bib}

\end{document}